\documentclass[fleqn, usenatbib]{mnras}
\usepackage{newtxtext,newtxmath}
\usepackage[T1]{fontenc}
\usepackage[utf8]{inputenc}
\usepackage{ae,aecompl}
\usepackage{graphicx}
\usepackage{xcolor}
\usepackage{soul}
\usepackage{amsmath}
\usepackage{commath}
\usepackage{siunitx}
\usepackage{nth}
\usepackage[capitalise, noabbrev]{cleveref}
\usepackage{enumitem}
\usepackage{csvsimple}
\usepackage{multirow}
\usepackage[acronym]{glossaries}
\hypersetup{unicode}

\crefname{equation}{equation}{equations}

\DeclareRobustCommand{\VAN}[3]{#2}
\let\VANthebibliography\thebibliography
\def\thebibliography{\DeclareRobustCommand{\VAN}[3]{##3}\VANthebibliography}

\bibpunct[]{(}{)}{;}{a}{}{,}
\definecolor{COS2987}{rgb}{0.8941176470588236, 0.10196078431372549, 0.10980392156862745}
\definecolor{COS3018}{rgb}{0.21568627450980393, 0.49411764705882355, 0.7215686274509804}
\definecolor{UVIS001}{rgb}{0.30196078431372547, 0.6862745098039216, 0.2901960784313726}
\definecolor{UVIS007}{rgb}{0.596078431372549, 0.3058823529411765, 0.6392156862745098}
\definecolor{UVIS019}{rgb}{1.0, 0.4980392156862745, 0.0}

\newcommand{\program}{\textsc}
\newcommand{\ssim}{\sim \!}

\newcommand{\lymana}{{Lyman-\ensuremath{\upalpha}}}

\newcommand{\lya}{{Ly\ensuremath{\upalpha}}}

\newcommand{\HII}{{\ion{H}{II}}}
\newcommand{\CII}{{[\ion{C}{II}]}}
\newcommand{\OIII}{{[\ion{O}{III}]}}

\newcommand{\NII}{{[\ion{N}{II}]}}
\newcommand{\CIIIs}{{\ion{C}{III}]}}
\newcommand{\CIIIf}{{[\ion{C}{III}]}}
\newcommand{\Hbeta}{{\ensuremath{\mathrm{H \upbeta}}}}

\newcommand{\CIILam}{{\CII\ 158 $\upmu\mathrm{{m}}$}}
\newcommand{\OIIILam}{{\OIII\ 88 $\upmu\mathrm{{m}}$}}
\newcommand{\NIILam}{{\NII\ 205 $\upmu\mathrm{{m}}$}}

\newacronym{EoR}{EoR}{Epoch of Reionisation}
\newacronym{ISM}{ISM}{interstellar medium}
\newacronym{IGM}{IGM}{intergalactic medium}
\newacronym{PDR}{PDR}{photodissociation region}
\newacronym{XDR}{XDR}{X-ray dominated region}
\newacronym{QSO}{QSO}{quasar}
\newacronym[longplural={luminous infrared galaxies}]{LIRG}{LIRG}{luminous infrared galaxy}
\newacronym[longplural={(ultra-)luminous infrared galaxies}]{ULIRG}{(U)LIRG}{(ultra-)luminous infrared galaxy}
\newacronym[longplural={hyper-luminous infrared galaxies}]{HyLIRG}{HyLIRG}{hyper-luminous infrared galaxy}
\newacronym[longplural={submillimetre galaxies}]{SMG}{SMG}{submillimetre galaxy}
\newacronym[longplural={Lyman-break galaxies}]{LBG}{LBG}{Lyman-break galaxy}
\newacronym[longplural={\lya\ emitting galaxies}]{LAE}{LAE}{\lya\ emitting galaxy}
\newacronym[shortplural={SNe}, longplural={supernovae}]{SN}{SN}{supernova}
\newacronym{CMB}{CMB}{Cosmic Microwave Background}
\newacronym{LMC}{LMC}{Large Magellanic Cloud}

\newacronym{LyC}{LyC}{Lyman-continuum}
\newacronym{RJ}{RJ}{Rayleigh-Jeans}

\newacronym{SFR}{SFR}{star formation rate}
\newacronym{sSFR}{sSFR}{specific star formation rate}
\newacronym{UV}{UV}{ultraviolet}
\newacronym{IR}{IR}{infrared}
\newacronym{NIR}{NIR}{near-infrared}
\newacronym{FIR}{FIR}{far-infrared}
\newacronym{SED}{SED}{spectral energy distribution}
\newacronym{EW}{EW}{equivalent width}
\newacronym{IMF}{IMF}{initial mass function}
\newacronym{AGB}{AGB}{asymptotic giant branch}
\newacronym{MZR}{MZR}{mass-metallicity relation}
\newacronym{IRX}{IRX}{infrared excess}

\newacronym{FWHM}{FWHM}{full width at half maximum}
\newacronym{SNR}{SNR}{signal-to-noise ratio}
\newacronym{RMS}{RMS}{root mean square}
\newacronym{PSF}{PSF}{point spread function}

\newacronym{ALMA}{ALMA}{Atacama Large Millimeter/submillimeter Array}
\newacronym{HST}{\textit{HST}}{\textit{Hubble Space Telescope}}
\newacronym{WFC3}{WFC3}{Wide Field Camera 3}
\newacronym{ACS}{ACS}{Advanced Camera for Surveys}
\newacronym{JWST}{\textit{JWST}}{\textit{James Webb Space Telescope}}
\newacronym{STScI}{STScI}{Space Telescope Science Institute}
\newacronym{CASA}{\program{casa}}{Common Astronomy Software Application}
\newacronym{REBELS}{REBELS}{Reionization Era Bright Emission Line Survey}
\newacronym{DGS}{DGS}{Dwarf Galaxy Survey}
\newacronym{GOALS}{GOALS}{Great Observatories All-sky LIRG Survey}
\newacronym{SPT}{SPT}{South Pole Telescope}

\newacronym{MERCURIUS}{\program{mercurius}}{Multimodal Estimation Routine for the Cosmological Unravelling of Rest-frame Infrared Uniformised Spectra}

\glsdisablehyper

\title[ALMA band-8 observations of luminous $z \sim 7$ LBGs]{Dual constraints with ALMA: new \OIIILam\ and dust-continuum observations reveal the ISM conditions of luminous LBGs at $z \sim 7$}

\author[J. Witstok et al.]{{Joris Witstok$^{1,2}$\thanks{E-mail: \href{mailto:jnw30@cam.ac.uk}{jnw30@cam.ac.uk}},
    Renske Smit$^{1,2,3}$\thanks{E-mail: \href{mailto:r.smit@ljmu.ac.uk}{r.smit@ljmu.ac.uk}},
    Roberto Maiolino$^{1,2,4}$, 
    Nimisha Kumari$^{5}$,
    }
    \newauthor{Manuel Aravena$^{6}$,
        Leindert Boogaard$^{7}$,
        Rychard Bouwens$^{9}$,
        Stefano Carniani$^{8}$,
    }
    \newauthor{Jacqueline A. Hodge$^{9}$,
        Gareth C. Jones$^{1,2,10}$,
        Mauro Stefanon$^{9}$,
        Paul van der Werf$^{9}$,
    }
    \newauthor{and Sander Schouws$^{9}$
    }
    \\
    $^{1}$Kavli Institute for Cosmology, University of Cambridge, Madingley Road, Cambridge CB3 0HA, UK
    \\
    $^{2}$Cavendish Laboratory, University of Cambridge, 19 JJ Thomson Avenue, Cambridge CB3 0HE, UK
    \\
    $^{3}$Astrophysics Research Institute, Liverpool John Moores University, 146 Brownlow Hill, Liverpool L3 5RF, UK
    \\
    $^{4}$Department of Physics and Astronomy, University College London, Gower Street, London WC1E 6BT, UK
    \\
    $^{5}$AURA for the European Space Agency, Space Telescope Science Institute, 3700 San Martin Drive, Baltimore, MD 21218, USA
    \\
    $^{6}$N\'ucleo de Astronom\'ia, Facultad de Ingenier\'ia y Ciencias, Universidad Diego Portales, Av. Ej\'ercito 441, Santiago, Chile
    \\
    $^{7}$Max Planck Institute for Astronomy, K\"{o}nigstuhl 17, 69117 Heidelberg, Germany
    \\
    $^{8}$Scuola Normale Superiore, Piazza dei Cavalieri 7, I-56126 Pisa, Italy
    \\
    $^{9}$Leiden Observatory, Leiden University, NL-2300 RA Leiden, Netherlands
    \\
    $^{10}$Department of Physics, University of Oxford, Denys Wilkinson Building, Keble Road, Oxford OX1 3RH, UK
}

\date{Accepted ---. Received ---; in original form ---}

\pubyear{2022}

\AtBeginDocument{
  \hypersetup{
    pdftitle={Dual constraints with ALMA},
    pdfauthor={J. Witstok et al.},
    pdfsubject={New [OIII] 88 μm and dust-continuum observations reveal the ISM conditions of luminous LBGs at z ~ 7},
    pdfkeywords={{galaxies: high-redshift} -- {dark ages, reionization, first stars} -- {methods: observational} -- {submillimetre: ISM} -- {techniques: spectroscopic} -- {ISM: dust}}
  }
}

\begin{document}
\label{firstpage}
\pagerange{\pageref{firstpage}--\pageref{lastpage}}
\maketitle

\glsunset{GOALS}

\begin{abstract}
    We present new \OIIILam\ observations of five bright $z \sim 7$ Lyman-break galaxies spectroscopically confirmed by ALMA through \CIILam, unlike recent \OIII\ detections where \lymana\ was used. This nearly doubles the sample of Epoch of Reionisation galaxies with robust ($5 \sigma$) \CII\ and \OIII\ detections. We perform a multi-wavelength comparison with new deep \textit{HST} images of the rest-frame UV, whose compact morphology aligns well with \OIII\ tracing ionised gas. By contrast, we find more spatially extended \CII\ emission likely produced in neutral gas, as indicated by a \NIILam\ non-detection in one source. We find a correlation between the optical $\OIII + \Hbeta$ equivalent width and \OIII/\CII, as seen in local metal-poor dwarf galaxies. \program{Cloudy} models of a nebula of typical density harbouring a young stellar population with a high ionisation parameter adequately reproduce the observed lines. Surprisingly, however, our models fail to reproduce the strength of \OIIILam, unless we assume an $\mathrm{\upalpha/Fe}$ enhancement and near-solar nebular oxygen abundance. On spatially resolved scales, we find \OIII/\CII\ shows a tentative anti-correlation with infrared excess, $L_\text{IR}/L_\text{UV}$, also seen on global scales in the local Universe. Finally, we introduce the far-infrared spectral energy distribution fitting code \program{mercurius} to show that dust-continuum measurements of one source appear to favour a low dust temperature and correspondingly high dust mass. This implies a high stellar metallicity yield and may point towards the need of dust production or grain-growth mechanisms beyond supernovae.
\end{abstract}


\section{Introduction}
\label{sec:Introduction}

The \gls{EoR} marks a critical turning point in the early Universe and its study is one of the frontiers in modern astrophysics. During the \gls{EoR}, the first galaxies emerged and started to rapidly form stars, which in turn began to ionise the surrounding gas -- first the \gls{ISM}, and eventually the \glsentrylong{IGM} \citep[\glsentryshort{IGM}\glsunset{IGM};][]{2018PhR...780....1D, 2021arXiv211013160R}. Compared to their present-day counterparts, these galaxies have not had much time to build up an abundance of metals \citep[e.g.][]{2019A&ARv..27....3M}. This suggests they have metal-poor stellar populations, which results in an enhanced output of \gls{UV} photons due to the hardened spectra of O- and B-type stars, in addition to containing a reduced amount of dust, implying these early galaxies generally experience less absorption of \gls{UV} radiation \citep{2016ARA&A..54..761S}.

Regions where star formation takes place, especially in the very early Universe, therefore harbour stellar radiation fields with a strong intrinsic flux of ionising \gls{UV} photons. Deep broadband surveys in the optical and \gls{IR} by the \gls{HST} have indeed proven an effective method of finding star-forming \gls{EoR} galaxies via their rest-frame \gls{UV} emission, having identified a considerable number through the Lyman-break technique \citep[currently nearly \num{2000} candidates at $z \geq 6$; see e.g.][]{2021AJ....162...47B}.

Even at the earliest stages of galaxy evolution, however, dust can be of major influence. Dust grains not only catalyse star formation through the formation of molecules \citep{2018MNRAS.474.1545C} and fragmentation of gas \citep{2006MNRAS.369.1437S}, but they also obscure our view of these \gls{UV}-bright star-forming regions. Dust is able to absorb a significant proportion of stellar optical and \gls{UV} light, re-emitting the absorbed energy as thermal \gls{IR} radiation. It therefore poses observational challenges to the inference of \glspl{SFR} solely from \gls{UV} and optical measurements \citep{2012ARA&A..50..531K, 2014ARA&A..52..415M, 2020ApJ...902..112B}. Indeed, recent work renewed attention on an existing notion \citep{2020RSOS....700556H}: that a non-trivial fraction of high-redshift galaxies ($z \gtrsim 4$) may be so-called ``\gls{HST}-dark'' systems which, even in the deepest \gls{HST} imaging (reaching $H \sim 27 \, \mathrm{mag}$), appear completely obscured at observed optical and \gls{NIR} wavelengths \citep{2018A&A...620A.152F, 2019ApJ...884..154W, 2019Natur.572..211W, 2021Natur.597..489F, 2021ApJ...923..215C, 2022ApJ...925...23M}. But while dust complicates the interpretation in the rest-frame \gls{UV} and optical, its thermal emission also serves as a probe of the mass of the \gls{ISM} \citep[e.g.][]{2017ApJ...837..150S}.

In recent years, the \gls{ALMA} has opened a new observational window for the study of star-forming galaxies at high redshift \citep[see][ for a review]{2020RSOS....700556H}. \gls{ALMA} is uniquely positioned to observe their \gls{FIR} emission at (sub)millimetre wavelengths, which serves a twofold purpose. First, these efforts allow \gls{ALMA} to directly detect the dust continuum emission. The first explorations of the \gls{FIR} \gls{SED} of galaxies in the \gls{EoR} with \gls{ALMA} indicate that they may rapidly build up considerable amounts of dust \citep[e.g.][]{2015Natur.519..327W, 2017ApJ...837L..21L}. Second, nebular emission lines enable crucial spectroscopic redshift confirmations and offer valuable insights into \gls{ISM} conditions by probing, for instance, the ionisation state, metal enrichment, and kinematics of the gas \citep[see][ for a review]{2019ARA&A..57..511K}.

Nebular line emission typically arises in gas surrounding hot and massive stars. Gas in close vicinity forms a photoionised \HII\ region populated by species with an ionisation potential higher than hydrogen (e.g. O$^{2+}$, which requires $\ssim 35 \, \mathrm{eV}$ to form), whereas shielded gas further out will contain low-ionisation species such as C$^+$ (the ionisation potential of neutral carbon, $11.3 \, \mathrm{eV}$, is just below that of hydrogen, $13.6 \, \mathrm{eV}$; \citealt{2005AAS...207.8117A}). These regions can transition into \glspl{PDR} -- defined as mostly neutral gas where \gls{UV} photons still play a significant role in the chemistry and/or heating -- in which carbon is partly photoionised \citep[instead of hydrogen, as in \HII\ regions;][]{1999RvMP...71..173H}. The variety of ionised and neutral gas reservoirs comprising the \gls{ISM} can be traced by the many \gls{ISM} coolants in the \gls{FIR}, notably the \CIILam\ and \OIIILam\ fine-structure lines \citep[\CII\ and \OIII\ hereafter; see e.g.][]{2019PASJ...71...71H, 2019MNRAS.487.5902K, 2019MNRAS.487.1689P, 2020ApJ...896...93H, 2022ApJ...931..160B}. \OIII\ is almost exclusively produced in \HII\ regions, while \CII\ emission can arise in both \HII\ regions and \glspl{PDR}. Other \gls{FIR} emission lines further complement the picture: for instance, the \NIILam\ transition is a powerful proxy of the ionisation state of hydrogen in the \CII-producing \gls{ISM} \citep[e.g.][]{2012A&A...542L..34N, 2014ApJ...782L..17D}.

The \CII\ and \OIII\ lines shift into high-frequency coverage of \gls{ALMA} with sufficient atmospheric transmission (i.e. band 8) above redshifts $z \sim 2.8$ and $z \sim 5.8$, respectively. For galaxies in the \gls{EoR}, where ground-based observing facilities are restricted to spectroscopic studies of weaker rest-frame \gls{UV} lines such as the $\CIIIs \, \lambda \, 1907, \CIIIf \, \lambda \, 1909 \, \Angstrom$ doublet \citep[e.g.][]{2021ApJ...917L..36T}, observations of the \CII\ and \OIII\ with \gls{ALMA} have first of all proven an effective spectroscopic confirmation tool, with almost as many UV-bright galaxies at $z > 6.5$ now having been spectroscopically confirmed via the \CII\ line as with \ion{H}{I} \lymana\ \citep[\lya;][]{2022ApJ...931..160B}. Moreover, the \OIII\ and \CII\ lines offer a powerful way of exploring \gls{ISM} properties at high redshift, even in ``normal'' star-forming galaxies (i.e. $\text{\gls{SFR}} \lesssim 100 \, \mathrm{M_\odot \, yr^{-1}}$).\footnote{For example, \citet{2018Natur.553..178S} presented \CII\ detections of two $z \sim 7$ photometric galaxy candidates using a combined on-source time of less than an hour, while typical spectroscopic observations of rest-frame \gls{UV} lines require multiple hours of integration time on a $10 \, \mathrm{m}$-class telescope.} Finally, owing to its interferometric nature, \gls{ALMA} produces spatially resolved spectroscopic measurements, in contrast to unresolved slit spectroscopy, which additionally faces the undesirable effect of slit losses.

In local starburst galaxies, \CII\ is observed to be the dominant \gls{FIR} line, while in metal-poor dwarfs \OIII\ takes over this role ($L_\mathrm{[OIII]}/L_\mathrm{[CII]} > 1$; e.g. \citealt{2020ApJ...896...93H}). Critically, a large filling factor of diffuse, ionised gas emitting \OIIILam\ means metal-poor systems have a more porous \gls{ISM} through which ionising radiation could ``leak'' \citep{2015A&A...578A..53C}. The relative strengths of emission lines, in particular the \OIII\ and \CII\ lines, are thus a powerful indicator of the physical state of the \gls{ISM}. Interestingly, recent observations of galaxies in the \gls{EoR} have systematically revealed \OIII/\CII\ ratios similar to or even exceeding those in local metal-poor dwarf galaxies \citep[e.g.][]{2020MNRAS.499.5136C}.

In summary, specific properties of the \gls{ISM} that can be derived from combined observations of line and continuum emission with \gls{ALMA} include the dust mass and temperature, probed through its thermal emission \citep[e.g.][]{2020MNRAS.493.4294B, 2021MNRAS.508L..58B, 2022ApJ...928...31S}, as well as the temperature, density, ionisation \citep[e.g.][]{2019MNRAS.489....1F, 2019MNRAS.487.1689P, 2021MNRAS.505.5543V}, metal enrichment \citep[e.g.][]{2015ApJ...813...36V}, and gas kinematics \citep[e.g.][]{2021MNRAS.507.3540J}, all of which can be inferred from emission line strengths and spectral profiles. In turn, a robust understanding of the \gls{ISM} conditions in typical \gls{EoR} star-forming galaxies -- that is, both of their dust and gas content -- is crucial in constructing a complete physical picture of star formation in this earliest epoch and therefore also of the process of cosmic reionisation.

\begingroup
    \renewcommand{\arraystretch}{1.25} % Default value: 1
    \begin{table*}
        \centering
        \caption[Sources studied in this work.]
        {Sources studied in this work (colours for reference). Coordinates are given as right ascension ($\alpha_\text{J2000}$), declination ($\delta_\text{J2000}$), and redshift ($z$). These are followed by general properties of the galaxies: the apparent magnitude ($m_\text{\gls{UV}}$) and luminosity ($L_\text{\gls{UV}}$) of the rest-frame \gls{UV}, the stellar mass ($M_*$), and the \gls{EW} of the optical \OIII\ and \Hbeta\ lines, $\text{\gls{EW}} ( \OIII + \Hbeta )$, as presented in the works by \citet{2015ApJ...801..122S, 2018Natur.553..178S} and \citet{2022arXiv220204080S, 2022ApJ...928...31S}.}
        \label{tab:Targets}
        \begin{tabular}{llllllll}
            \hline
            Source & $\alpha_\text{J2000} \, (\mathrm{h})$ & $\delta_\text{J2000} \, (\mathrm{deg})$ & $z$ & $m_\text{\gls{UV}} \, (\mathrm{mag})$ & $L_\text{\gls{UV}} \, (10^{11} \, \mathrm{L_\odot})$ & $M_* \, (10^9 \, \mathrm{M_\odot})$ & $\text{\gls{EW}} ( \OIII + \Hbeta ) \, (\Angstrom)$ \\
            \hline
            % \textit{COSMOS} & & & & & & & \\
            \begin{filecontents*}{ALMA_COSMOS_targets.csv}
field|object|RA|Dec|z|mUV|LUV|M|EW
COSMOS|\setulcolor{COS2987}\ul{COS-2987030247}|10:00:29.8700|02:13:02.470|6.8076|$24.8 \pm 0.1$|$1.3 \pm 0.1$|$1.70_{-0.18}^{+0.49}$|$1128_{-166}^{+166}$
COSMOS|\setulcolor{COS3018}\ul{COS-3018555981}|10:00:30.1850|02:15:59.810|6.8540|$24.9 \pm 0.1$|$1.1 \pm 0.1$|$1.38_{-0.21}^{+0.71}$|$1424_{-143}^{+143}$
\end{filecontents*}
\csvreader[separator=pipe, late after line=\\, head to column names]{ALMA_COSMOS_targets.csv}{}{\object & \RA & \Dec & \z & \mUV & \LUV & \M & \EW}
            % \textit{UVISTA} & & & & & & & \\
            \begin{filecontents*}{ALMA_UVISTA_targets.csv}
field|object|RA|Dec|z|mUV|LUV|M|EW
UVISTA|\setulcolor{UVIS001}\ul{UVISTA-Z-001}|10:00:43.3600|02:37:51.300|7.0611|$23.9 \pm 0.1$|$2.9 \pm 0.1$|$3.80_{-2.10}^{+0.88}$|$1004_{-206}^{+442}$
UVISTA|\setulcolor{UVIS007}\ul{UVISTA-Z-007}|09:58:46.2100|02:28:45.800|6.7498|$24.5 \pm 0.1$|$1.5 \pm 0.2$|$3.72_{-2.37}^{+4.60}$|$761_{-168}^{+530}$
UVISTA|\setulcolor{UVIS019}\ul{UVISTA-Z-019}|10:00:29.8900|01:46:46.400|6.7544|$25.1 \pm 0.2$|$1.0 \pm 0.1$|$3.24_{-1.10}^{+1.78}$|$628_{-99}^{+226}$
\end{filecontents*}
\csvreader[separator=pipe, late after line=\\, head to column names]{ALMA_UVISTA_targets.csv}{}{\object & \RA & \Dec & \z & \mUV & \LUV & \M & \EW}
            \hline
        \end{tabular}
    \end{table*}
\endgroup

\begingroup
    \setlength{\tabcolsep}{5pt} % Default value: 6pt
    \renewcommand{\arraystretch}{1} % Default value: 1
    \begin{table*}
        \centering
        \caption[Overview of \gls{ALMA} observations of the five $z \sim 7$ galaxies.]
        {Overview of \gls{ALMA} observations of the five $z \sim 7$ galaxies. For each emission line observed, the observed frequency ($\nu_\text{obs}$), total on-source integration time ($t_\text{int}$), and channel width ($\Delta \nu_\text{obs}$) are shown. The first indicated beam size, $A_\text{beam}$ (given as the \glspl{FWHM} of the major and minor axes), is the one tuned to match the beam between \CII\ and \OIII\ as closely as possible, which is achieved by the listed weighting scheme of either natural or Briggs weighting, an (optional) $uv$ taper, and the robust parameter (for Briggs weighting; see also \cref{sssec:Observations:ALMA_data_reduction}). The second beam size indicated for a line is one using natural weighting without tapering. The \gls{RMS} noise (per channel of the given width) in naturally weighted images is shown in the second to last column, in good agreement with the theoretically predicted sensitivity (in brackets) computed for use in the \program{tclean} task (\cref{sssec:Observations:ALMA_data_reduction}). The final column lists the \gls{ALMA} project codes (\cref{sssec:Observations:ALMA_programmes}).}
        \label{tab:Observations}
        \begin{tabular}{lllllp{2.95cm}p{2cm}p{3.75cm}}
            \hline
            Source & Emission line & $\nu_\text{obs}$ & $t_\text{int}$ & $\Delta \nu_\text{obs}$ & Weighting scheme: $A_\text{beam}$ & RMS (sensitivity) & ALMA project \\
            & & ($\mathrm{GHz}$) & ($\mathrm{h}$) & ($\mathrm{MHz}$) & & ($\mathrm{\upmu Jy/beam}$) & code(s) \\
            \hline
            \begin{filecontents*}{ALMA_COSMOS_observations.csv}
field|object|z|emline|MS|freq|chanwidth|int|ant|lbl|res|beamsize|RMS|sens|codes
COSMOS|\setulcolor{COS2987}\ul{COS-2987030247}|6.8076|{[\ion{C}{II}]} 158 $\upmu\mathrm{m}$|5|243.429|40|3.6|45|1397.8|0.182|Natural ($0.4 \arcsec$): $0.7 \arcsec \times 0.6 \arcsec$\newline Natural: $0.4 \arcsec \times 0.4 \arcsec$|\newline 76.3|89.5|2015.1.01111.S, 2018.1.01359.S
|||{[\ion{O}{III}]} 88 $\upmu\mathrm{m}$|2|434.575|70|1.1|48|360.6|0.395|Briggs ($0.5$): $0.7 \arcsec \times 0.5 \arcsec$\newline Natural: $0.8 \arcsec \times 0.6 \arcsec$|\newline 487|548|2018.1.00429.S
|||{[\ion{N}{II}]} 205 $\upmu\mathrm{m}$|2|186.996|30|1.3|46|783.5|0.422|Natural: $0.8 \arcsec \times 0.7 \arcsec$|166|202|2018.1.01551.S
COSMOS|\setulcolor{COS3018}\ul{COS-3018555981}|6.8540|{[\ion{C}{II}]} 158 $\upmu\mathrm{m}$|17|241.991|40|14|44|1397.8|0.183|Natural ($0.4 \arcsec$): $0.6 \arcsec \times 0.6 \arcsec$\newline Natural: $0.4 \arcsec \times 0.4 \arcsec$|\newline 45.0|48.7|2015.1.01111.S, 2017.1.00604.S
|||{[\ion{O}{III}]} 88 $\upmu\mathrm{m}$|2|432.008|70|0.99|47|360.6|0.397|Briggs ($0.5$): $0.7 \arcsec \times 0.5 \arcsec$\newline Natural: $0.9 \arcsec \times 0.6 \arcsec$|\newline 429|456|2018.1.00429.S
\end{filecontents*}
\csvreader[separator=pipe, late after line=\\, head to column names]{ALMA_COSMOS_observations.csv}{}{\ifcsvstrcmp{\field}{COSMOS}{\csviffirstrow{}{& & & & & & & \\}}{} \object & \emline & \freq & \int & \chanwidth & \beamsize & $\RMS \, (\sens)$ & \codes}
            \begin{filecontents*}{ALMA_UVISTA_observations.csv}
field|object|z|emline|MS|freq|chanwidth|int|ant|lbl|res|beamsize|RMS|sens|codes
UVISTA|\setulcolor{UVIS001}\ul{UVISTA-Z-001}|7.0611|{[\ion{C}{II}]} 158 $\upmu\mathrm{m}$|20|235.774|40|4.2|47|1397.8|0.188|Natural: $0.6 \arcsec \times 0.5 \arcsec$|234|119|2015.1.00540.S, 2018.1.00085.S, 2018.1.00933.S, 2019.1.01611.S
|||{[\ion{O}{III}]} 88 $\upmu\mathrm{m}$|2|420.909|70|1.3|48|500.2|0.294|Natural ($0.2 \arcsec$): $0.6 \arcsec \times 0.5 \arcsec$\newline Natural: $0.5 \arcsec \times 0.5 \arcsec$|\newline 435|465|2019.1.01524.S
UVISTA|\setulcolor{UVIS007}\ul{UVISTA-Z-007}|6.7498|{[\ion{C}{II}]} 158 $\upmu\mathrm{m}$|2|245.244|40|0.44|46|313.7|0.804|Briggs ($1.0$): $1.3 \arcsec \times 1.1 \arcsec$\newline Natural: $1.4 \arcsec \times 1.2 \arcsec$|\newline 272|286|2018.1.00085.S
|||{[\ion{O}{III}]} 88 $\upmu\mathrm{m}$|2|437.817|70|1.6|46|500.2|0.282|Natural ($1.0 \arcsec$): $1.1 \arcsec \times 1.1 \arcsec$\newline Natural: $0.6 \arcsec \times 0.5 \arcsec$|\newline 738|641|2019.1.01524.S
UVISTA|\setulcolor{UVIS019}\ul{UVISTA-Z-019}|6.7544|{[\ion{C}{II}]} 158 $\upmu\mathrm{m}$|7|245.099|40|4.4|45|1547.1|0.163|Natural ($0.2 \arcsec$): $0.4 \arcsec \times 0.4 \arcsec$\newline Natural: $0.4 \arcsec \times 0.3 \arcsec$|\newline 81.2|92.3|2018.1.00085.S, 2019.1.01611.S
|||{[\ion{O}{III}]} 88 $\upmu\mathrm{m}$|2|437.557|70|0.89|45|500.2|0.283|Briggs ($1.5$): $0.5 \arcsec \times 0.4 \arcsec$\newline Natural: $0.5 \arcsec \times 0.4 \arcsec$|\newline 1015|680|2019.1.01524.S
\end{filecontents*}
\csvreader[separator=pipe, late after line=\\, head to column names]{ALMA_UVISTA_observations.csv}{}{\ifcsvstrcmp{\field}{UVISTA}{& & & & & & & \\}{} \object & \emline & \freq & \int & \chanwidth & \beamsize & $\RMS \, (\sens)$ & \codes}
            \hline
        \end{tabular}
    \end{table*}
\endgroup

The main aim of this work is to analyse \gls{ALMA} observations of the \CIILam, \OIIILam, \NIILam, and underlying dust continuum emission for a sample of five luminous, star-forming \glspl{LBG} at $z \sim 7$. Specifically, we present new observations of the \OIIILam\ line with \gls{ALMA} in addition to new \gls{HST} imaging, which significantly increases the number of \gls{EoR} sources detected in both \CII\ and \OIII\ \citep[currently $9$, while only $5$ have detections of at least $5 \sigma$ in both lines;][]{2020MNRAS.499.5136C}. Moreover, the sample considered in this work arguably has a weaker selection bias since, in contrast to the existing studies, all galaxies have previously been spectroscopically confirmed through \CII\ instead of \lya. Before reionisation is completed, the latter can namely only escape from galaxies occupying ionised bubbles, which are indicative of overdense regions \citep[e.g.][]{2021arXiv211207675L}.

Our outline is as follows. \Cref{sec:Observations} discusses the available \gls{ALMA} and \gls{HST} data and \cref{sec:Results} briefly discusses the results of these observations. In \cref{sec:Discussion:Dust_properties,sec:Discussion:Emission_line_properties}, we compare our findings concerning respectively the dust continuum and emission lines to other works in the literature and discuss our findings. \Cref{sec:Summary} provides a summary.

In our analysis, we adopt the cosmological parameters $\Omega_\text{m} = 0.3$, $\Omega_\Lambda = 0.7$, and $H_0 = 70 \, \mathrm{km \, s^{-1} \, Mpc^{-1}}$ throughout (implying an angular scale of $5.2 \, \mathrm{kpc/arcsec}$ at $z=7$). All magnitudes are in the AB system \citep{1983ApJ...266..713O}.

\section{Observations}
\label{sec:Observations}

\subsection{Target selection}
\label{ssec:Observations:Target_selection}

Five galaxies at $z \sim 7$, all located in the $\ssim 2 \, \mathrm{deg^2}$ UltraVISTA field \citep[UVISTA;][]{2012A&A...544A.156M}, are considered in this work: COS-2987030247, COS-3018555981, UVISTA-Z-001, UVISTA-Z-007, and UVISTA-Z-019 (\cref{tab:Targets}). The first two, COS-2987030247 and COS-3018555981, are contained in the \gls{HST} CANDELS field \citep{2011ApJS..197...35G} and were spectroscopically confirmed with \gls{ALMA} observations of \CIILam\ in Cycle 3 \citep[\gls{ALMA} project code 2015.1.01111.S;][]{2018Natur.553..178S}. The two sources were selected for their large optical emission line \glspl{EW}, which enables an accurate photometric redshift determination using \textit{Spitzer}/IRAC broadband photometry at $3.6 \, \upmu\mathrm{{m}}$ and $4.5 \, \upmu\mathrm{{m}}$ \citep[see][ for details]{2015ApJ...801..122S}. The \glspl{EW} of the optical \OIII\ (notably at $4960 \, \Angstrom$ and $5008 \, \Angstrom$) and \Hbeta\ lines (hereafter simply $\text{\gls{EW}} ( \OIII + \Hbeta )$), presented in \citet{2015ApJ...801..122S}, are $\text{\gls{EW}} ( \OIII + \Hbeta ) > \num{1000} \, \Angstrom$, while the median \gls{EW} at $z \sim 7$ is around $\num{1000} \, \Angstrom$ \citep[see e.g.][]{2021MNRAS.500.5229E}. All five sources are bright in the \gls{UV}, with apparent magnitudes between $24$ and $25 \, \mathrm{mag}$, implying $L_\text{\gls{UV}} \sim 2 L_\text{\gls{UV}}^* (z \sim 7)$. \gls{SED} fitting with a \citet{2003PASP..115..763C} \gls{IMF} yields stellar masses between $1$ and $4$ times $10^9 \, \mathrm{M_\odot}$ \citep[; see \cref{tab:Targets}]{2018Natur.553..178S, 2022ApJ...928...31S}.

The precise photometric redshift allowed \gls{ALMA} to perform efficient blind spectral scans to search for the \CII\ line without first requiring a spectroscopic confirmation through \lya. Its success encouraged a second programme in Cycle 6 \citep[2018.1.00085.S;][]{2022arXiv220204080S} that has successfully detected the \CII\ line in three out of six additional galaxies, selected from ground-based imaging in the wider COSMOS field. Among these are the latter three sources considered here (UVISTA-Z-001, UVISTA-Z-007, and UVISTA-Z-019), two of which in comparison have modest \glspl{EW} \citep[presented in][]{2022arXiv220204080S}, $\text{\gls{EW}} ( \OIII + \Hbeta ) \sim 600$-$700 \, \Angstrom$, more typical for \gls{UV}-bright LBGs at $z\sim7$ \citep[e.g.][]{2021MNRAS.500.5229E}. For an in-depth description of \gls{ALMA} target selection strategies and scanning efficacy of \CII\ we refer to \citet{2022ApJ...931..160B}, introducing the successor of these two pilot programmes, the \glsentryfull{REBELS}\glsunset{REBELS}. This \gls{ALMA} Large Program has targeted \CII\ and \OIII\ in $40$ of the brightest \gls{UV}-selected star-forming galaxy candidates at $z > 6.5$.

\subsection{ALMA data}
\label{ssec:Observations:ALMA}

An overview of the \gls{ALMA} observations considered in this work is shown in \cref{tab:Observations}. In the following sections, we will briefly describe the existing \gls{ALMA} observations of the \CIILam\ line, and the data reduction process of new measurements of the \OIIILam\ and \NIILam\ lines and the underlying dust continuum.

\subsubsection{ALMA programmes used in this study}
\label{sssec:Observations:ALMA_programmes}

In this study, we consider three main follow-up programmes of the bright \glspl{LBG} that were confirmed with \CII\ described in \cref{ssec:Observations:Target_selection}. COS-2987030247 was observed in band 5, targeting the \NIILam\ line with programme 2018.1.01551.S in Cycle 6 (note that for COS-3018555981 \NIILam\ falls into the atmospheric feature at the centre of band 5 and could not be included in this programme). All five bright \CII\ emitting galaxies described in \cref{ssec:Observations:Target_selection}, were approved for A-ranked observations in band 8 in Cycle 6 (project code 2018.1.00429.S) and 7 (2019.1.01524.S) for the CANDELS-COSMOS and UVISTA sources, respectively.

Finally, we also make use of additional \CII\ data that was taken at high angular resolution ($\theta_\text{beam} < 0.2 \arcsec$), co-adding these with the low-resolution data for all sources except UVISTA-Z-007 (2017.1.00604.S, 2018.1.01359.S, 2019.1.01611.S). Similarly, we include additional band-6 observations of the dust continuum in UVISTA-Z-001 (2015.1.00540.S, 2018.1.00933.S) presented in \citet{2018MNRAS.481.1631B, 2022MNRAS.510.5088B}.

\subsubsection{ALMA data reduction}
\label{sssec:Observations:ALMA_data_reduction}

All data, including previous observations discussed above (see \cref{tab:Observations}), were calibrated and reduced with the automated pipeline of the \glsentrylong{CASA} \citep[\glsentryshort{CASA}\glsunset{CASA};][]{2007ASPC..376..127M} version 5.6 (with the exception of 2015.1.00540.S and 2015.1.01111.S, which required version 4.7).

Before the imaging step, we performed the \program{uvcontsub} task to separate the continuum and line visibilities specifically for imaging the \CII\ emission in COS-3018555981, UVISTA-Z-001, and UVISTA-Z-019 as well as the \OIII\ emission in UVISTA-Z-001, where the underlying dust continuum is confidently detected \citep[\cref{ssec:Results:Dust_continuum}; see also][]{2022ApJ...928...31S}. Images were made with the \program{tclean} task for a large range of parameter sets, using all available data sets combined. We created images using both natural and Briggs weightings; under the natural weighting scheme, several different $uv$ tapers were considered as well as no taper. For measuring the total \OIII\ and \CII\ line fluxes, a moderately ($0.6 \arcsec$) tapered image was chosen to avoid resolving out extended emission \citep[see e.g.][]{2018MNRAS.478.1170C, 2020MNRAS.499.5136C} as best as possible while maintaining a good \gls{SNR}. However, a natural weighting without any tapering was considered for the \NII\ line to maximise the \gls{SNR}. Continuum images were created by averaging (i.e. the \program{mfs} mode in \program{tclean}) over all available spectral channels that were well outside the relevant emission line range (i.e. $\abs{v} > \num{1000} \, \mathrm{km/s}$). Natural weighting was again used for optimal \gls{SNR}, however given the heterogeneity of the \gls{ALMA} data, the following tapers were applied on a case-by-case basis to match the beams across various bands: $0.4 \arcsec$ for the band-6 data of COS-2987030247, $0.6 \arcsec$ for the band-6 data of COS-3018555981, $0.8 \arcsec$ for the band-8 data of UVISTA-Z-001, $1.0 \arcsec$ for the band-8 data of UVISTA-Z-007, and $0.2 \arcsec$ for the band-6 data of UVISTA-Z-019.

Finally, when comparing the \CII\ and \OIII\ lines, we also tuned the weighting or taper scheme to match the beam sizes as closely as possible. We use Briggs weighting for the line observed with lowest spatial resolution with a robust parameter tuned to the highest resolution achievable while maintaining a reasonable \gls{SNR}. The other line is imaged with natural weighting and, if required, a small taper. \Cref{tab:Observations} lists the resulting matched beam sizes for \CII\ and \OIII\ (as well as beam sizes obtained with an untapered, natural-weighting scheme). Furthermore, the measured \glsdisp*{RMS}{root-mean-square (RMS)} noise in naturally weighted images is compared in \cref{tab:Observations} to the theoretically expected value of the sensitivity of an interferometric image \citep{2017isra.book.....T},\footnote{We note this is a simplified formula which leaves out a few additional system efficiencies; see Section 9.2.1 of the \gls{ALMA} Technical Handbook \citep{ALMA_technical_handbook} for a more detailed form.}
\begin{equation}
    \label{eq:Interferometry_sensitivity}
    \delta S_\nu = \frac{2 k_\text{B} T_\text{sys}}{A_\text{eff} \sqrt{N_\text{ant} (N_\text{ant} - 1) t_\text{int} \Delta \nu}} ,
\end{equation}

\noindent where $k_\text{B}$ is the Boltzmann constant, $N_\text{ant}$ is the number of antennas (where we take the mean if observations obtained with multiple configurations are combined), $t_\text{int}$ is the total on-source integration time, $\Delta \nu$ is the channel width, $T_\text{sys}$ is the system temperature, and $A_\text{eff}$ is the effective collective area of each dish. Given the dish radius $r_\text{ant}$,
\begin{equation*}
    A_\text{eff} = \eta \pi r_\text{ant}^2,
\end{equation*}

\noindent which is equal to $\ssim 80 \, \mathrm{m^2}$ for the $12$-meter ALMA antennae assuming an efficiency factor of $\eta \sim 0.7$ \citep{ALMA_technical_handbook}. These sensitivities, which agree well with the empirically measured \gls{RMS} noise, were computed for use in the \program{tclean} task, where the threshold parameter was set to three times this value.

\begin{figure*}
	\centering
	\includegraphics[width=\linewidth]{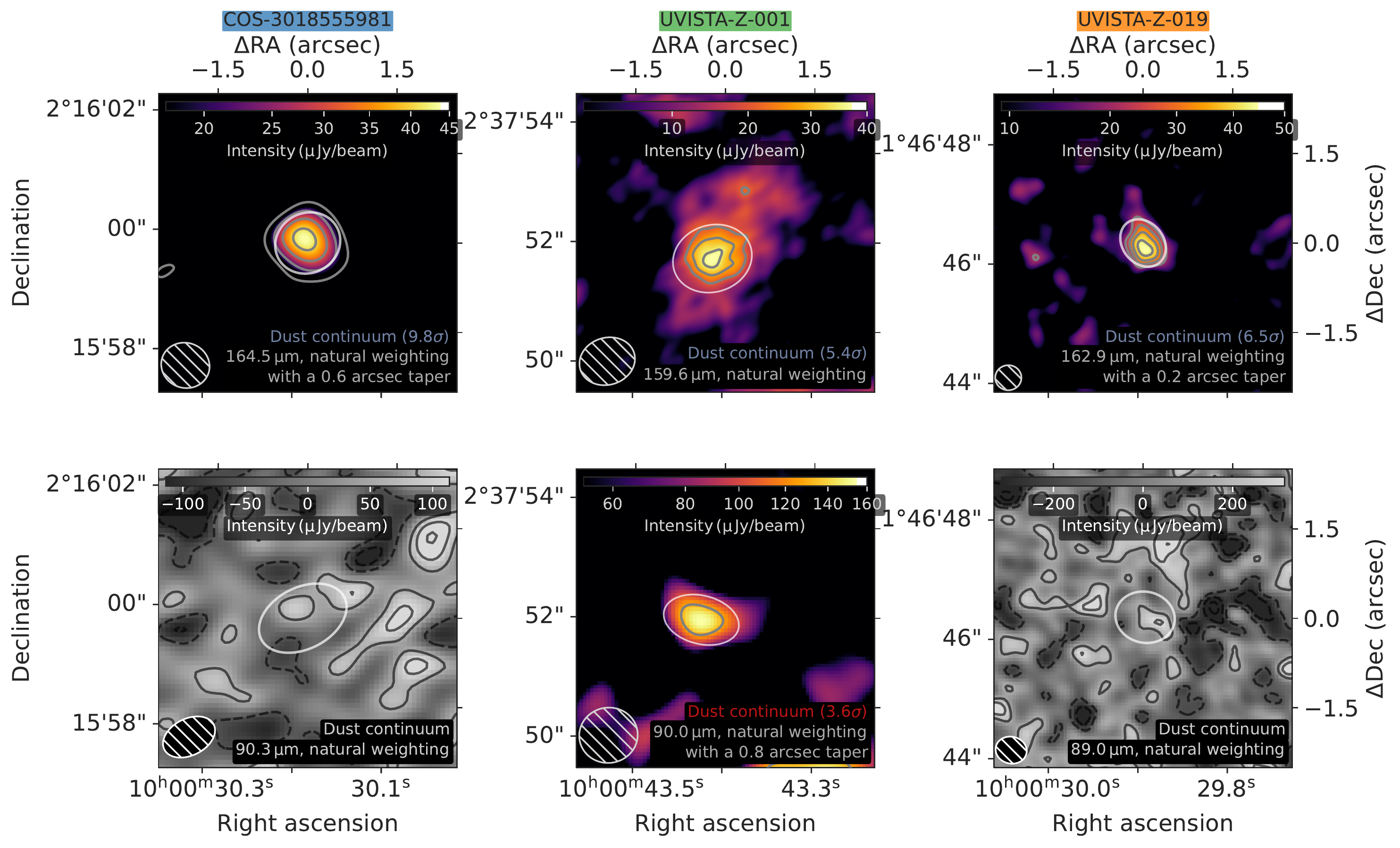}
	\caption[Naturally weighted dust continuum maps of COS-3018555981, UVISTA-Z-001, and UVISTA-Z-019.]{Naturally weighted dust continuum maps of COS-3018555981 (first column), UVISTA-Z-001 (second column), and UVISTA-Z-019 (third column); beam sizes are indicated in the bottom left. Top row: clear detections at $\lambda_\text{emit} \sim 160 \, \mathrm{\upmu m}$ (imaged with varying degrees of taper; see \cref{sssec:Observations:ALMA_data_reduction,ssec:Results:Dust_continuum}). Grey contours show subsequent significance levels in steps of $2 \sigma$ (COS-3018555981) or $1 \sigma$ (UVISTA-Z-001 and UVISTA-Z-019), starting at $3 \sigma$. White ellipses show the two-dimensional Gaussian fit obtained by \program{imfit} procedure in \gls{CASA}. Bottom row: (non-)detections at $\lambda_\text{emit} \sim 90 \, \mathrm{\upmu m}$. Contours are the same as the top row for UVISTA-Z-001, otherwise solid (dashed) grey lines show positive (negative) $1 \sigma$, $2 \sigma$, and $3 \sigma$ contours and white ellipses show the aperture used for determining a flux density upper limit (\cref{ssec:Results:Dust_continuum}). Since there is significant extended emission at $\ssim 160 \, \mathrm{\upmu m}$ in UVISTA-Z-001, the white ellipses show the aperture used to measure the flux of the compact component in both the top and bottom panels (\cref{ssec:Results:Dust_continuum}).
	}
	\label{fig:Dust_continuum_maps}
\end{figure*}

\subsection{\textit{HST} imaging}
\label{ssec:Observations:HST}

For sources at a redshift of $z \sim 7$, the rest-frame \gls{UV} is observed in the \gls{NIR}, ideally suited for observations with the infrared channel of the \gls{WFC3} on board \gls{HST}. For the two sources in the COSMOS field, COS-2987030247 and COS-3018555981, \gls{HST} imaging is available in the F125W ($J_{125}$), F140W ($JH_{140}$), and F160W ($H_{160}$) filters as part of CANDELS (GO 12440, PI: Faber) and the 3D-HST Treasury Programs (GO 12328, PI: Van Dokkum).\footnote{High-level science products are available on \url{https://archive.stsci.edu/prepds/3d-hst/}.} Combined, these observations reach a median depth of $26.1 \, \mathrm{mag}$, $25.5 \, \mathrm{mag}$, and $25.8 \, \mathrm{mag}$ ($5\sigma$ for a $1 \arcsec$-diameter aperture) in the $J_{125}$, $JH_{140}$, and $H_{160}$ bands, respectively \citep{2011ApJS..197...35G, 2011ApJS..197...36K, 2014ApJS..214...24S}. Throughout this work, we use a single stacked image, created by weighting the three filters by their inverse variance (covering $1400 \, \Angstrom \lesssim \lambda_\text{emit} \lesssim 2200 \, \Angstrom$). For UVISTA-Z-001, we use \gls{HST} imaging in the F140W filter (GO 13793, PI: Bowler) presented in \citet{2017MNRAS.466.3612B}.\footnote{Data may be obtained from the MAST at \href{https://dx.doi.org/10.17909/6gya-3b10}{10.17909/6gya-3b10}.}

For COS-2987030247, we manually remove a foreground source to the north-west. The foreground source is offset by just under $1 \arcsec$ and clearly (above $5 \sigma$) detected in the \gls{ACS} F606W and F814W filters, while for sources at $z \sim 6.8$, the \gls{IGM} would absorb any emission below the Lyman-continuum limit at a rest-frame wavelength of $912 \, \Angstrom$ (observed at $\ssim 0.7 \, \mathrm{\upmu m}$). In our stacked image, we replaced pixels with artificial noise if they are detected above $4 \sigma$ in a weighted \gls{ACS} image that has been smoothed to match the slightly more extended \gls{PSF} of \gls{WFC3}.

In addition, \gls{HST} observations to acquire rest-frame \gls{UV} imaging of UVISTA-Z-007 and UVISTA-Z-019 were awarded in Cycle 28 in the Mid-Cycle General Observer programme ID 16506 (PI: Witstok).\footnote{Data may be obtained from the MAST at \href{https://dx.doi.org/10.17909/t9-jhsf-m392}{10.17909/T9-JHSF-M392}.} Observations were performed with \gls{WFC3} using the F140W ($JH_{140}$) filter with a $5 \, \mathrm{ks}$ exposure per target, motivated by the goal of properly resolving the spatial substructure of the rest-frame \gls{UV}. We divided each orbit into 4 exposures, which allows for a four-point dithering pattern to improve sampling of the \gls{PSF} and to remove bad pixels, cosmic ray impacts, and detector artefacts. We used the \program{SPARS50} sampling sequence with $\program{nsamp}=13$ and $\program{nsamp}=14$ in two subsequent orbits for both targets.

Calibrated data products were combined using the \program{AstroDrizzle} task \citep[e.g.][]{2002PASP..114..144F} within the \program{DrizzlePac} package\footnote{See \url{https://www.stsci.edu/scientific-community/software/drizzlepac.html}.} setting the \program{final\_pixfrac} parameter to $0.8$ and choosing a pixel size of $0.065 \arcsec$. The resulting images reach a $5\sigma$ depth of $27.7 \, \mathrm{mag}$ for a $0.5 \arcsec$-diameter aperture (i.e. $25.9 \, \mathrm{mag/arcsec^2}$).

The astrometry of all images was calibrated to \textit{Gaia} data \citep{2016A&A...595A...1G, 2021A&A...649A...1G} by fitting a two-dimensional Gaussian to the \gls{HST} images of a number of nearby stars with a $g$-band magnitude of $g \lesssim 21 \, \mathrm{mag}$ ($7$, $5$, $6$, $4$, and $2$ for COS-2987030247, COS-3018555981, UVISTA-Z-001, UVISTA-Z-007, and UVISTA-Z-019, respectively). Among the stars in each field, we found the measured peak offsets relative to \textit{Gaia} positions agreed very well within less than $0.05 \arcsec$; taken together, they resulted in a required median correction of less than about $0.1 \arcsec$ in all cases.

\begingroup
    \setlength{\tabcolsep}{6pt} % Default value: 6pt
    \renewcommand{\arraystretch}{1.25} % Default value: 1
    \begin{table*}
        \centering
        \caption[Continuum fluxes]
        {Average, minimum, and maximum channel rest-frame wavelength and observed frequency used for the aggregate continuum images and corresponding (upper limits on) continuum fluxes. For non-detections, $3 \sigma$ upper limits are given (with theoretical $3 \sigma$ sensitivities, given by \cref{eq:Interferometry_sensitivity}, in brackets). For detections, deconvolved sizes $A_\text{dust}$ as measured by the \gls{CASA} \program{imfit} procedure are given. Dust-related quantities listed are the mass $M_\text{dust}$, mass surface density $\Sigma_\text{dust}$, and yield $y_\text{dust}$, discussed in \cref{ssec:Discussion:Dust_masses}; the intrinsic SED temperature $T_\text{dust}$ and peak temperature $T_\text{peak}$, discussed in \cref{ssec:Discussion:Dust_temperatures}; the integrated \gls{IR} and \gls{FIR} luminosities ($L_\text{(F)IR}$); and finally, the obscured and total \glspl{SFR}, $\text{\gls{SFR}}_\text{\gls{IR}}$ and $\text{\gls{SFR}}_\text{tot}$. These properties are inferred from \glsentryshort{MERCURIUS}\glsreset{MERCURIUS} fits where possible (i.e. for COS-3018555981, UVISTA-Z-001, and UVISTA-Z-019), under the fiducial self-consistent opacity model (for a more in-depth discussion, see \cref{ssec:Discussion:Dust_SED_fitting_procedure}). Otherwise, they are approximated under a fully optically thin \gls{SED} with the fiducial assumptions of $T_\text{dust} = 50 \, \mathrm{K}$, $\beta_\text{\gls{IR}} = 1.5$, which however introduces large systematic uncertainties (see \cref{ssec:Results:Dust_continuum} for details). For COS-3018555981, an upper limit ($95 \%$ confidence) on the dust temperatures is reported in brackets.}
        \label{tab:Continuum_fluxes_and_dust_properties}
        \begin{tabular}{llp{2.05cm}p{2.05cm}p{2.05cm}p{2.05cm}p{2.05cm}}
            \hline
            Regime & Quantity & \setulcolor{COS2987}\ul{COS-2987030247} & \setulcolor{COS3018}\ul{COS-3018555981} & \setulcolor{UVIS001}\ul{UVISTA-Z-001} & \setulcolor{UVIS007}\ul{UVISTA-Z-007} & \setulcolor{UVIS019}\ul{UVISTA-Z-019}
\\
\hline
\multirow{4}{*}{Band 8} & $\nu_\text{obs} \, (\mathrm{GHz})$ & $426.2_{ -6.7 }^{ +9.3 }$ & $422.5_{ -5.4 }^{ +8.4 }$ & $413.0_{ -6.9 }^{ +8.9 }$ & $434.5_{ -7.7 }^{ +6.2 }$ & $434.4_{ -7.8 }^{ +6.1 }$
\\
& $\lambda_\text{{emit}} \, (\mathrm{{\upmu m}})$ & $90.1_{ -1.9 }^{ +1.4 }$ & $90.3_{ -1.8 }^{ +1.2 }$ & $90.0_{ -1.9 }^{ +1.5 }$ & $89.0_{ -1.3 }^{ +1.6 }$ & $89.0_{ -1.2 }^{ +1.6 }$
\\
& $S_{\nu\text{, obs}} \, (\mathrm{\upmu Jy})$ & $<163 \, (160)$ & $<210 \, (240)$ & $189 \pm 100$ & $<361 \, (260)$ & $<840 \, (625)$
\\
\vspace{1.5ex}
& $A_\text{dust} \, (\mathrm{kpc^2})$ & \dots & \dots & $(5.2 \pm 2.4) \times$ \newline $(2.0 \pm 1.3)$ & \dots & \dots
\\
\multirow{4}{*}{Band 6} & $\nu_\text{obs} \, (\mathrm{GHz})$ & $250.3_{ -9.8 }^{ +7.3 }$ & $232.1_{ -8.0 }^{ +13.3 }$ & $232.9_{ -16.0 }^{ +10.0 }$ & $242.4_{ -2.1 }^{ +2.0 }$ & $237.3_{ -11.0 }^{ +8.8 }$
\\
& $\lambda_\text{{emit}} \, (\mathrm{{\upmu m}})$ & $153.4_{ -4.4 }^{ +6.2 }$ & $164.5_{ -8.9 }^{ +5.9 }$ & $159.6_{ -6.6 }^{ +11.8 }$ & $159.6_{ -1.3 }^{ +1.4 }$ & $162.9_{ -5.8 }^{ +7.9 }$
\\
& $S_{\nu\text{, obs}} \, (\mathrm{\upmu Jy})$ & $<22.5 \, (18.4)$ & $76 \pm 13$ & $97 \pm 30$ & $<69.6 \, (81.8)$ & $131 \pm 36$
\\
\vspace{1.5ex}
& $A_\text{dust} \, (\mathrm{kpc^2})$ & \dots & $(4.3 \pm 0.8) \times$ \newline $(3.6 \pm 0.7)$ & $(10.3 \pm 3.4) \times$ \newline $(3.7 \pm 3.0)$ & \dots & $(4.1 \pm 1.1) \times$ \newline $(3.0 \pm 0.9)$
\\
\multirow{3}{*}{Band 5} & $\nu_\text{obs} \, (\mathrm{GHz})$ & $195.6_{ -9.4 }^{ +6.4 }$ & \dots & \dots & \dots & \dots
\\
& $\lambda_\text{{emit}} \, (\mathrm{{\upmu m}})$ & $196.3_{ -6.3 }^{ +9.9 }$ & \dots & \dots & \dots & \dots
\\
& $S_{\nu\text{, obs}} \, (\mathrm{\upmu Jy})$ & $<19.4 \, (22.8)$ & \dots & \dots & \dots & \dots
\\
&  &  &  &  &  & 
\\
\cline{2-7}
& $T_\text{dust}$ & Fixed: $50 \, \mathrm{K}$ & \glsentryshort{MERCURIUS} fit & \glsentryshort{MERCURIUS} fit & Fixed: $50 \, \mathrm{K}$ & \glsentryshort{MERCURIUS} fit
\\
& $\lambda_0$ (opacity) & Optically thin & Self-consistent: $8.4_{ -4.4 }^{ +11.8 } \, \mathrm{ \upmu m }$ & Self-consistent: $0.6_{ -0.3 }^{ +0.7 } \, \mathrm{ \upmu m }$ & Optically thin & Self-consistent: $4.2_{ -2.7 }^{ +7.5 } \, \mathrm{ \upmu m }$
\\
& $\beta_\text{\gls{IR}}$ & Fixed: $1.5$ & Fixed: $1.5$ & Fixed: $1.5$ & Fixed: $1.5$ & Fixed: $1.5$
\\
\cline{2-7}
&  &  &  &  &  & 
\\
& $M_\text{dust} \, (\mathrm{M_\odot})$ & $\lesssim4 \cdot 10^{6}$ & $7_{-5}^{+19} \cdot 10^{7}$ & $4_{-3}^{+8} \cdot 10^{6}$ & $\lesssim1 \cdot 10^{7}$ & $2.1_{-1.6}^{+7.6} \cdot 10^{7}$
\\
& $\Sigma_\text{dust} \, (\mathrm{M_\odot \, pc^{-2}})$ & \dots & $7_{-4}^{+18}$ & $0.1_{-0.1}^{+0.3}$ & \dots & $2.3_{-1.8}^{+8.5}$
\\
& $M_\text{dust} / M_*$ & $\lesssim0.002$ & $0.05_{-0.04}^{+0.16}$ & $0.001_{-0.0007}^{+0.003}$ & $\lesssim0.003$ & $0.007_{-0.006}^{+0.03}$
\\
& $y_\text{dust, AGB} \, (\mathrm{M_\odot})$ & $\lesssim0.1$ & $1.5_{-1.1}^{+4.6}$ & $0.03_{-0.02}^{+0.08}$ & $\lesssim0.1$ & $0.2_{-0.2}^{+0.9}$
\\
\vspace{1.5ex}
& $y_\text{dust, SN} \, (\mathrm{M_\odot})$ & $\lesssim0.2$ & $4.5_{-3.1}^{+13.3}$ & $0.09_{-0.06}^{+0.2}$ & $\lesssim0.3$ & $0.6_{-0.5}^{+2}$
\\
& $T_\text{dust} \, (\mathrm{K})$ & \dots & $29_{-5}^{+9}$ ($<48$) & $59_{-20}^{+41}$ & \dots & $47_{-17}^{+40}$
\\
\vspace{1.5ex}
& $T_\text{peak} \, (\mathrm{K})$ & \dots & $26_{-5}^{+8}$ ($<43$) & $53_{-17}^{+37}$ & \dots & $42_{-15}^{+35}$
\\
$8$-$1000 \, \mathrm{\upmu m}$ & $L_\mathrm{IR} \, (\mathrm{L_\odot})$ & $\lesssim2.1 \cdot 10^{11}$ & $9.9_{-2.3}^{+6.8} \cdot 10^{10}$ & $2.0_{-1.3}^{+9.5} \cdot 10^{11}$ & $\lesssim6.9 \cdot 10^{11}$ & $3.1_{-1.8}^{+18.1} \cdot 10^{11}$
\\
\vspace{1.5ex}
$42.5$-$122.5 \, \mathrm{\upmu m}$ & $L_\mathrm{FIR} \, (\mathrm{L_\odot})$ & $\lesssim1.4 \cdot 10^{11}$ & $6.4_{-1.6}^{+4.8} \cdot 10^{10}$ & $1.1_{-0.6}^{+1.1} \cdot 10^{11}$ & $\lesssim4.7 \cdot 10^{11}$ & $2.1_{-1.3}^{+4.2} \cdot 10^{11}$
\\
$\mathrm{IR}$ ($8$-$1000 \, \mathrm{\upmu m}$) & $\text{SFR}_\text{IR} \, (\mathrm{M_\odot \, yr^{-1}})$ & $\lesssim32$ & $15_{-3}^{+10}$ & $30_{-20}^{+142}$ & $\lesssim102$ & $46_{-27}^{+270}$
\\
$\mathrm{UV+IR}$ & $\text{SFR}_\text{tot} \, (\mathrm{M_\odot \, yr^{-1}})$ & $\lesssim54$ & $34_{-4}^{+10}$ & $79_{-20}^{+142}$ & $\lesssim128$ & $63_{-27}^{+270}$
\\
            \hline
        \end{tabular}
    \end{table*}
\endgroup

\section{Results}
\label{sec:Results}

\subsection{Dust continuum}
\label{ssec:Results:Dust_continuum}

Except for a $\ssim 4 \sigma$ detection in UVISTA-Z-001, the dust continuum at a rest-frame wavelength of $\lambda_\text{emit} \sim 90 \, \mathrm{\upmu m}$ remains undetected for the other sources. The same holds for the dust continuum of COS-2987030247 around the \NII\ line at $205 \, \mathrm{\upmu m}$. The continuum at $\lambda_\text{emit} \sim 160 \, \mathrm{\upmu m}$, however, is detected in COS-3018555981, UVISTA-Z-001, and UVISTA-Z-019 \citep[see also][]{2022ApJ...928...31S, 2022MNRAS.510.5088B}. Dust continuum maps of these sources are shown in \cref{fig:Dust_continuum_maps}.

For consistency in the flux measurements across different \gls{ALMA} bands, their beams are matched by applying a slight taper to higher resolution data (see \cref{sssec:Observations:ALMA_data_reduction}). For COS-3018555981 and UVISTA-Z-019, we place $3 \sigma$ upper limits at $\ssim 90 \, \mathrm{\upmu m}$ by empirically measuring the \gls{RMS} noise in a beam-shaped aperture matched in (convolved) size to the source extent at $\ssim 160 \, \mathrm{\upmu m}$; for COS-2987030247 and UVISTA-Z-007, a (\gls{FWHM}) beam-sized aperture is used. Since there is significant extended emission at $\ssim 160 \, \mathrm{\upmu m}$ in UVISTA-Z-001, we measure fluxes of the compact component in an aperture centred on the peak of the emission and matched in (convolved) size to the source extent at $\ssim 90 \, \mathrm{\upmu m}$ (implications for the inferred dust properties will be discussed in \cref{sec:Discussion:Dust_properties}). In estimating the uncertainties on our continuum detections as well as upper limits, we conservatively take a systematic flux calibration uncertainty into account ($10 \%$ in band 5 and 6, $20 \%$ in band 8).\footnote{See Section A.9.2 of the \gls{ALMA} Proposers' Guide \citep{ALMA_proposers_guide}.}

In \cref{tab:Continuum_fluxes_and_dust_properties}, all continuum measurements around the three emission lines (including $A_\text{dust}$, the deconvolved source extent in $\mathrm{kpc^2}$, if detected) are summarised. Also presented in \cref{tab:Continuum_fluxes_and_dust_properties} are other characteristics of the sources like their corresponding total \gls{IR} and \gls{FIR} luminosities (defined as the integrated luminosity between $8$ and $\num{1000} \, \mathrm{\upmu m}$ and between $42.5$ and $122.5 \, \mathrm{\upmu m}$ in the rest frame, respectively; see e.g. \citealt{2020ApJ...902...78R}) and other dust properties, leveraging constraints from the combined continuum measurements. These properties have been derived from a best-fit ``greybody'' spectrum \citep[e.g.][]{2014PhR...541...45C} if possible and will be discussed in more detail in \cref{sec:Discussion:Dust_properties}. When only upper limits are available (i.e. for COS-2987030247 and UVISTA-Z-007; see \cref{ssec:Discussion:Dust_SED_fitting_procedure}), we report $3 \sigma$ upper limits on the \gls{IR} and \gls{FIR} luminosities by assuming a fiducial $\beta_\text{\gls{IR}} = 1.5$ and $T_\text{dust} = 50 \, \mathrm{K}$; we acknowledge the uncertainty on these parameters by adding an additional $0.4 \, \mathrm{dex}$ systematic uncertainty to the integrated luminosities and corresponding obscured \gls{SFR} \citep[cf.][]{2020MNRAS.499.5136C} before setting upper limits.

\subsection{\texorpdfstring{\NIILam}{[NII] 205 μm}}
\label{ssec:Results:NII}

We placed a $3 \sigma$ upper limit on the \NII\ luminosity of COS-2987030247 by first measuring the \gls{RMS} noise in the untapered, naturally-weighted datacube over a single beam (assuming the source is unresolved). We obtained an upper limit on the line flux in $\mathrm{Jy \, km/s}$ by scaling this noise to a range of channels covering twice the maximum \gls{FWHM} between the \OIII\ and \CII\ lines (\cref{ssec:Results:OIII_CII}), which we then converted to a luminosity, as shown in \cref{tab:Line_fluxes}. The lower limit on the \CIILam\ to \NIILam\ ratio is $L_\CII/L_\NII > 4.8$ ($3 \sigma$).

\begin{figure*}
	\centering
	\includegraphics[width=\linewidth]{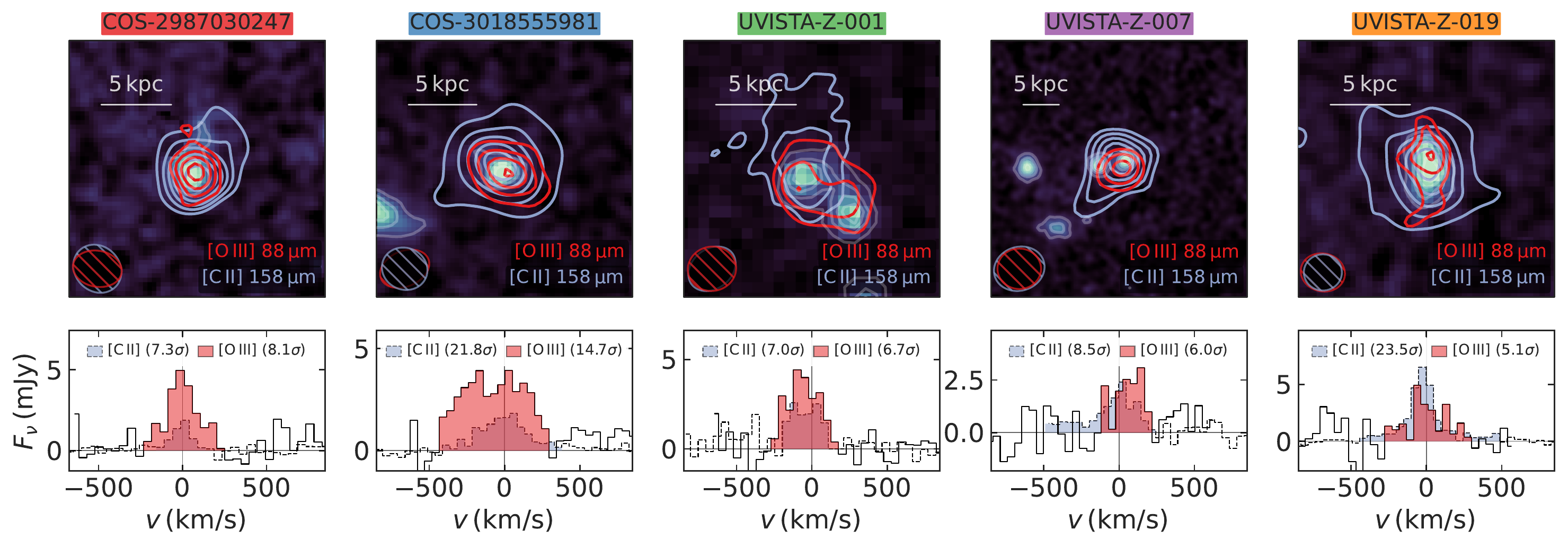}
	\caption[Detections of the \CII\ and \OIII\ lines.]{Detections of the \CIILam\ and \OIIILam\ lines. Top row: contour images of the \CII\ and \OIII\ lines overlaid on a background of \gls{HST} rest-frame \gls{UV} images (see \cref{ssec:Observations:HST}). The line images have matched beam sizes (as listed in \cref{tab:Observations}) and are produced by collapsing all channels within the \gls{FWHM} of the line (as those of the \gls{SNR} determination, which however use natural weighting without tapering; \cref{ssec:Results:OIII_CII}). Contours start at $3 \sigma$ and increase in steps of $5 \sigma$ for the \CII\ contours in COS-3018555981 and UVISTA-Z-019, $3 \sigma$ for the \OIII\ contours in COS-3018555981, $2 \sigma$ for both lines in UVISTA-Z-001, and $1 \sigma$ otherwise. Bottom row: \gls{SNR}-weighted line spectra extracted from a $2 \sigma$ region in a naturally weighted image integrated over the \gls{FWHM} (cf. \cref{ssec:Results:OIII_CII}). Coloured channels, which were used to produce a subsequent moment-zero map for measuring the total line flux, indicate the line detection. The significance of each line detection is shown in the legend (also see \cref{ssec:Results:OIII_CII} for details).
	}
	\label{fig:CII_and_OIII_maps_and_spectra}
\end{figure*}

\begin{table*}
    \centering
    \caption[Observed line fluxes of \CIILam, \OIIILam, and \NIILam.]
    {Observed line fluxes ($F_\text{line}$) of \CIILam, \OIIILam, and \NIILam, given in units of $\mathrm{Jy \, km/s}$ (i.e. as $S_\nu \Delta v$; see \cref{ssec:Results:NII,ssec:Results:OIII_CII} for details). Also shown are corresponding line luminosities and the \OIII/\CII\ luminosity ratio. For non-detections, $3 \sigma$ upper limits are given.}
    \label{tab:Line_fluxes}
    \begin{tabular}{llllllll}
        \hline
        Source & $F_\CII \, (\mathrm{Jy \, km/s})$ & $L_\CII \, (10^8 \, \mathrm{L_\odot})$ & $F_\OIII \, (\mathrm{Jy \, km/s})$ & $L_\OIII \, (10^8 \, \mathrm{L_\odot})$ & $L_\OIII/L_\CII$ & $F_\NII \, (\mathrm{Jy \, km/s})$ & $L_\NII \, (10^8 \, \mathrm{L_\odot})$ \\
        \hline
        \begin{filecontents*}{ALMA_line_fluxes.csv}
field|object|SnudvCII|LCII|SnudvOIII|LOIII|SnudvNII|LNII|OIIICII
COSMOS|\setulcolor{COS2987}\ul{COS-2987030247}|$0.250 \pm 0.061$|$2.83 \pm 0.69$|$0.79 \pm 0.20$|$16.0 \pm 4.1$|$<0.0679$|$<0.591$|$5.7 \pm 1.8$
COSMOS|\setulcolor{COS3018}\ul{COS-3018555981}|$0.347 \pm 0.027$|$3.97 \pm 0.31$|$1.58 \pm 0.29$|$32.3 \pm 5.9$|\dots|\dots|$8.13 \pm 0.92$
UVISTA|\setulcolor{UVIS001}\ul{UVISTA-Z-001}|$0.564 \pm 0.095$|$6.7 \pm 1.1$|$0.92 \pm 0.22$|$19.6 \pm 4.7$|\dots|\dots|$2.91 \pm 0.66$
UVISTA|\setulcolor{UVIS007}\ul{UVISTA-Z-007}|$0.408 \pm 0.065$|$4.57 \pm 0.73$|$0.65 \pm 0.30$|$12.9 \pm 5.9$|\dots|\dots|$2.8 \pm 1.0$
UVISTA|\setulcolor{UVIS019}\ul{UVISTA-Z-019}|$0.686 \pm 0.049$|$7.69 \pm 0.55$|$0.83 \pm 0.35$|$16.6 \pm 7.0$|\dots|\dots|$2.16 \pm 0.54$
\end{filecontents*}
\csvreader[separator=pipe, late after line=\\, head to column names]{ALMA_line_fluxes.csv}{}{\object & \SnudvCII & \LCII & \SnudvOIII & \LOIII & \OIIICII & \SnudvNII & \LNII}
        \hline
    \end{tabular}
\end{table*}

\subsection{\texorpdfstring{\OIIILam}{[OIII] 88 μm} and \texorpdfstring{\CIILam}{[CII] 158 μm}}
\label{ssec:Results:OIII_CII}

We obtained moment-zero maps by integrating a naturally weighted, clean datacube over the \gls{FWHM} around the line centre. Defining the \gls{SNR} as that measured for the peak pixel, the \OIII\ and \CII\ lines are detected with a significance of at least $\text{\gls{SNR}} > 5$ in all five sources. Total line fluxes, however, are measured on cleaned cubes with a $0.6 \arcsec$ taper to recover extended emission as well as possible (\cref{sssec:Observations:ALMA_data_reduction}), integrating all channels where line flux is detected (coloured channels in \cref{fig:CII_and_OIII_maps_and_spectra}) in a spectrum extracted from a contiguous region of spaxels reaching at least $2 \sigma$ in the moment-zero map, the spectrum each spaxel weighted by its \gls{SNR}. In turn, this initial moment-zero map was created by integrating channels within half of the \gls{FWHM} from the line centre (used in the \gls{SNR} determination described above). Again, we take a conservative systematic flux calibration uncertainty into account, as discussed in \cref{ssec:Results:Dust_continuum}. The resulting line fluxes, luminosities, and ratios are listed in \cref{tab:Line_fluxes}.

\section{Dust properties}
\label{sec:Discussion:Dust_properties}

Although accounting for only a small fraction of baryonic matter in galaxies, dust is highly relevant in the context of galaxy evolution \citep[e.g.][]{2019MNRAS.489.1397L}. Representative dust properties of \gls{EoR} galaxies, however, are still difficult to constrain with current observations partly due to the degeneracy between dust mass and temperature, complicated further by the dust opacity. Only in a few cases, multiple constraints on the \gls{FIR} \gls{SED} of normal star-forming galaxies ($\text{\gls{SFR}} \lesssim 100 \, \mathrm{M_\odot \, yr^{-1}}$) in the \gls{EoR} are available \citep[e.g.][]{2019PASJ...71...71H, 2020MNRAS.493.4294B, 2021MNRAS.508L..58B}. Here, we use the available band-6 detections in combination with the non-detections in band 8 to obtain further insights into the (F)IR luminosity, $L_\text{(F)IR}$, corresponding obscured and total \gls{SFR},\footnote{Obtained from the \gls{UV} and \gls{IR} luminosities using the conversions in \citet{2012ARA&A..50..531K}.} dust mass (\cref{ssec:Discussion:Dust_masses}), and dust temperature (\cref{ssec:Discussion:Dust_temperatures}) of typical \gls{UV}-bright \gls{EoR} galaxies.

\subsection{Dust \texorpdfstring{\glsentryshort{SED}}{SED} fitting procedure}
\label{ssec:Discussion:Dust_SED_fitting_procedure}

A dust-continuum detection at $\lambda_\text{emit} \sim 160 \, \mathrm{\upmu m}$ combined with a detection (UVISTA-Z-001) or even an upper limit (COS-3018555981 and UVISTA-Z-019) at $\lambda_\text{emit} \sim 90 \, \mathrm{\upmu m}$ can provide insight into the dust properties beyond those derived with a fixed temperature, emissivity, and opacity model. To investigate this in detail, we performed a fitting routine using the multimodal nested sampling algorithm \program{multinest} \citep[described in][]{2009MNRAS.398.1601F} implemented in \program{python} as the \program{pymultinest} package \citep{2014A&A...564A.125B}, as described below. The full code of this routine which treats detections and upper limits uniformly and we therefore refer to as \glsentryshort{MERCURIUS} (\glsentrylong{MERCURIUS}\glsunset{MERCURIUS}), is available online.\footnote{See \url{https://github.com/joriswitstok/mercurius/}.} Apart from the main fitting procedure, it also includes a greybody \gls{SED} exploration visualisation tool; both are illustrated in a documented example.

Radiative transfer predicts the intensity emerging from a region of dust at a temperature $T_\text{dust}$ becomes a modified black body \citep[often referred to as a greybody; e.g.][]{2020MNRAS.498.4109J},
\begin{equation}
    \label{eq:Dust_radiative_transfer}
    J_\nu = \left( 1 - e^{-\tau (\nu)} \right) B_\nu \left( T_\text{dust} \right),
\end{equation}

\noindent where $B_\nu$ is the Planck function \citep{1901AnP...309..553P}, a specific intensity in units of $\mathrm{erg \, s^{-1} \, cm^{-2} \, sr^{-1} \, Hz^{-1}}$ that is attenuated by the opacity term containing the optical depth, $\tau (\nu)$. However, the observed flux needs to be corrected for the effect of observing against the isotropic \gls{CMB}, which involves subtracting the Planck function for $T_\text{\gls{CMB}} (z)$, the \gls{CMB} temperature at redshift $z$, from the dust blackbody term \citep[as in equation (18) in][]{2013ApJ...766...13D}. The flux density observed at $\nu_\text{obs} = \nu / (1+z)$,\footnote{We note that in quantities related to the emission process (such as $B_\nu$), $\nu$ is short for the rest-frame frequency $\nu_\text{emit}$, whereas for the observable quantity $S_{\nu\text{, obs}}$ -- the redshifted flux density in the observer's frame -- $\nu$ implicitly stands for the observed frequency, introducing the factor $(1+z)$.} denoted $S_{\nu\text{, obs}}$ and given in $\mathrm{erg \, s^{-1} \, cm^{-2} \, Hz^{-1}}$, is then given by the following general form of the observed \gls{SED} \citep[see also][]{2020MNRAS.498.4109J},
\begin{equation}
    \label{eq:General_dust_SED_flux_density}
    S_{\nu\text{, obs}} = \frac{A_\text{dust} \left( 1+z \right)}{D_L^2 (z)} \left( 1 - e^{-\tau (\nu)} \right) \left[ B_\nu \left(T_\text{dust} \right) - B_\nu \left(T_\text{\gls{CMB}} (z) \right) \right],
\end{equation}

\noindent where $A_\text{dust}$ is the area subtended by the source. In our fiducial model, the optical depth is taken to be proportional to the dust mass surface density $\Sigma_\text{dust}$ \citep{2014PhR...541...45C} via
\begin{equation}
    \label{eq:Dust_mass_absorption_coefficient_definition}
    \tau (\nu) \equiv \kappa_\nu \, \Sigma_\text{dust} = \kappa_\nu \, C \, \frac{M_\text{dust}}{A_\text{dust}} \text{, with } C \geq 1.
\end{equation}

Here, we substituted $\Sigma_\text{dust}$ by a combination of $C$, a clustering factor,\footnote{We introduce the parameterisation of $C$ to relate the true area of dust emission $A_\text{dust}^*$ (which can only be measured with infinitely high resolution) to its galaxy-averaged value, $A_\text{dust}$, via $\Sigma_\text{dust} = M_\text{dust} / A_\text{dust}^* = C \, M_\text{dust} / A_\text{dust}$, such that $1/C$ represents the covering fraction of dust (we note $C \geq 1$ since $\Sigma_\text{dust} \geq M_\text{dust} / A_\text{dust}$). Unless mentioned otherwise, we assumed $C = 1$ throughout as would be expected for a homogeneous distribution, although a clustered dust distribution (on scales below the resolution of interferometric observations), resulting in $C > 1$, may be more realistic (see \cref{ssec:Discussion:Dust_temperatures}).} the dust mass $M_\text{dust}$, and $A_\text{dust}$, which we take to be the deconvolved size of the dust emission found by the \program{imfit} procedure in \gls{CASA} (given in \cref{tab:Continuum_fluxes_and_dust_properties}; see e.g. \cref{fig:Dust_continuum_maps}). The absorption cross section (per unit mass), $\kappa_\nu$, is the frequency-dependent constant of proportionality. This frequency dependence is parametrised via a power law with the dust emissivity $\beta_\text{\gls{IR}}$ as its exponent \citep{2006ApJ...636.1114D},
\begin{equation}
    \label{eq:Dust_mass_absorption_coefficient_form}
    \kappa_\nu = \kappa_{\nu, \text{ ref}} \left( \frac{\nu}{\nu_\text{ref}} \right)^{\beta_\text{\gls{IR}}}.
\end{equation}

Following \citet{2022ApJ...928...31S}, we used $\kappa_{\nu, \text{ ref}} = 8.94 \, \mathrm{cm^2 \, g^{-1}}$ at $\lambda_\text{ref} = 158 \, \mathrm{\upmu m}$ (i.e. $\nu_\text{ref} = \nu_\CII \simeq 1.90 \, \mathrm{THz}$), appropriate for dust ejected by \glspl{SN} after reverse shock destruction (values for different dust grain compositions range from $5 \, \mathrm{cm^2 \, g^{-1}} \lesssim \kappa_{\nu, \text{ ref}} \lesssim 30 \, \mathrm{cm^2 \, g^{-1}}$; see \citealt{2014MNRAS.443.1704H}, and references therein). As will be discussed in \cref{ssec:Discussion:Dust_masses}, \glspl{SN} are the most likely origin of dust in these young, metal-poor star-forming galaxies. However, since the detailed dust composition is in principle unknown, $\kappa_{\nu, \text{ ref}}$ carries with it a systematic uncertainty that can lower dust masses by $\ssim 3 \times$ or increase them by $\ssim 1.5 \times$.

For convenience, the optical depth can be expressed in the form
\begin{equation}
    \label{eq:Optical_depth}
    \tau (\nu) = \left( \frac{\nu}{\nu_0} \right)^{\beta_\text{\gls{IR}}} = \left( \frac{\lambda_0}{\lambda} \right)^{\beta_\text{\gls{IR}}},
\end{equation}

\noindent which clearly marks the point at which the dust transitions from optically thin to thick, at $\lambda_0 = c/\nu_0$. For \glspl{LBG} at high redshift, $\lambda_0$ is typically assumed to be well below the sampled wavelength regime so that it is safe to approximate the entire \gls{SED} as being optically thin \citep[$\tau (\nu) \ll 1$; e.g.][]{2021MNRAS.508L..58B} which simply reduces \cref{eq:General_dust_SED_flux_density} to
\begin{equation}
    \label{eq:OT_dust_SED_flux_density}
    S_{\nu\text{, obs}} = M_\text{dust} \, \frac{1+z}{D_L^2 (z)} \, \kappa_\nu \left[ B_\nu \left(T_\text{dust} \right) - B_\nu \left(T_\text{\gls{CMB}} (z) \right) \right].
\end{equation}

However, this assumption can lead to a significantly underestimated dust temperature and overestimated dust mass if incorrectly applied on measurements that sample a region where the approximation does not hold \citep{2020A&A...634L..14C, 2020MNRAS.498.4109J}.\footnote{Together with the fact that the intrinsic dust temperature has a large impact on the derived dust masses at shorter wavelengths, this is why dust masses should ideally be inferred from the highest \gls{FIR} wavelengths ($\lambda_\text{emit} > 450 \, \mathrm{\upmu m}$), where the \gls{SED} certainly is optically thin \citep{2012MNRAS.425.3094C}.} For a fixed $\lambda_0$, an a posteriori consistency check on the assumed opacity model can be performed, since it follows from \cref{eq:Dust_mass_absorption_coefficient_definition,eq:Dust_mass_absorption_coefficient_form,eq:Optical_depth} that
\begin{equation}
    \label{eq:Consistent_l0}
    \lambda_0 = \left( \kappa_{\nu, \text{ ref}} \, \Sigma_\text{dust} \right)^{1/\beta_\text{\gls{IR}}} \lambda_\text{ref}.
\end{equation}

Alternatively, \cref{eq:Consistent_l0} allows for a self-consistent framework around the general greybody \gls{SED} model in \cref{eq:General_dust_SED_flux_density} where, given a dust mass, we infer $\lambda_0$ a priori, which we took as our fiducial opacity model.

Under a set opacity model, we then used a freely varying dust temperature and logarithmic dust mass. As the number of free parameters should not exceed the number of constraints and different assumptions of $\lambda_0$ typically dominate over those of the dust emissivity \citep{2014PhR...541...45C}, we -- conservatively, as will become clear in \cref{ssec:Discussion:Dust_temperatures} -- fixed the dust emissivity to $\beta_\text{\gls{IR}} = 1.5$. A flat prior within the range $10^4 \, \mathrm{M_\odot} < M_\text{dust} < M_*$ was assumed on $\log_{10} M_\text{dust}$; a gamma distribution with shape parameter $a = 1.5$ and shifted to start at $T_\text{\gls{CMB}}(z)$ was used for the dust temperature (a standard choice for a parameter with a non-negative continuous domain as it is the conjugate prior to many likelihood distributions; we note that there is little difference when assuming a flat prior, but a maximum temperature has to be assumed). The model's predicted observed flux density at a given wavelength are compared with the actual observed flux densities and a likelihood is assigned based on the squared residuals between model and observations (weighted by the inverse variance). Following the formalism in \citet{2012PASP..124.1208S}, if the model greybody curve exceeds (falls below) the upper limits, the likelihood is lowered (increased) according to the significance of the discrepancy (agreement).

\begin{figure*}
	\centering
	\includegraphics[width=\linewidth]{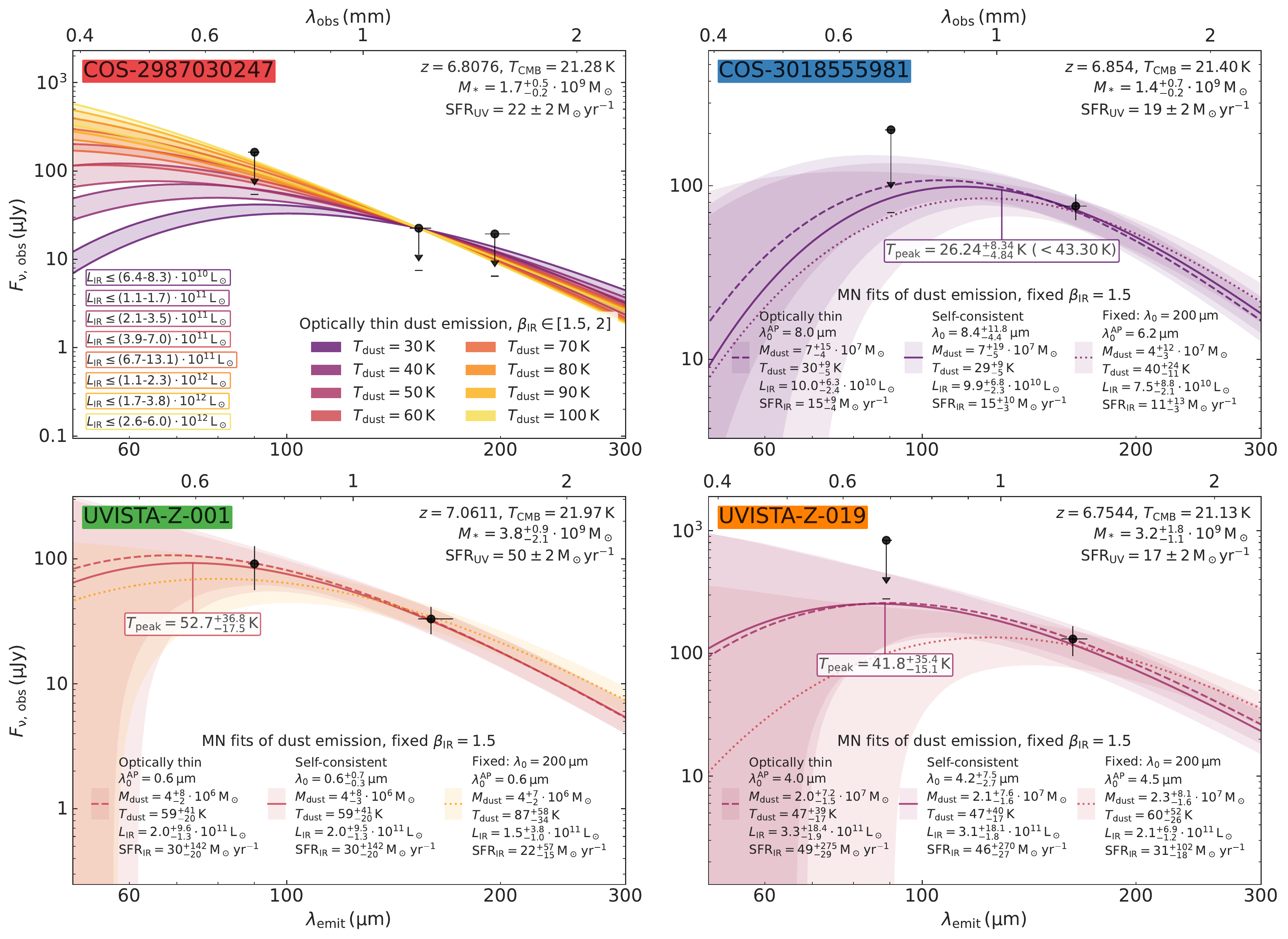}
	\caption[Overview of the dust continuum detections and upper limits.]{Overview of the dust continuum detections and upper limits for four galaxies investigated in this work. Upper limits are drawn at $3 \sigma$, with a line beneath indicating a $1 \sigma$ level. Where only upper limits are available (i.e. for COS-2987030247), we plot a wide variety of greybody dust \glspl{SED}, for $30 \, \mathrm{K} \leq T_\text{dust} \leq 100 \, \mathrm{K}$ and $1.5 \leq \beta_\text{\gls{IR}} \leq 2$ (under an entirely optically thin opacity model). The resulting range of \gls{IR} luminosities is indicated in the bottom left, showing $L_\text{\gls{IR}}$ can vary by more than an order of magnitude between the most extreme choices of $T_\text{dust}$ and $\beta_\text{\gls{IR}}$. For COS-3018555981, UVISTA-Z-001, and UVISTA-Z-019, on the other hand, the \gls{MERCURIUS} fitting routine described in \cref{ssec:Discussion:Dust_SED_fitting_procedure} is applied, fixing $\beta_\text{\gls{IR}} = 1.5$ but considering various assumptions on the opacity model: either an entirely optically thin \gls{SED} or the general opacity model given in \cref{eq:Optical_depth} with either a self-consistent or fixed $\lambda_0$, the wavelength separating the optically thick and thin regimes. The ``best-fit'' line shows the \gls{SED} curve for the maximum likelihood in the $(M_\text{dust}, T_\text{dust})$ plane, shaded regions show the \nth{16} and \nth{84} percentiles at each wavelength of all curves produced according to the posterior distribution. We note the best-fit model of COS-3018555981 and UVISTA-Z-019 is expected not to pass through the $\lambda_\text{emit} \sim 160 \, \mathrm{\upmu m}$ detection, since the upper limit at $\ssim 90 \, \mathrm{\upmu m}$ is also taken into account. The resulting dust mass and temperature are indicated for each fit, as is $\lambda_0^\text{AP}$, the value of $\lambda_0$ that is inferred a posteriori (if applicable). The inferred peak temperature (though measured on the \gls{CMB}-corrected spectrum; \cref{ap:Dust peak temperature measurements}) for the fiducial self-consistent opacity model is annotated (see also \cref{tab:Continuum_fluxes_and_dust_properties}).
	}
	\label{fig:L_IR_constraints}
\end{figure*}

The full \glspl{SED} of four galaxies considered in this work are shown in \cref{fig:L_IR_constraints}. We show one of the two sources for which only upper limits on the dust continuum are available (COS-2987030247), where a range of dust temperatures ($30 \, \mathrm{K} \leq T_\text{dust} \leq 100 \, \mathrm{K}$) and emissivity parameters ($1.5 \leq \beta_\text{\gls{IR}} \leq 2$) for templates that fit the constraints are considered instead of a \gls{MERCURIUS} fit (idem for UVISTA-Z-007 not shown here). These templates are created under an entirely optically thin opacity model (i.e. an \gls{SED} described by \cref{eq:OT_dust_SED_flux_density}), which is a valid assumption since their dust masses are too small to be confidently detected (as shown in \cref{tab:Continuum_fluxes_and_dust_properties}). As can be seen from these templates, the constraints cannot be used to deduce the best-fit parameters, which is why we report (an upper limit on) the (F)IR luminosity and other parameters with a fiducial temperature $T_\text{dust} = 50 \, \mathrm{K}$ and $\beta_\text{\gls{IR}} = 1.5$ (\cref{ssec:Results:Dust_continuum}). From \cref{fig:L_IR_constraints}, it can furthermore be seen that the \gls{IR} luminosity can vary by more than an order of magnitude between the most extreme choices of $T_\text{dust}$ and $\beta_\text{\gls{IR}}$ \citep[see also][]{2020RSOS....700556H}. As described in \cref{ssec:Results:Dust_continuum}, we take the systematic uncertainty resulting from fixing these parameters into account in our estimates of the (F)IR luminosity and obscured \gls{SFR}.

For the three sources for which we have at least one $5 \sigma$ detection (COS-3018555981, UVISTA-Z-001, and UVISTA-Z-019), on the other hand, the \gls{MERCURIUS} fitting routine described above is applied, fixing $\beta_\text{\gls{IR}} = 1.5$ but considering different assumptions on the opacity model: either an entirely optically thin \gls{SED} or the general opacity model, with $\lambda_0$ set self-consistently or fixed to an extreme $200 \, \mathrm{\upmu m}$. The ``best-fit'' line shows the \gls{SED} curve for the maximum likelihood in the $(M_\text{dust}, T_\text{dust})$ plane, shaded regions show the deviation of the \nth{16} and \nth{84} percentiles from the median (i.e. \nth{50} percentile) at each wavelength of all curves produced according to the posterior distribution. We note the best-fit model of COS-3018555981 and UVISTA-Z-019 is expected not to pass through the $\lambda_\text{emit} \sim 160 \, \mathrm{\upmu m}$ detection, since the upper limit at $\ssim 90 \, \mathrm{\upmu m}$ is also taken into account. In the next two sections, we will discuss the inferred dust properties in more detail.\footnote{Unless mentioned otherwise, reported quantities are given as the \nth{50} (i.e. median), \nth{16}, and \nth{84} percentiles (as a $\pm 1 \sigma$ confidence range) of the parameter's marginalised posterior distribution.}

\subsection{The build-up of dust: masses and yields}
\label{ssec:Discussion:Dust_masses}

First, we briefly discuss the estimates of the dust mass of our sources, while stressing these are fairly uncertain, firstly due to the range of possible $\kappa_\nu$ depending on the dust composition as discussed in \cref{ssec:Discussion:Dust_SED_fitting_procedure}. Secondly, our measurements do not probe the \gls{RJ} tail of the \gls{SED}, meaning that the dust mass becomes sensitive on the intrinsic dust temperature: sampling at a rest-frame wavelength $200 \, \mathrm{\upmu m}$ introduces a factor $\ssim 5$ difference in $M_\text{dust}$ for a $2 \times$ difference in $T_\text{dust}$, for instance \citep{2020A&A...634L..14C}. The latter effect, however, is reflected in the uncertainty estimates from the posterior distributions obtained in the \gls{MERCURIUS} fitting routine, which probes a range of temperatures. As will be discussed below, the distributions are based purely on the likelihood of underlying dust masses and temperatures producing the observed continuum (non-)detections and in principle could result in unphysical scenarios (save dust masses above the stellar mass and temperatures below the \gls{CMB} temperature, whose solutions are discarded; \cref{ssec:Discussion:Dust_SED_fitting_procedure}).

For the five sources considered in this work, we find the dust mass typically range upwards of a few million solar masses. Interestingly, the tentatively low dust temperature of COS-3018555981 (discussed in more detail in \cref{ssec:Discussion:Dust_temperatures}) under the fiducial self-consistent general opacity model (where $\lambda_0 \simeq 10 \, \mathrm{\upmu m}$) nominally implies a relatively high dust mass, $M_\text{dust} = 7_{-5}^{+19} \cdot 10^7 \, \mathrm{M_\odot}$. Since the self-consistently derived $\lambda_0$ is well below the lowest sampled wavelength ($\lambda_\text{emit} \sim 90 \, \mathrm{\upmu m}$), an optically thin \gls{SED} appears a rather good approximation producing a similar dust mass and temperature, although the estimated uncertainty in the self-consistent model is larger and thus more conservative (mainly allowing for even higher dust masses that are compensated for by an increasing $\lambda_0$).

The estimated dust mass translates into a substantial dust-to-stellar mass ratio, $M_\text{dust}/M_* = 0.05_{-0.04}^{+0.16}$ (\cref{tab:Continuum_fluxes_and_dust_properties}). Although we stress the current evidence is limited, the formal likelihood treatment evidently yields rather extreme solutions; might we expect the true values be more modest from a physical perspective? We consider dust grains comprise of metals mainly produced by short-lived, massive stars, while most of the stellar mass instead is dominated by low-mass stars, so that $M_\text{dust}/M_*$ cannot exceed the stellar metallicity yield -- the mass-fraction of metals released into the \gls{ISM} compared to the total (remaining) stellar mass -- typically assumed to be a few percent \citep{2019A&ARv..27....3M}. However, higher stellar yields are certainly plausible for a (more) top-heavy \gls{IMF}: for instance, a \citet{2001MNRAS.322..231K} or \citet{2003PASP..115..763C} \gls{IMF} already results in stellar metallicity yields that are twice as high than for a \citet{1955ApJ...121..161S} \gls{IMF} \citep{2016MNRAS.455.4183V}. Moreover, we note the stellar mass estimate of $M_* = 1.4_{-0.2}^{+0.7} \cdot 10^9 \, \mathrm{M_\odot}$ \citep[taken from][]{2018Natur.553..178S} is a conservative one in this context as it is $\ssim 0.5 \, \mathrm{dex}$ higher than the best fit found by the \gls{REBELS} collaboration, who explored several photometric fitting codes \citetext{private communication, REBELS collaboration 2021; see also \citealt{2022ApJ...931..160B}}. Upcoming \gls{JWST} observations (GTO: 1217, GO: 1837) will provide a highly accurate stellar mass and give insight into the star formation history of this source, and hence also its age.

By assuming a stellar \gls{IMF}, the dust-to-stellar mass ratio can be converted into a dust yield, denoted $y_\text{dust}$, which represents the average dust mass formed per dust-producing star. Here, we follow the method of \citet{2015A&A...577A..80M}, who considered \glspl{SN} and \gls{AGB} stars as sources of dust production, assuming their relevant stellar mass ranges are $8$-$40 \, \mathrm{M_\odot}$ and $3$-$8 \, \mathrm{M_\odot}$, respectively. Both yields from \glspl{SN} and \gls{AGB} stars for a \citet{2003PASP..115..763C} \gls{IMF} are shown in \cref{tab:Continuum_fluxes_and_dust_properties}. Average \gls{AGB} dust yields that are obtained under \citeauthor{1955ApJ...121..161S} and top-heavy \glspl{IMF} can be $\ssim 2 \times$ higher, but this only strengthens the interpretation discussed below. More critically, however, the inferred \gls{SN} dust yields could decrease (increase) by a factor of $\ssim 1.5$ for a top-heavy (\citeauthor{1955ApJ...121..161S}) \gls{IMF} since these \glspl{IMF} result in a higher (lower) number of \gls{SN} events, thereby lowering (raising) the required average yield per \gls{SN}.

The yield that would be required per \gls{AGB} star, $y_\text{dust, AGB} \sim 1.5_{-1.1}^{+4.6} \, \mathrm{M_\odot}$ for COS-3018555981, far exceeds the theoretically expected value, $\ssim 0.02 \, \mathrm{M_\odot}$, even more so when the yield is increased up to $2 \times$ under a different \gls{IMF} \citep[as already shown for other \gls{EoR} galaxies in e.g.][]{2015A&A...577A..80M, 2019A&A...624L..13L, 2022ApJ...928...31S}. This renders \gls{AGB} stars an unlikely candidate as the main source of dust formation in this early cosmic epoch (even within the systematic uncertainty of $\kappa_{\nu, \text{ ref}}$). Instead, when \glspl{SN} are considered, a nominally gigantic yield of $y_\text{dust, \gls{SN}} = 4.5_{-3.1}^{+13.3} \, \mathrm{M_\odot}$ per dust-producing star would be required. We note that the uncertainty estimates, apart from incorporating a systematic $10 \%$ flux calibration uncertainty (\cref{ssec:Results:Dust_continuum}), are purely statistical and the \gls{SN} dust yield could systematically be reduced by a high $\kappa_{\nu, \text{ ref}}$ resulting in a lower dust mass (see \cref{ssec:Discussion:Dust_SED_fitting_procedure}). However, as also discussed in \cref{ssec:Discussion:Dust_SED_fitting_procedure}, the current assumption of $\kappa_{\nu, \text{ ref}} = 8.94 \, \mathrm{cm^2 \, g^{-1}}$ is valid for dust produced by \glspl{SN} and thus other choices would be inconsistent for estimating the \gls{SN} dust yield.

Even the lowest required yield within the estimated $1 \sigma$ uncertainties, $\ssim 1.4 \, \mathrm{M_\odot}$, is already in slight tension with the maximum theoretical yield per \gls{SN}, which is $1$ to $2 \, \mathrm{M_\odot}$ without the subsequent dust destruction by the \gls{SN} \citep[e.g.][]{2003ApJ...598..785N}. The discrepancy might even be more serious for various reasons: if dust destruction does play a significant role, this lowers the theoretical maximum yield, while the required dust yield per \gls{SN} is increased under a different \gls{IMF}. Finally, we find yet a lower dust temperature and higher mass and yield increase if we instead assume a higher dust emissivity $1.8 \lesssim \beta_\text{\gls{IR}} \lesssim 2$, the value typically found at high redshift for more massive systems \citep[e.g.][]{2020MNRAS.498.4109J, 2020A&A...634L..14C, 2020MNRAS.498.4192F, 2021ApJ...923..215C} -- this is why we conservatively opt for a fiducial $\beta_\text{\gls{IR}} = 1.5$ here. Hence, there is tentative evidence that other evolutionary mechanisms, such as the growth of dust grains in the \gls{ISM} \citep{2019A&A...624L..13L}, need to be invoked. A final caveat to this result is that the dust mass decreases if $\lambda_0$ is fixed to $200 \, \mathrm{\upmu m}$ (since higher temperatures are feasible in this scenario; \cref{ssec:Discussion:Dust_temperatures}), thereby bringing the dust yield down to $y_\text{dust, \gls{SN}} \sim 3 \, \mathrm{M_\odot}$, which however is still inconsistent with theoretical expectations. Yet this scenario would perhaps be equally surprising, as will be discussed further in \cref{ssec:Discussion:Dust_temperatures}.

\begin{figure*}
	\centering
	\includegraphics[width=0.8\linewidth]{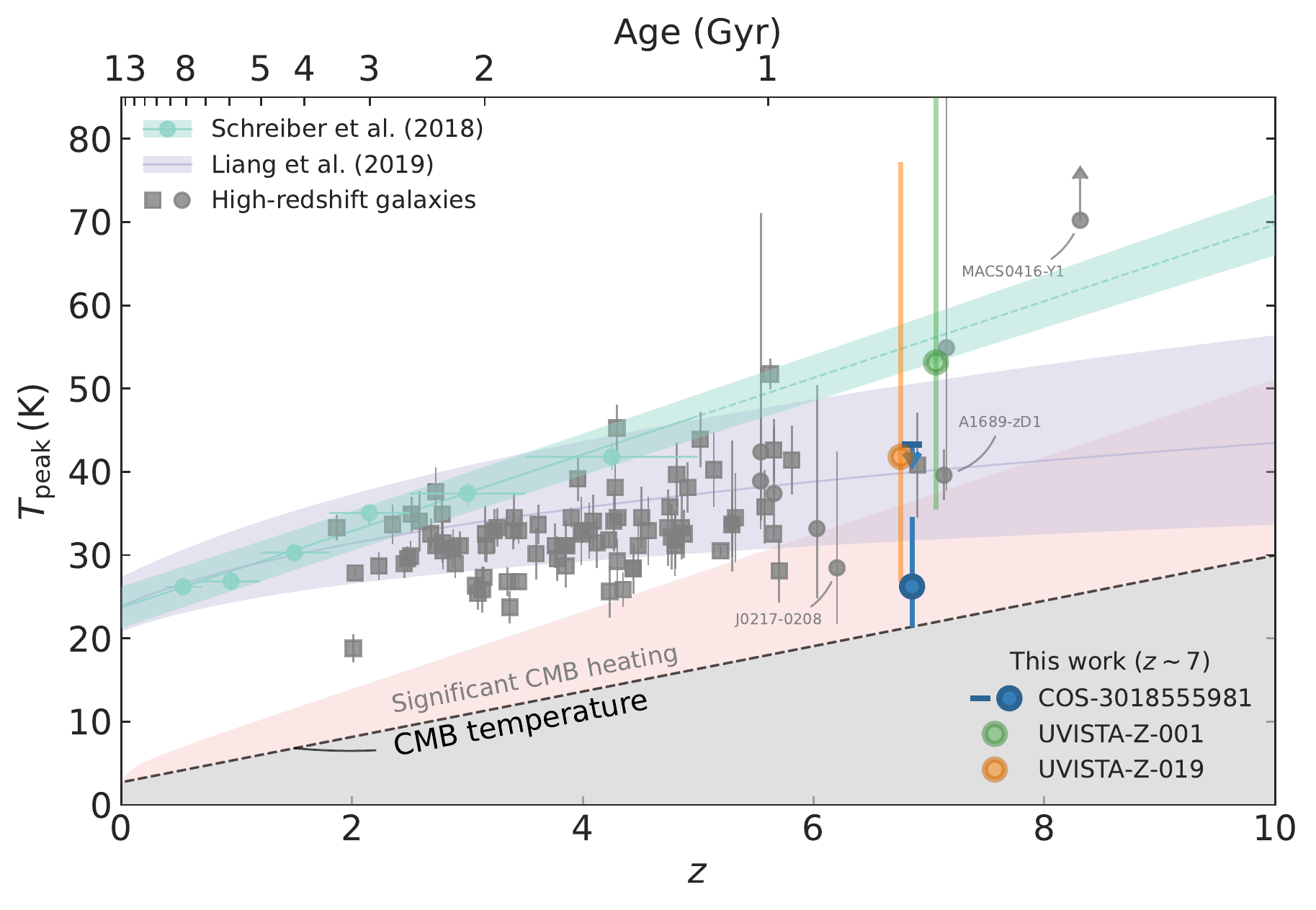}
	\caption[Dust peak temperature as a function of redshift.]{Dust peak temperature, $T_\text{peak}$, as a function of cosmic time. Several peak temperatures derived from measurements presented in this work under the self-consistent general opacity model are drawn: for UVISTA-Z-019, we show its median and $68\%$ confidence range (largely invariant under the chosen opacity model), while for COS-3018555981 an upper limit ($95\%$ confidence) is shown in addition. Further, measurements inferred from stacked spectra and a (partially extrapolated) linear fit by \citet{2018A&A...609A..30S} are shown, as is the power-law fit to the peak-temperature evolution of simulated galaxies by \citet{2019MNRAS.489.1397L}. Details on the literature data points can be found in \cref{ap:Dust peak temperature measurements} (\glspl{SMG} are shown as squares, other high-redshift galaxies as circles with errorbars). Three sources are highlighted with labels, having either a notably low (J0217-0208), high (MACS0416-Y1), or accurately constrained dust temperature (A1689-zD1). The grey dashed line shows the \gls{CMB} temperature over cosmic time, the grey area below representing astrophysically unattainable temperatures -- we note, however, that effective peak temperatures of greybody \glspl{SED} can in principle fall below this boundary as they are lower than the dust temperature itself (\cref{ssec:Discussion:Dust_temperatures}). The light-red area above indicates the region where \gls{CMB} heating has a significant (greater than $1\%$) effect compared to $z=0$ (see \cref{ssec:Discussion:Dust_temperatures} for details).
	}
	\label{fig:T_peak_evolution}
\end{figure*}

\subsection{Cosmic evolution of dust temperature: evidence for extremely cold or highly clustered dust?}
\label{ssec:Discussion:Dust_temperatures}

In the following, we move on to discuss the dust temperature, arguably one of its most important properties as it is essential in deriving other observable quantities such as the infrared luminosity and dust mass from a limited number of photometric data points \citep{2020MNRAS.497..956S, 2022MNRAS.513.3122S}. Unfortunately, establishing the dust temperature is not trivial. Directly fitting observed \glspl{SED} necessarily yields a luminosity-weighted dust temperature, which may be skewed towards high temperatures by hot dust in star-forming regions \citep{2019MNRAS.489.1397L}. Hot dust can only be observed directly, however, if the emission is optically thin \citep{2020MNRAS.498.4192F}. When little information is available, these effects cause the dust temperature, mass, and opacity to become degenerate.

In \cref{fig:T_peak_evolution}, measurements of the \gls{SED} peak temperature across cosmic time are shown. The peak temperature is defined as a purely observational quantity, defined similarly to Wien's displacement law using the \gls{SED} peak wavelength\footnote{The peak of the intrinsic \gls{SED}, i.e. after it has been corrected for observing the dust emission against the \gls{CMB} (\cref{ssec:Discussion:Dust_SED_fitting_procedure}).} in the rest frame, $\lambda_\text{emit, peak}$:
\begin{equation}
    \label{eq:T_peak}
    T_\text{peak} = \frac{b}{\lambda_\text{emit, peak}} \text{, where } b \simeq 2898 \, \mathrm{K / \upmu m}.
\end{equation}

Details on the motivation for comparing peak temperatures, rather than $T_\text{dust}$, and the precise data from the literature that is included can be found in \cref{ap:Dust peak temperature measurements}. We report ``intrinsic'' peak temperatures, in the sense that there is no conversion to the current epoch ($z=0$) to nullify the effect of \gls{CMB} heating at the relevant epoch, for two reasons. Firstly, peak temperatures are an observational construct (lower than the intrinsic dust temperature), and secondly, in most cases, the dust has already largely decoupled from the \gls{CMB} and heating \citep[described by equation (12) of][]{2013ApJ...766...13D} starts to become relevant only when going back in cosmic time significantly or for extremely cold dust that has cooled down nearly to the \gls{CMB} temperature. This is illustrated by the red shaded region, which marks an increase of more than $1\%$ in dust temperature between $z=0$ and the relevant redshift (under $\beta_\text{\gls{IR}} = 1.5$) and falls below almost all measurements shown here (although neglected in \cref{fig:T_peak_evolution}, we will apply this correction when relevant in the following).

Although a consistent picture starts to emerge at $z \lesssim 4$ \citep[e.g.][]{2018A&A...609A..30S, 2022ApJ...930..142D}, it is unclear whether this also applies to galaxies in the \gls{EoR} due to limited observational data and the highly uncertain relative importance of the possible mechanisms of dust formation \citep{2015MNRAS.451L..70M}. Theoretical models predict elevated dust temperatures compared to the local Universe \citep[e.g.][]{2019MNRAS.487.1844M, 2020MNRAS.497..956S, 2022MNRAS.511.4999V}, which is indeed confirmed by observations in some cases \citep[with extreme examples given in][]{2018MNRAS.477..552B, 2020MNRAS.493.4294B, 2022arXiv220314312V}. However, the peak temperatures for several objects -- most notably $T_\text{peak} \sim 28 \, \mathrm{K}$ seen in J0217-0208 \citep{2020ApJ...896...93H} -- are lower than what may be expected from an extrapolation of the trend at low redshift (such as the shown fit by \citealt{2018A&A...609A..30S}; see also \citealt{2020ApJ...902..112B}) and theoretical predictions from cosmological hydrodynamic simulations \citep[e.g.][]{2019MNRAS.489.1397L, 2022MNRAS.511.4999V}. As pointed out by \citet{2022MNRAS.513.3122S}, a large scatter in dust temperatures may be expected at early times owing to strongly varying physical conditions in the complex high-redshift \gls{ISM}. This encourages detailed follow-up studies of objects with seemingly extreme dust properties such as COS-3018555981, as we will discuss in the following.

We place a constraint on the peak temperature through the \gls{MERCURIUS} fitting procedure for COS-3018555981, UVISTA-Z-001, and UVISTA-Z-019, the three sources for which we have at least one confident detection. However, despite a $\ssim 7 \sigma$ detection at $\ssim 160 \, \mathrm{\upmu m}$ (\cref{fig:Dust_continuum_maps}), the upper limit at $\lambda_\text{emit} \sim 90 \, \mathrm{\upmu m}$ is not stringent enough to provide precise constraints on the dust temperature of UVISTA-Z-019. Therefore, we show its median value and $68 \%$ confidence range in \cref{fig:T_peak_evolution} for the fiducial opacity model with self-consistent $\lambda_0$ (as listed in \cref{tab:Continuum_fluxes_and_dust_properties}). While this result does not change significantly under a different opacity model, the estimated uncertainty is so high that within $1 \sigma$ deviations, it is consistent with strongly varying scenarios, confirming the temperature is largely unconstrained. With two detections, we estimate the dust temperature in UVISTA-Z-001 to be $T_\text{dust} = 59_{-20}^{+41} \, \mathrm{K}$. However, due to the matched apertures we have chosen (see \cref{ssec:Results:Dust_continuum}), we note this temperature is representative for the compact component only; the data at $\ssim 160 \, \mathrm{\upmu m}$ reveals an extended dust-continuum emission that is not detected at $\ssim 90 \, \mathrm{\upmu m}$, suggesting a colder dust reservoir exists on larger scales \citep[see also][]{2022arXiv220606939A}.

Similarly, the dust-continuum measurements of COS-3018555981 -- a deep $\ssim 90 \, \mathrm{\upmu m}$ non-detection combined with a significant $\ssim 160 \, \mathrm{\upmu m}$ detection -- favour a low dust temperature for this object. In the fiducial self-consistent opacity model (where we find $\lambda_0 \simeq 10 \, \mathrm{\upmu m}$, i.e. effectively almost optically thin), we even find a nominal intrinsic temperature of $T_\text{dust} = 29_{-5}^{+9} \, \mathrm{K}$ or alternatively, an upper limit of $T_\text{dust} < 48 \, \mathrm{K}$ ($95 \%$ confidence). The median estimate begins to approach $\ssim 21 \, \mathrm{K}$, the \gls{CMB} temperature at $z \sim 7$.\footnote{The lower end of the $68\%$ confidence range actually enters the regime where \gls{CMB} heating plays a significant role such that $T_\text{dust}^{z=0} = 28_{-7}^{+10} \, \mathrm{K}$ when translated to $z=0$.} As noted in \cref{ssec:Discussion:Dust_masses}, the temperature becomes even lower (likewise, the dust mass and dust yield increase) if we instead assume a higher dust emissivity such as $\beta_\text{\gls{IR}} = 2$, which effectively causes the optical depth to drop more quickly towards higher wavelengths. The peak temperature is $T_\text{peak} = 26_{-5}^{+8} \, \mathrm{K}$, shown in \cref{fig:T_peak_evolution} as the blue circle along with an upper limit: $T_\text{peak} < 43 \, \mathrm{K}$ ($95 \%$ confidence). Such low temperatures are at odds with $T_\text{peak}$ measurements of other galaxies at $z \sim 7$ -- with the notable exception of J0217-0208 \citep{2020ApJ...896...93H} -- as well as with theoretical predictions \citep[e.g.][]{2019MNRAS.489.1397L} and extrapolated observed behaviour at low redshift \citep{2018A&A...609A..30S} of the peak-temperature evolution.

In the general opacity case, though, the likelihood of higher peak temperature increases since greybody \gls{SED} templates can, if they are sufficiently optically thick, become possible within the constraints even when their peak wavelength would otherwise fall below $90 \, \mathrm{\upmu m}$ (i.e. $T_\text{peak} \gtrsim 32 \, \mathrm{K}$). When $\lambda_0 = 200 \, \mathrm{\upmu m}$, the upper limit relaxes slightly to $T_\text{peak} < 53 \, \mathrm{K}$ (though the median and $1 \sigma$ confidence ranges are mostly unaffected, as expected when comparing the peak temperature; see \cref{ap:Dust peak temperature measurements} for details). Moreover, this optically thick scenario favours a fit with a significantly higher intrinsic temperature of $T_\text{dust} = 40_{-11}^{+24} \, \mathrm{K}$, albeit somewhat uncertain since a larger range of temperatures is now allowed.

At first glance, an \gls{SED} that only turns optically thin at $\ssim 200 \, \mathrm{\upmu m}$ seems to alleviate the tentative tension of an extremely low dust temperature (and high dust mass and yield, cf. \cref{ssec:Discussion:Dust_masses}). This in itself is a surprising finding, though, since such a high $\lambda_0$ is typically seen in \glsdisp{ULIRG}{ULIRGs} or even \glsdisp{HyLIRG}{HyLIRGs} \citep[e.g.][]{2020MNRAS.498.4192F, 2020A&A...634L..14C}. \glsdisp{LIRG}{LIRGs (luminous infrared galaxies)} are a class of galaxies defined by \gls{IR} luminosity exceeding $10^{11} \, \mathrm{L_\odot}$ (and would therefore include UVISTA-Z-019 and within the uncertainty, possibly even the other three sources; \cref{tab:Continuum_fluxes_and_dust_properties}). The ultra- and hyper-variants, \glsdisp{ULIRG}{ULIRGs} and \glsdisp{HyLIRG}{HyLIRGs}, exhibit an $L_\text{\gls{IR}}$ of more than $10^{12} \, \mathrm{L_\odot}$ and $10^{13} \, \mathrm{L_\odot}$, respectively \citep{1996ARA&A..34..749S}; such extreme systems commonly have dust reservoirs of $10^{8} \, \mathrm{M_\odot}$ or even larger and contain very compact star-forming regions accompanied by high surface densities of \gls{SFR}, gas, and dust \citep{2010MNRAS.403..274C}.

Indeed, we find a comparatively low dust surface mass density of a few $\mathrm{M_\odot \, pc^{-2}}$ (\cref{tab:Continuum_fluxes_and_dust_properties}). To put this in perspective, this is two orders of magnitude lower than $\ssim 500 \, \mathrm{M_\odot \, pc^{-2}}$ found for a massive \gls{HyLIRG} starburst at $z \sim 4$ for which $\lambda_0 = 170 \pm 23 \, \mathrm{\upmu m}$ \citep{2020A&A...634L..14C}. Hence, the resulting a-posteriori estimates for $\lambda_0$ in an opacity model with fixed $\lambda_0 = 200 \, \mathrm{\upmu m}$ are more than an order of magnitude lower than what is assumed ($\lambda_0^\text{AP} \sim 5 \, \mathrm{\upmu m}$; see \cref{fig:L_IR_constraints}). This discrepancy would point towards the need of a highly clustered distribution of dust in order to produce a self-consistent optical depth: from \cref{eq:Dust_mass_absorption_coefficient_definition,eq:Consistent_l0}, this would require $C \gg 1$ ($C \sim 10^3$ to increase $\lambda_0^\text{AP}$ by a factor of $100$, for $\beta_\text{\gls{IR}} = 1.5$).

In conclusion, the results lead to a paradox with a remarkably low dust temperature in combination with high dust mass and yield on the one hand (optically thin \gls{SED}), or the need of a highly clustered dust distribution (optically thick \gls{SED}) on the other. In either case, follow-up observations are needed to clarify the tantalising dust properties of COS-3018555981.

\begin{figure}
	\centering
	\includegraphics[width=\linewidth]{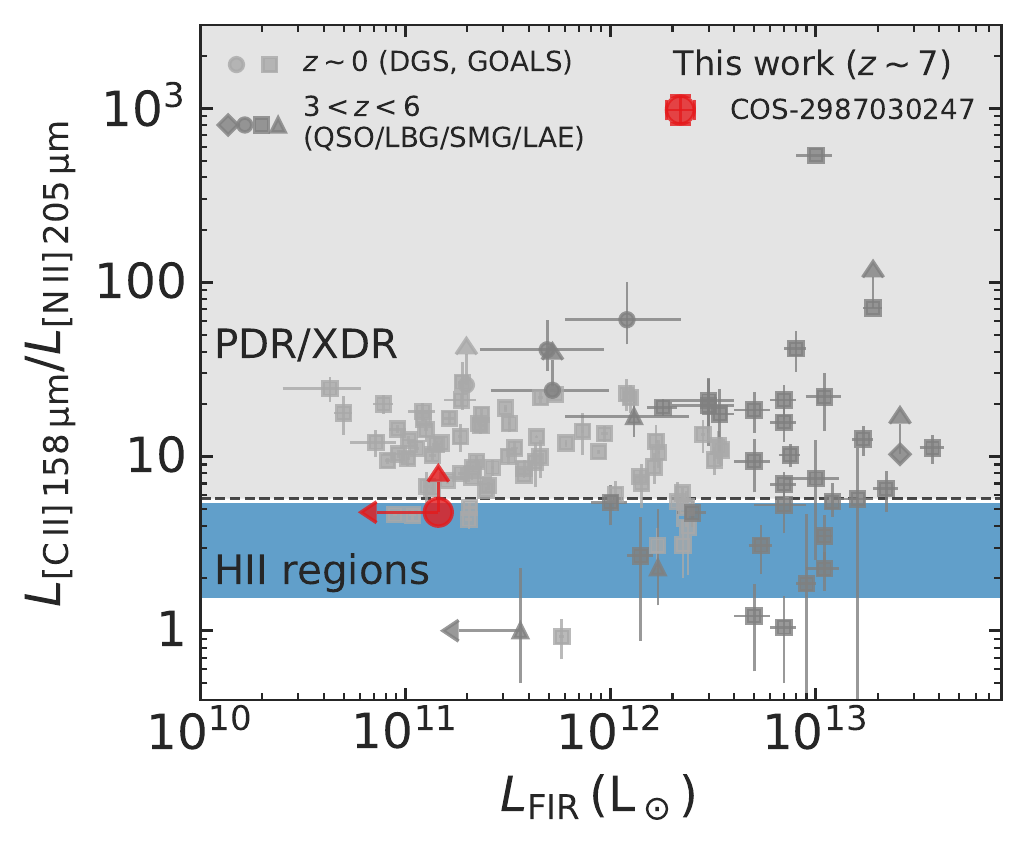}
	\caption[Luminosity ratio of \CII\ over \NII\ as a function of $L_\text{\gls{FIR}}$.]{Luminosity ratio of \CIILam\ over \NIILam\ as a function of the \gls{FIR} luminosity. Local galaxies shown are from \gls{GOALS} \citep{2017ApJ...846...32D} and \gls{DGS} \citep{2015A&A...578A..53C, 2019A&A...626A..23C}; constraints on high-redshift galaxies \citep{2014ApJ...782L..17D, 2016ApJ...832..151P, 2019ApJ...882..168P} and \glspl{SMG} as part of the \gls{SPT} survey \citep{2020MNRAS.494.4090C} are included as well (see \cref{ssec:Discussion:CII/NII} for details). \glspl{LBG} are indicated with circles, \glspl{SMG} with squares, \glsentryshort{LAE}s\glsreset{LAE} with upward facing triangles, and the \gls{QSO}\glsreset{QSO} as a diamond. The luminosity ratio expected for \CII\ emission originating in \HII\ regions for typical physical conditions (solar metallicity and $\log_{10} U < -3$, $n_\text{H} < 10^2 \, \mathrm{cm^{-3}}$; see \cref{ssec:Discussion:CII/NII}) is shown as the blue shaded area. A dashed line demarcates $L_\CII/L_\NII \simeq 6$, the minimum value found across a large range of \gls{PDR} models \citep[grey shaded area; a similar neutral medium can be created in the presence of \glspl{XDR}\glsreset{XDR} or shocked gas, see e.g.][]{2014ApJ...782L..17D}. For COS-2987030247, the $3 \sigma$ upper limit on \NII\ is fully consistent with \CII\ production in a \gls{PDR}-like medium.
	}
	\label{fig:CII/NII_ratio}
\end{figure}

\section{Emission line properties}
\label{sec:Discussion:Emission_line_properties}

\subsection{On the origin of \texorpdfstring{\NIILam}{[NII] 205 μm} and \texorpdfstring{\CIILam}{[CII] 158 μm} emission}
\label{ssec:Discussion:CII/NII}

The line ratio of \CII\ to \NII\ is a well-known probe of the ionisation state of the \gls{ISM} \citep[e.g.][]{2012A&A...542L..34N}. The \NIILam\ transition namely has a critical density very close to that of \CIILam\ ($\ssim 50 \, \mathrm{cm^{-3}}$ in ionised gas; e.g. \citealt{2013ARA&A..51..105C}), while the ionisation potentials of C$^+$ ($24.4 \, \mathrm{eV}$) and N$^+$ ($29.6 \, \mathrm{eV}$) are also comparable \citep[e.g.][]{2014ApJ...782L..17D}. In addition, the ionisation potential of neutral carbon ($11.3 \, \mathrm{eV}$) and nitrogen ($14.5 \, \mathrm{eV}$) are similar, but crucially just below and above that of hydrogen ($13.6 \, \mathrm{eV}$). Therefore, while \NII\ strictly originates in regions where hydrogen is (largely) ionised, \CII\ emission can arise in both \HII\ regions and \glspl{PDR}. For this reason, the line ratio of \CII\ to \NII\ is a powerful proxy of the ionisation state of the \CII-producing gas, independent of, for example, the ionisation parameter \citep[e.g.][]{2012A&A...542L..34N}. In \cref{fig:CII/NII_ratio}, we compare the \CII/\NII\ ratio of COS-2987030247 (lower limit of $L_\CII/L_\NII > 4.8$ at $3 \sigma$, see \cref{ssec:Results:NII}) against a number of sources from the literature, as a function of the \gls{FIR} luminosity.

In this figure, several local galaxies are shown, all observed by \textit{Herschel}/PACS: local \glspl{ULIRG} from the \gls{GOALS} survey \citep{2017ApJ...846...32D}, as well as one lower limit that is part of the \glsentrylong{DGS} \citep[\glsentryshort{DGS}\glsunset{DGS};][]{2015A&A...578A..53C, 2019A&A...626A..23C}; both surveys will be discussed in more detail in \cref{ssec:Discussion:OIII/CII_in_the_local_and_high-redshift_Universe}. Furthermore, a sample of $3 \lesssim z \lesssim 6$ \glspl{SMG} observed by the \gls{SPT} is shown (\citealt{2020MNRAS.494.4090C}; \gls{FIR} luminosities are presented in \citealt{2020ApJ...902...78R}). We also include several individual high-redshift sources: a \gls{QSO}, \gls{SMG}, and two \glspl{LAE} presented in \citet{2014ApJ...782L..17D}, and several \glspl{LBG} from \citet{2016ApJ...832..151P, 2019ApJ...882..168P}.

While \citet{2014ApJ...782L..17D} report the typical luminosity ratio expected for \CII\ emission originating in \HII\ regions to be $L_\CII/L_\NII \sim 2$, we indicate here the range of line ratios found across a large range of photoionisation models, which will be introduced in \cref{ssec:Discussion:OIII/CII_photoionisation_models}. We show the luminosity ratio expected for pure \HII\ regions with typical gas conditions (solar metallicity and $\log_{10} U < -3$, $n_\text{H} < 10^2 \, \mathrm{cm^{-3}}$), but we note the ratio can increase with a lower metallicity, higher ionisation parameter, and/or higher gas density (in addition to a sub-solar C/N ratio). The line ratios predicted for models incorporating a \gls{PDR} (see \cref{ssec:Discussion:OIII/CII_photoionisation_models} for details) can be substantially higher, suggesting the line ratios of the majority of the observations are consistent with an additional component of \CII\ being produced in a neutral \glspl{PDR} environment \citep[or similarly in \glspl{XDR} or shocked gas;][, and references therein]{2014ApJ...782L..17D}. For COS-2987030247, the $3 \sigma$ upper limit on \NII\ is fully consistent with \CII\ production in a \gls{PDR}-like medium. While it does simultaneously allow for \HII\ regions as the primary origin of the \CII\ emission, the upper limit suggests this scenario is unlikely (for reasonable physical conditions) at a $\ssim 2 \sigma$ level.

\begingroup
    \renewcommand{\arraystretch}{1.25} % Default value: 1
    \begin{table}
        \centering
        \caption[Molecular gas masses.]
        {Molecular gas masses, $M_\text{mol}$, estimated from the \CIILam\ luminosity (\cref{tab:Line_fluxes}) using the conversion given in \citet{2018MNRAS.481.1976Z}. Also given are $M_\text{mol}/M_\text{dust}$ and $M_\text{mol}/M_*$, (lower limits of) the gas-to-dust and gas-to-stellar mass ratios (see \cref{ssec:Discussion:CII/NII} for details).}
        \label{tab:Gas_masses}
        \begin{tabular}{llll}
            \hline
            Source & $M_\text{mol} \, (10^9 \, \mathrm{M_\odot})$ & $M_\text{mol}/M_\text{dust}$ & $M_\text{mol}/M_*$ \\
            \hline
            \begin{filecontents*}{ALMA_molecular_gas_mass.csv}
field|object|LCII|Mgas|MgMd|MgMs
COSMOS|\setulcolor{COS2987}\ul{COS-2987030247}|$2.8 \pm 0.7$|$8.8 \pm 4.8$|$\gtrsim 2.5 \cdot 10^{3}$|$5.2_{-3.1}^{+2.9}$
COSMOS|\setulcolor{COS3018}\ul{COS-3018555981}|$4.0 \pm 0.3$|$12.3 \pm 2.2$|$1.7_{-1.3}^{+3.7} \cdot 10^{2}$|$8.9_{-3.4}^{+2.2}$
UVISTA|\setulcolor{UVIS001}\ul{UVISTA-Z-001}|$6.7 \pm 1.1$|$20.9 \pm 7.9$|$5.8_{-4.5}^{+13.3} \cdot 10^{3}$|$5.5_{-2.3}^{+7.1}$
UVISTA|\setulcolor{UVIS007}\ul{UVISTA-Z-007}|$4.6 \pm 0.7$|$14.2 \pm 5.0$|$\gtrsim 1.2 \cdot 10^{3}$|$3.8_{-2.5}^{+6.8}$
UVISTA|\setulcolor{UVIS019}\ul{UVISTA-Z-019}|$7.7 \pm 0.6$|$23.8 \pm 3.8$|$1.2_{-0.9}^{+4.5} \cdot 10^{3}$|$7.4_{-2.9}^{+4.0}$
\end{filecontents*}
\csvreader[separator=pipe, late after line=\\, head to column names]{ALMA_molecular_gas_mass.csv}{}{\object & \Mgas & \MgMd & \MgMs}
            \hline
        \end{tabular}
    \end{table}
\endgroup

Having established that a contribution to the \CIILam\ emission of COS-2987030247 is likely traced to a neutral, \gls{PDR}-like medium, we expect the same to be true for the other three $z \sim 7$ star-forming galaxies with similar physical properties. Indeed, this is in line with recent findings of high \ion{H}{I} gas mass fractions at early times \citep{2021ApJ...922..147H, 2022arXiv220607763H}. Moreover, \citet{2018MNRAS.481.1976Z} have showed there exists a linear relation between the \CII\ luminosity and the molecular gas mass of a galaxy, regardless of its gas depletion time (i.e. starburst or main-sequence phase), redshift, and metallicity. Since all galaxies have confident \CII\ detections, we can easily convert $L_\CII$ into an estimate of $M_\text{mol}$, the molecular gas mass, though we caution our sources are beyond the redshift (and possibly metallicity) regime at which \citeauthor{2018MNRAS.481.1976Z} established the calibration. We adopted their median conversion factor of $\alpha_\CII = 31 \, \mathrm{M_\odot \, L_\odot^{-1}}$ and a systematic $0.3 \, \mathrm{dex}$ spread. The results are shown in \cref{tab:Gas_masses}. We also provide estimates on the gas-to-dust and gas-to-stellar mass ratios, which, in the form of $M_\text{mol}/M_\text{dust}$ and $M_\text{mol}/M_*$ respectively, are strictly lower limits since we do not take into account the \ion{H}{I} gas mass. Interestingly, we find a modest gas-to-dust ratio for COS-3018555981 (albeit a lower limit), which implies a high metallicity \citep{2014A&A...563A..31R}. We note this would be consistent with the findings in \cref{ssec:Discussion:Dust_masses} requiring a high metal yield.

We now turn to a discussion focused on the \OIIILam\ line in comparison to \CIILam\ and in the context of the optical \OIII\ lines (inferred from \textit{Spitzer} broadband photometry so that these include $\OIII \, \lambda \lambda \, 4960, \, 5008 \, \Angstrom$ and \Hbeta; see \cref{ssec:Observations:Target_selection}).

\begin{figure}
	\centering
	\includegraphics[width=\linewidth]{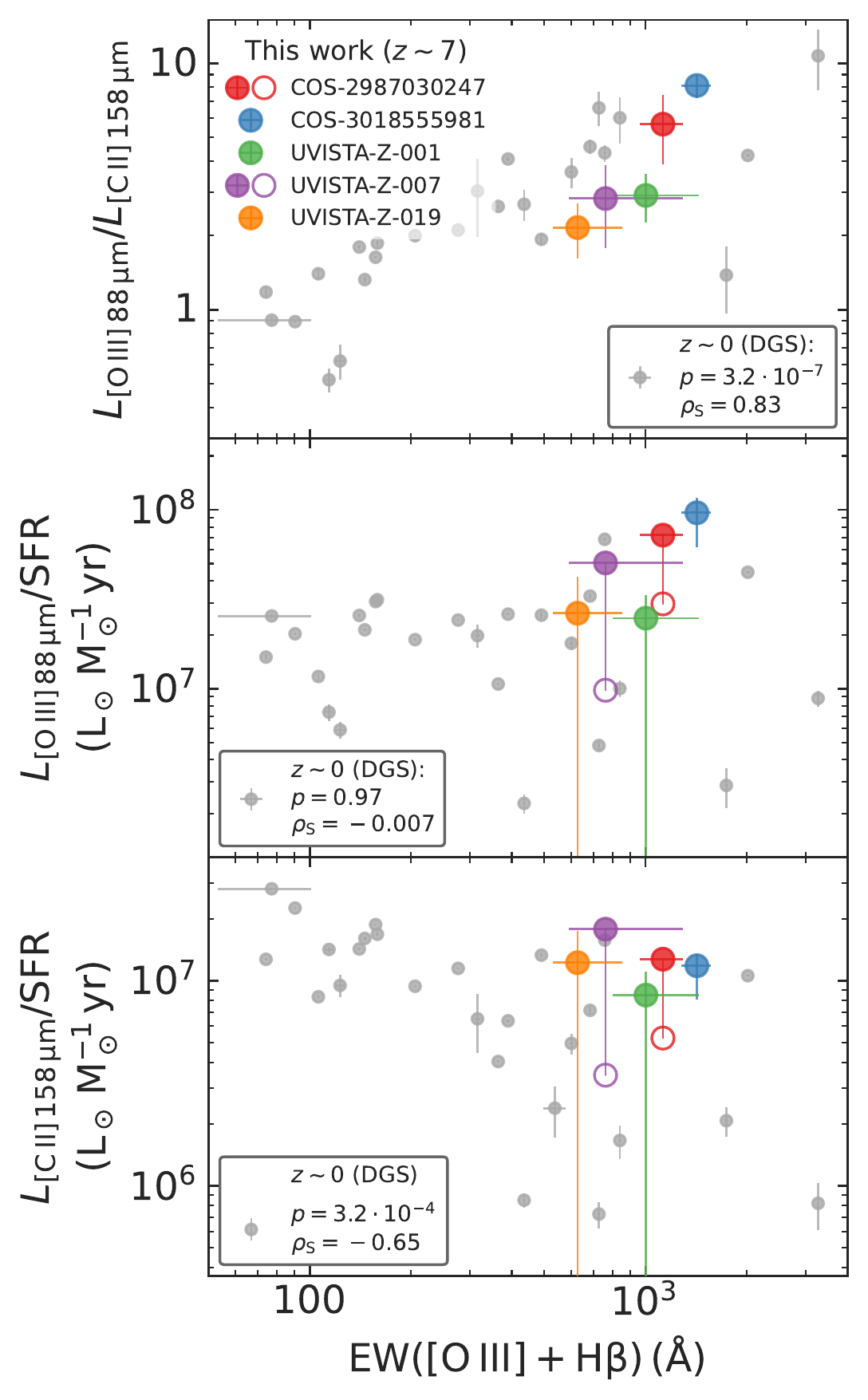}
	\caption[Relative luminosities of \OIIILam\ and \CIILam\ as a function of $\text{\gls{EW}} ( \OIII + \Hbeta )$.]{Ratios of the \OIIILam\ and \CIILam\ luminosities with respect to each other (top panel) and with respect to the \gls{SFR} (middle panel for \OIII, bottom panel for \CII) as a function of the \gls{EW} of the optical \OIII\ and \Hbeta\ lines. If the measured \gls{SFR} is an upper limit (due to a non-detection of the dust continuum), unobscured \gls{SFR} measurements are shown with an open circle. Small grey circles are galaxies from \gls{DGS} (see \cref{ssec:Discussion:OIII/CII_in_the_local_and_high-redshift_Universe}). The $p$-value, for a null hypothesis that the data are uncorrelated, and Spearman's rank correlation coefficient ($\rho_\text{S}$) of these points measured in logarithmic space are indicated in each panel. Coloured circles show the five $z \sim 7$ galaxies considered in this work. There is a significant positive correlation between the \OIII/\CII\ ratio and their optical \gls{EW} among the \gls{DGS} galaxies, which seems to derive mostly from a negative correlation between \CII/\gls{SFR} and the \gls{EW}; however, the correlation of the \OIII/\CII\ luminosity ratio is stronger than that of the (relative) \CII\ luminosity on its own. Furthermore, it is unclear whether the same is true for the small sample of high-redshift galaxies.
	}
	\label{fig:OIII/CII_correlations}
\end{figure}

\subsection{\texorpdfstring{\OIIILam}{[OIII] 88 μm} and \texorpdfstring{\CIILam}{[CII] 158 μm} in the local and high-redshift Universe}
\label{ssec:Discussion:OIII/CII_in_the_local_and_high-redshift_Universe}

In the local Universe, two main samples of galaxies with measurements of several \gls{FIR} lines (including \OIII\ and \CII) exist, as mentioned in \cref{ssec:Discussion:CII/NII}: firstly, \gls{GOALS} has observed over $200$ \glspl{ULIRG} at $z \sim 0$ with \textit{Herschel}/PACS and \textit{Herschel}/SPIRE \citep{2017ApJ...846...32D}; secondly, the \gls{DGS} uses \textit{Herschel}/PACS to investigate just under $50$ metal-poor dwarf galaxies \citep{2013PASP..125..600M, 2014PASP..126.1079M, 2014A&A...568A..62D, 2015A&A...578A..53C, 2019A&A...626A..23C}. Among the dwarf galaxies of the \gls{DGS}, the luminosity ratio of \OIIILam\ to \CIILam\ varies between $0.5$ and $13$ with a median of $2$, whereas the median ratio for $159$ of the \glspl{ULIRG} where both lines are detected is an order of magnitude smaller (namely $\ssim 0.2$, varying between $0.04$ and $3$). \citet{2015A&A...578A..53C} argue that a large filling factor of diffuse, ionised gas emitting \OIIILam\ is needed to explain the observed ratios in dwarf galaxies. Crucially, metal-poor systems may therefore have a more porous \gls{ISM} through which ionising radiation could escape, which implies that the \OIII/\CII\ ratio can indicate leakage of \gls{LyC} radiation \citep[see also][]{2016Sci...352.1559I, 2020MNRAS.498..164K}.

Similar to the metal-poor dwarf galaxies, star-forming galaxies in the \gls{EoR} have been shown to exhibit systematically high \OIII/\CII\ ratios between $1$ and $10$ \citep{2020ApJ...896...93H}, which is confirmed by our findings of (integrated) line ratios between around $2$ and $8$ (\cref{tab:Line_fluxes}). Given the limited amount of significant detections and ancillary observations, however, it is not straightforward to disentangle the contributions of various physical properties that influence the observed ratios of galaxies in the early Universe. First of all, there may be certain observational systematics or selection biases. For instance, flux measurements in interferometric data sets have to be treated carefully, since the spatial extent of the \CII\ emission has been shown to be significantly larger than that of \OIII\ \citep[][; see also \cref{sssec:Observations:ALMA_data_reduction}]{2018MNRAS.478.1170C, 2020MNRAS.499.5136C}. Even when accounting for the extended \CII\ emission, various studies have suggested that the typical \gls{EoR} galaxy has even higher \OIII/\CII\ ratios than found even in local metal-poor dwarf galaxies \citep{2020MNRAS.499.5136C, 2020ApJ...896...93H}. High \OIII/\CII\ ratios are generally indicative of strong bursts of star formation with short gas depletion times ($t_\text{dep} \lesssim 50 \, \mathrm{Myr}$; \citealt{2021MNRAS.505.5543V}), which may be common among the \gls{UV}-bright galaxies at high redshift that have typically been selected for follow-up spectroscopy with \gls{ALMA}.

The \gls{sSFR} is expected to scale inversely with the age of the stellar population in a galaxy, implying that such strong, recent bursts of star formation should be indicated by a high \gls{sSFR} \citep{2021MNRAS.505.5543V}. Indeed, galaxies at high redshift ($z \gtrsim 6$) that have been observed in both \OIIILam\ and \CIILam\ \citep[see e.g.][]{2020MNRAS.499.5136C} are characterised by higher \glspl{sSFR} (roughly $1$ to $2 \, \mathrm{dex}$) than galaxies from the local \gls{DGS} and \gls{GOALS} samples.
However, the \gls{DGS} and \gls{GOALS} samples share a similar median \gls{sSFR} of $0.31 \, \mathrm{Gyr^{-1}}$ and $0.36 \, \mathrm{Gyr^{-1}}$ respectively, despite having a factor of $\ssim 10$ difference in their \OIII/\CII\ ratios, making the global \gls{sSFR} of galaxies a poor indicator of high \OIII/\CII\ emission. Potentially, the local dwarf galaxies have a (much) older stellar component \citep[e.g.][]{1998ARA&A..36..435M}, which lowers their global stellar mass and overall \gls{sSFR}, while merely their young star-forming regions -- possibly fuelled by recent inflows of gas -- are thought to have the most similar \gls{ISM} conditions to high-redshift galaxies. Alternatively, the time scales on which \glspl{SFR} are measured may be too long to give an accurate indication of their \OIII/\CII\ ratio, since most galaxies in \gls{DGS} and \gls{GOALS} have \glspl{SFR} measured through their (\gls{UV} and) \gls{IR} luminosity (probing time scales on the order of $100 \, \mathrm{Myr}$; e.g. \citealt{2012ARA&A..50..531K}).

We therefore consider an alternative (inverse) tracer of starburst age, the \gls{EW} of the optical \OIII\ and \Hbeta\ lines, which makes a fairer comparison between local and distant galaxies specifically when the \gls{EW} is measured across the young star-forming regions of local dwarf galaxies.\footnote{The \gls{EW} measurements of \gls{DGS} sample will be presented in Kumari et al. (in prep.).} In the top panel of \cref{fig:OIII/CII_correlations}, we show the \OIII/\CII\ luminosity ratio as a function of the \gls{EW} of the optical \OIII\ and \Hbeta\ lines both for the \gls{DGS} and $z \sim 7$ galaxies considered in this work.\footnote{We disregard the \gls{GOALS} sample here, since their \OIII\ and \Hbeta\ \glspl{EW} are not available.} Among the metal-poor dwarf galaxies, there is a clear positive correlation: the Spearman's rank correlation coefficient measured in logarithmic space is $\rho_\text{S} \simeq 0.83$ (with a $p$-value of $3.2 \cdot 10^{-7}$); this is noticeably higher than when \gls{EW} is swapped for \gls{sSFR}, where $\rho_\text{S} \simeq 0.35$ ($p \simeq 0.056$). In the bottom two panels of \cref{fig:OIII/CII_correlations}, we investigate whether this effect is mostly due to brighter \OIIILam\ line or a \CIILam\ deficit towards higher \glspl{EW} (and hence younger ages). The correlation coefficients suggest it derives mostly from a negative correlation between \CII/\gls{SFR} and the \gls{EW} ($\rho_\text{S} \simeq -0.65$ with $p \simeq 3.2 \cdot 10^{-4}$) rather than between \OIII/\gls{SFR} and \gls{EW} ($\rho_\text{S} \simeq -0.007$ with $p \simeq 0.97$); however, the correlation of the \OIII/\CII\ luminosity ratio is stronger than that of the (relative) \CII\ luminosity on its own. Furthermore, it is unclear whether the same is true for the small sample of high-redshift galaxies (where indeed the opposite seems true).

To investigate whether the observed line intensities line up with theoretical expectations, we will introduce a set of photoionisation models aimed to mimic the expected physical condition of the \gls{ISM} in the following.

\subsection{Comparing observed emission line strengths to the predictions of photoionisation models}
\label{ssec:Discussion:OIII/CII_photoionisation_models}

To gain further insight into the physical conditions of the \gls{ISM}, we employed one-dimensional, radiative-transfer models of a plane-parallel nebula in \program{Cloudy} \citep[v17.02;][]{2017RMxAA..53..385F} to simulate an \HII\ region that smoothly transitions into a \gls{PDR}. As the incident radiation field, we used a single burst of star formation with varying ages generated by \program{bpass} v2.1 \citep{2017PASA...34...58E} under a \citeauthor{1955ApJ...121..161S} \gls{IMF}, ranging in stellar mass from $1$ to $100 \, \mathrm{M_\odot}$. Motivated by for example the observed extreme optical \glspl{EW}, we choose the set of \program{bpass} models where interacting binary stars are included \citep[significantly boosting the ionising flux; see e.g.][]{2016MNRAS.456..485S}. In addition, the \gls{CMB} at $z=7$ and a cosmic ray background \citep[crucial to the \gls{PDR} physics; see][]{2005AAS...207.8117A} are included in all models. Following \citet{2019A&A...626A..23C}, we chose a density law that increases linearly with column density, resulting in an exponential density profile that is nearly constant in the \HII\ region.\footnote{This circumvents the more extreme assumptions of a constant density throughout the \HII\ region and \gls{PDR} or constant pressure, which gives rise to a density discontinuity at the boundary layer between them.} As in \citet{2020ApJ...896...93H}, we explored stellar metallicities of $Z_*/\mathrm{Z_\odot} \in \left\{0.05, 0.2, 1 \right\}$ where, as a baseline model, we set the gas metallicity $Z_\text{neb}$ equal to that of the stars under the default abundance pattern in \program{Cloudy} (which has $12 + \log \left ( \text{O/H} \right)_\odot = 8.69$; \citealt{2009ARA&A..47..481A}), except for helium which we set according to equation (1) in \citet{2004ApJS..153...75G}.

We varied the ionisation parameter between $\log_{10} U = -4$ and $-0.5$, in steps of $0.5$, while fixing the hydrogen density to $n_\text{H} \sim 10^2 \, \mathrm{cm^{-3}}$ (both defined at the illuminated face of the cloud). Though we do explore different gas densities (as discussed below), the choice for this particular value was motivated both by observations at high redshift \citep[e.g.][]{2016ApJ...816...23S} and typical densities found in local dwarf galaxies that show similarly strong \OIIILam\ emission relative to their \gls{SFR}, such as Mrk 209 \citep{1997ApJS..108....1I} and \textsc{\lowercase{II}} Zw 40 \citep{2000ApJ...531..776G}. Graphite and silicate dust grains with Orion-like size distribution and abundance were included, along with a self-consistent abundance correction to account for the depletion of metals onto grains. Finally, we set the turbulent velocity of the gas to $1.5 \, \mathrm{km/s}$ \citep{2019A&A...626A..23C}. The calculation is run until a visual extinction $A_V = 10 \, \mathrm{mag}$ is reached to ensure the full \CIILam\ luminosity is captured, since \CII\ predominantly arises in \glspl{PDR} \citep[e.g.][]{2005AAS...207.8117A}. The \gls{SFR} is measured -- as a surface density $\Sigma_\text{\gls{SFR}}$ in $\mathrm{M_\odot \, yr^{-1} \, kpc^{-2}}$, since the geometry is plane parallel -- through the incident \gls{UV} continuum (i.e. the stellar radiation field at the illuminated face of the cloud, which is not attenuated by dust) at $1550 \, \Angstrom$ via the conversion given by \citet{2012ARA&A..50..531K}.\footnote{We note that the conversion of \citet{2012ARA&A..50..531K} assumes a \citet{2003ApJ...598.1076K} \gls{IMF}, differing from \program{bpass} that generates the incident radiation field with a \citeauthor{1955ApJ...121..161S} \gls{IMF}. However, this is consistent with the conversion used for the observed \glspl{SFR} (see \cref{sec:Discussion:Dust_properties}); an additional translation between \glspl{IMF} would therefore shift both models and observations by an equal amount.}

In addition to these ``vanilla'' simulations of a \HII\ region combined with a \gls{PDR}, we also consider various deviations from the default scenario, whose impact will be discussed further in the next sections. Firstly, we varied the hydrogen density between $\log_{10} ( n_\text{H} \, [\mathrm{cm^{-3}}] ) = 0.5$ and $3$, in steps of $0.5$. Secondly, we ran simulations that stop when molecules first start to form, changing the stopping criterion in \program{Cloudy} to reaching a molecular hydrogen fraction of $10^{-6}$ instead of reaching a certain visual extinction, thereby effectively stripping away the \gls{PDR} and leaving only the bare \HII\ region. Thirdly, motivated by the observed $\mathrm{\upalpha/Fe}$ ratio at high redshift \citep[e.g.][]{2016ApJ...826..159S}, we considered models with an enhanced nebular $\mathrm{\upalpha/Fe}$ ratio, accomplished by scaling up the individual abundances of $\upalpha$ elements (C, O, Ne, Mg, Si, and S) in the gas by a given factor $X_\mathrm{\upalpha/Fe, \, neb}$. This results in an enhancement by the same factor of the $\mathrm{\upalpha/Fe}$ ratio relative to the default solar abundances:
\begin{equation*}
    X_\mathrm{\upalpha/Fe, \, neb} = \frac{\mathrm{\left( \upalpha/Fe \right)_{neb}}}{\mathrm{\left( \upalpha/Fe \right)}_\odot} = 10^{\left[ \mathrm{\upalpha/Fe} \right]_\text{neb}},
\end{equation*}

\noindent such that $X$ is simply the linear version of the logarithmic relative number density \citep[bracket notation;][]{2019A&ARv..27....3M}. In this case, we additionally swap the solar metallicity case for $Z_* = 0.005 \, \mathrm{Z_\odot}$. For our default $\upalpha$-enhancement of choice, $X_\mathrm{\upalpha/Fe, \, neb} = 4$ (as seems appropriate from the findings discussed in the following \cref{ssec:Discussion:OIII_lines}), the effective nebular metallicities for $\upalpha$ elements correspond to $Z_\text{neb}/\mathrm{Z_\odot} = 0.02$, $0.2$, and $0.8$ (for $Z_*/\mathrm{Z_\odot} = 0.005$, $0.05$, and $0.2$, respectively). Fourthly, we similarly scaled down the carbon abundances by a certain factor, reflecting a sub-solar C/O ratio that may be appropriate in the early Universe, as will be discussed in \cref{ssec:Discussion:OIII/CII_theoretical_insights}. Finally, we increased the rate of cosmic rays (to $10 \times$ the \program{Cloudy} default) that stimulates \gls{PDR} heating and line intensities \citep[notably that of \CIILam; e.g.][]{2019A&A...626A..23C}. In the following, we discuss the theoretical interpretations that can be applied to observed \OIIILam\ and \CIILam\ luminosities.

\begin{figure}
	\centering
	\includegraphics[width=\linewidth]{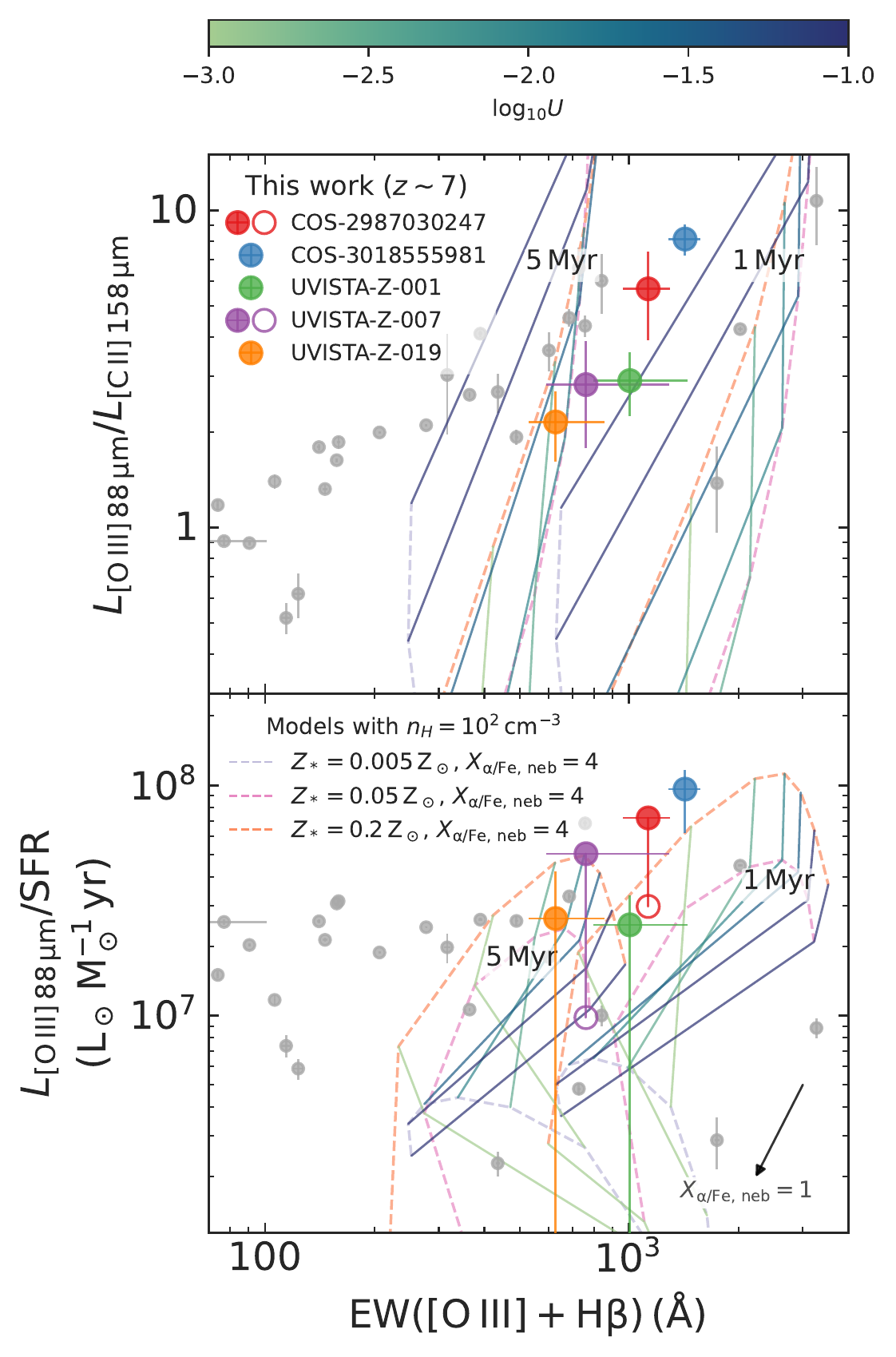}
	\caption[Ratios of \OIIILam\ over \CIILam\ and \OIIILam\ over \gls{SFR} as a function of $\text{\gls{EW}} ( \OIII + \Hbeta )$.]{Relative strength of \OIIILam\ with respect to \CIILam\ and \gls{SFR} as a function of the \gls{EW} of the optical \OIII\ and \Hbeta\ lines. Small grey circles are dwarf galaxies from the \gls{DGS} (see \cref{ssec:Discussion:OIII/CII_in_the_local_and_high-redshift_Universe}), coloured circles are the five $z \sim 7$ galaxies considered in this work. Top panel: luminosity ratio of \OIII\ over \CII. Grids of \program{Cloudy} models discussed in \cref{ssec:Discussion:OIII/CII_photoionisation_models} with $n_\text{H} = 10^2 \, \mathrm{cm^{-3}}$ and stellar ages of $1$ (right) and $5 \, \mathrm{Myr}$ (left) are shown as grids, where the coloured dashed lines indicate a constant metallicity relative to Solar (see legend in the bottom panel). The gas in the main grids shown here has an $\mathrm{\upalpha/Fe}$ ratio enhanced by a factor $4$ (see \cref{ssec:Discussion:OIII/CII_photoionisation_models} for details). Solid lines of fixed ionisation parameter are coloured according to the colourbar on the top. Bottom panel: ratio of \OIII\ luminosity over \gls{SFR}. The same \program{Cloudy} models as described above are overlaid. An arrow indicates the average shift of the models ($\ssim 0.5 \, \mathrm{dex}$) if a solar $\mathrm{\upalpha/Fe}$ ratio is considered (i.e. $X_\mathrm{\upalpha/Fe, \, neb} = 1$). The observations of \OIII/\gls{SFR} at $z \sim 7$ (which, if they are limits, show unobscured \gls{SFR} measurements with open circles) are mostly reproduced by \program{Cloudy} models, provided the metallicity is nearly at a solar level (\cref{ssec:Discussion:OIII_lines}).
	}
	\label{fig:OIII_and_CII_vs_EW(OIII+Hbeta)}
\end{figure}

\subsection{Insights on \texorpdfstring{\OIIILam}{[OIII] 88 μm} and \texorpdfstring{\CIILam}{[CII] 158 μm}}
\label{ssec:Discussion:OIII/CII_theoretical_insights}

As discussed in \cref{ssec:Discussion:OIII/CII_in_the_local_and_high-redshift_Universe}, \gls{ALMA} observations of the \OIIILam\ and \CIILam\ lines in several \gls{EoR} galaxies have hinted at high \OIII/\CII\ ratios compared to local \glspl{ULIRG} observed in \gls{GOALS} and similar to local metal-poor dwarf galaxies from \gls{DGS} with the highest $\text{\gls{EW}} ( \OIII + \Hbeta )$. From a theoretical perspective, there are many factors that could explain these high line ratios, if they are not purely caused by observational effects (as discussed in \cref{ssec:Discussion:OIII/CII_in_the_local_and_high-redshift_Universe}). \citet{2020ApJ...896...93H} used \program{Cloudy} photoionisation models to show that the \OIII/\CII\ luminosity ratio is expected to increase when increasing the ionisation parameter $U$, mostly through a decline of the \CII/\gls{SFR} ratio while \OIII/\gls{SFR} remains fairly unchanged, consistent with the observational trends found for DGS galaxies in \cref{ssec:Discussion:OIII/CII_in_the_local_and_high-redshift_Universe}. In addition, the hydrogen density $n_\text{H}$ is expected to play a role as the density approaches the critical density of \OIII\ ($510 \, \mathrm{cm^{-3}}$), which is lower than that of \CII\ produced in neutral gas ($\ssim 2.8 \cdot 10^3 \, \mathrm{cm^{-3}}$ when collision partners are atomic and molecular hydrogen; cf. \citealt{2013ARA&A..51..105C}). However, in reality, the \gls{ISM} is a complex, multi-phase environment; hence, \CII\ and \OIII\ are typically expected to emerge largely from distinct gas reservoirs characterised by different densities \citep[e.g.][]{2022MNRAS.510.5603K, 2022MNRAS.513.5621P}.

The effect of varying carbon and oxygen abundances are also difficult to isolate. On its own, altering both abundances equally would affect the individual line strengths, but to first order leave their ratio unchanged. A secondary effect, however, is that a lower metallicity also decreases the dust-to-gas ratio \citep{2018ARA&A..56..673G}. Less dust allows \gls{UV} photons to penetrate deeper into neutral gas, implying larger \glspl{PDR} can form. Hence, with a lower metallicity, \CII\ can become more luminous, which would lead to a suppressed \OIII/\CII\ ratio. On the other hand, the global metallicity is also linked to the hardness of the stellar radiation field and the ionisation parameter, such that a low metallicity would be expected to lead to more highly ionised gas and thus a higher \OIII/\CII\ ratio.

A recent study on a large number of simulated galaxies by \citet{2022MNRAS.510.5603K} instead argues for the need of a C/O abundance around $8 \times$ lower than solar in metal-poor systems, which would provide a direct explanation for the high \OIII/\CII\ ratios. This can be interpreted as a result of rapid metal enrichment by low-metallicity core-collapse \glspl{SN} responsible for producing the majority of oxygen, while type-Ia \glspl{SN} and AGB stars, an important source of carbon enrichment, act on longer timescales (\citealt{2020MNRAS.498.5541A}; see also \citealt{2019A&ARv..27....3M}). A caveat of these results is that dust is neglected; still, \citet{2020MNRAS.498.5541A} report a similar anti-correlation between metallicity and \OIII/\CII\ for simulated galaxies where dust is included, albeit in a smaller sample.

As discussed above, one-dimensional photoionisation codes are unlikely to capture the complexity of the \gls{ISM} in its entirety, particularly in the early Universe. However, they are still useful as a first-order approximation of the strengths of emission lines given varying physical conditions. Indeed, any discrepancies between the predictions of such simulations and observed quantities can point towards their shortcomings. Therefore, as a reference, \program{Cloudy} models described in \cref{ssec:Discussion:OIII/CII_photoionisation_models} are shown in the background of \cref{fig:OIII_and_CII_vs_EW(OIII+Hbeta)}. The models span the range of observed \OIII/\CII\ ratios, albeit for high values of the ionisation parameter ($\log_{10} U > -1.5$). However, we note that if the simulated \OIII\ luminosity is indeed underestimated as suggested by the findings in \cref{ssec:Discussion:OIII_lines}, the same \OIII/\CII\ ratio could be accomplished with a lower, more moderate ionisation parameter. In the case of the \OIII/\CII\ ratio, varying the gas density does not produce significantly different results; as expected, stripping the \gls{PDR} from the \HII\ region, increasing the $\upalpha$ enhancement, and lowering the C/O ratio all boost the \OIII/\CII\ ratio, whereas enhancing cosmic ray background lowers \OIII/\CII.

The high luminosity ratios seen at high redshift, and in a subset of metal-poor dwarf galaxies, can be explained as coming from an elevated \OIIILam\ luminosity, a \CIILam\ deficit, or a combination of both. \citet{2020MNRAS.499.5136C} argued there is no significant \CII\ deficit, as long as interferometric observations are properly set up to capture its full spatial extent. In line with these findings, we will show in the next section that photoionisation models appear to struggle to reproduce the (relative) strength of the observed \OIIILam\ line, pointing towards an \OIII\ enhancement causing the elevated \OIII/\CII\ ratios (even though \CII\ seems to be responsible for the correlation with \gls{EW} of the optical lines more than \OIII, as discussed in \cref{ssec:Discussion:OIII/CII_in_the_local_and_high-redshift_Universe}).

\subsection{Optical and \texorpdfstring{\glsentryshort{FIR}}{FIR} lines of \texorpdfstring{\OIII}{[OIII]}}
\label{ssec:Discussion:OIII_lines}

In the bottom panel of \cref{fig:OIII_and_CII_vs_EW(OIII+Hbeta)}, results from the \program{Cloudy} models discussed above (\cref{ssec:Discussion:OIII/CII_photoionisation_models}) are shown in the plane of relative strengths of key \OIII\ emission lines, \OIIILam\ and the $\OIII \, \lambda \lambda \, 4960, 5008 \, \Angstrom$ lines. Because we infer the rest-frame optical \OIII\ strength from the \textit{Spitzer}/IRAC $[3.6]-[4.5]$ colour, we estimate the line strength as the combined \gls{EW} of $\OIII \, \lambda \lambda \, 4960, 5008 \, \Angstrom$ and \Hbeta\ lines, denoted as $\text{\gls{EW}} ( \OIII + \Hbeta )$ here, while we estimate the \gls{FIR} \OIIILam\ line strength as $L_\text{\OIIILam}/\text{\gls{SFR}}$. Using these relative luminosities allows us to compare local and high-redshift galaxies on a more equal footing. Additionally, we note dust attenuation (and hence our choice of a $A_V = 10 \, \mathrm{mag}$ stopping criterion) is not of major importance here, since the $\OIII \, \lambda \, 5008 \, \Angstrom$ line and stellar continuum flux are, to first order, expected to be affected equally, which cancels out in the \gls{EW}, while the \OIIILam\ line is optically thin.

Observations of all metal-poor dwarf galaxies from the \gls{DGS} are included in addition to the five $z \sim 7$ star-forming galaxies considered in this work, for which we set lower limits using the unobscured \gls{SFR} only if no dust continuum was detected. All measurements shown in the bottom panel of \cref{fig:OIII_and_CII_vs_EW(OIII+Hbeta)}, except for a few dwarf galaxies with low \glspl{EW}, favour models with a young stellar population for the relative strengths of both \OIII\ lines to be successfully reproduced. Observations of the $z \sim 7$ galaxies in particular appear to require a young stellar population of between $1$ and $5 \, \mathrm{Myr}$. Lower gas densities (e.g. $n_\text{H} \sim 10^{0.5} \, \mathrm{cm^{-3}}$) slightly increase the relative \OIII\ luminosity and the \gls{EW} of the optical lines (both by around $0.2 \, \mathrm{dex}$), though the models with $n_\text{H} \sim 10^2 \, \mathrm{cm^{-3}}$ shown here coincide better with the data locus of the \gls{DGS} and high-redshift galaxies combined. Higher gas densities (e.g. $n_\text{H} \sim 10^{2.5} \, \mathrm{cm^{-3}}$), as sometimes observed in local high-redshift analogues \citep[e.g.][]{2016ApJ...827..126B} decrease the relative \OIII\ luminosity by $\ssim 0.3 \, \mathrm{dex}$. The presence of \glspl{PDR} and cosmic rays have a largely negligible impact.

Perhaps most remarkable is the fact that -- even with lower gas densities -- the observations seem to require a relatively high metallicity \citep[in line with the findings of][]{2020ApJ...896...93H}. We note that simulations with a very young stellar population of $1 \, \mathrm{Myr}$ and fully solar metallicity (i.e. $Z_* = Z_\text{neb} = \mathrm{Z_\odot}$; not shown here) are just able to recover the relative \OIIILam\ luminosity for very high ionisation parameters but fail to simultaneously replicate the \glspl{EW} of the optical lines for the two COSMOS sources in particular. Out of a large range of models, we find the most consistent are those where specifically $\upalpha$ elements have near-solar nebular abundances, i.e. $[ \mathrm{\upalpha/H} ]_\text{neb} \sim 0$, as illustrated in the bottom panel of \cref{fig:OIII_and_CII_vs_EW(OIII+Hbeta)} by the arrow, indicating the $\ssim 0.5 \, \mathrm{dex}$ downwards shift of the grid when $X_\mathrm{\upalpha/Fe, \, neb} = 1$ as opposed to $X_\mathrm{\upalpha/Fe, \, neb} = 4$ as shown. This scenario may be expected from a rapid metal enrichment of the \gls{ISM} that causes the stellar metallicity to lag behind the nebular metallicity. We note such high metal abundances corroborate the finding of a seemingly high dust-to-stellar mass ratio discussed in \cref{ssec:Discussion:Dust_masses} and modest gas-to-dust mass ratio discussed in \cref{ssec:Discussion:CII/NII}.

Interestingly, a metallicity calibration employing the \OIIILam\ line suggested by \citet{2020ApJ...903..150J} indeed predicts relatively high metallicities given the \OIIILam\ luminosities of the $z \sim 7$ galaxies; remarkably, this yields a super-solar value of $12 + \log \left ( \text{O/H} \right) \simeq 8.72$ for COS-3018555981 (solar abundance being $12 + \log \left ( \text{O/H} \right)_\odot = 8.69$; \citealt{2009ARA&A..47..481A}), as has been reported (albeit for higher-mass systems) in a number of recent works \citep[, and references therein]{2022ApJ...928..179L}. Yet such ($\upalpha$-element) abundances are exceptionally high compared to what may be expected for young star-forming galaxies less than a billion years after the Big Bang. Even when assuming there is no evolution in the \gls{MZR} beyond a redshift of $z \sim 3$ \citep[while evidence of an underlying evolution up to $z \sim 5$ is already emerging;][]{2021MNRAS.508.1686W}, a solar metallicity poses a significant offset (around $0.5 \, \mathrm{dex}$ or a factor of $\ssim 3$) from the predicted metallicity of a galaxy with $M_* \sim 10^9 \, \mathrm{M_\odot}$ \citep{2021ApJ...914...19S}.

Relatively high \OIIILam\ luminosities have previously been observed in local metal-poor galaxies, as noted by for example \citet{2012A&A...548A..91L} in the case of the \gls{LMC} and \citet{2015A&A...578A..53C, 2019A&A...626A..23C} for the \gls{DGS} galaxies. The models of \citet{2012A&A...548A..91L} point towards the need of a relatively cool ($T < \num{10000} \, \mathrm{K}$) ionised gas, consistent with our metal-rich models that have more efficient cooling: for solar metallicity, for instance, $T \sim \num{8000} \, \mathrm{K}$ in the \HII\ region, as opposed to $T \gtrsim \num{10000} \, \mathrm{K}$ for $Z < Z_\odot$. On the contrary, the metal-poor dwarf galaxies from the \gls{DGS} have directly observed nebular metallicities of $Z_\text{neb} \lesssim 0.3 \, \mathrm{Z_\odot}$, which are instead coupled with high temperatures, $\num{15000} \, \mathrm{K} \lesssim T \lesssim \num{20000} \, \mathrm{K}$ \citep[e.g.][]{1990Natur.343..238I, 1997ApJS..108....1I, 1998ApJ...500..188I, 2000ApJ...531..776G}, as would be expected given the principal role of heavy elements as a cooling agent. This indicates that photoionisation models seem to be missing a key ingredient to replicate the strength of the \OIIILam\ line.

One of the main assumptions of our simple, one-dimensional photoionisation model is to consider an ionisation-bounded nebula (with a smooth density profile and constant metallicity), while -- as noted in \cref{ssec:Discussion:OIII/CII_theoretical_insights} -- this is likely not entirely correct, given for instance the role star-forming galaxies are thought to play in reionising the neutral \gls{IGM} \citep{2021arXiv211013160R}. Indeed, \citet{2015A&A...578A..53C, 2019A&A...626A..23C} argue for the presence of escape channels which allow ionising radiation to escape \HII\ regions, leading to a large volume filling factor of diffuse (i.e. with density comparable to or lower than the critical density of \OIIILam, $510 \, \mathrm{cm^{-3}}$) ionised gas. Another possibility for the discrepancy would be that, in estimating the relative \OIII\ luminosity, the observed \gls{SFR} is underestimated, for example by considering dust temperatures lower than what they are in reality (in particular COS-3018555981; cf. \cref{ssec:Discussion:Dust_temperatures}); however, this seems unlikely in the case of COS-2987030247 and UVISTA-Z-007 for which we do not detect dust continuum emission (and thus have an upper limit on the obscured \gls{SFR}).

\begin{figure}
	\centering
	\includegraphics[width=\linewidth]{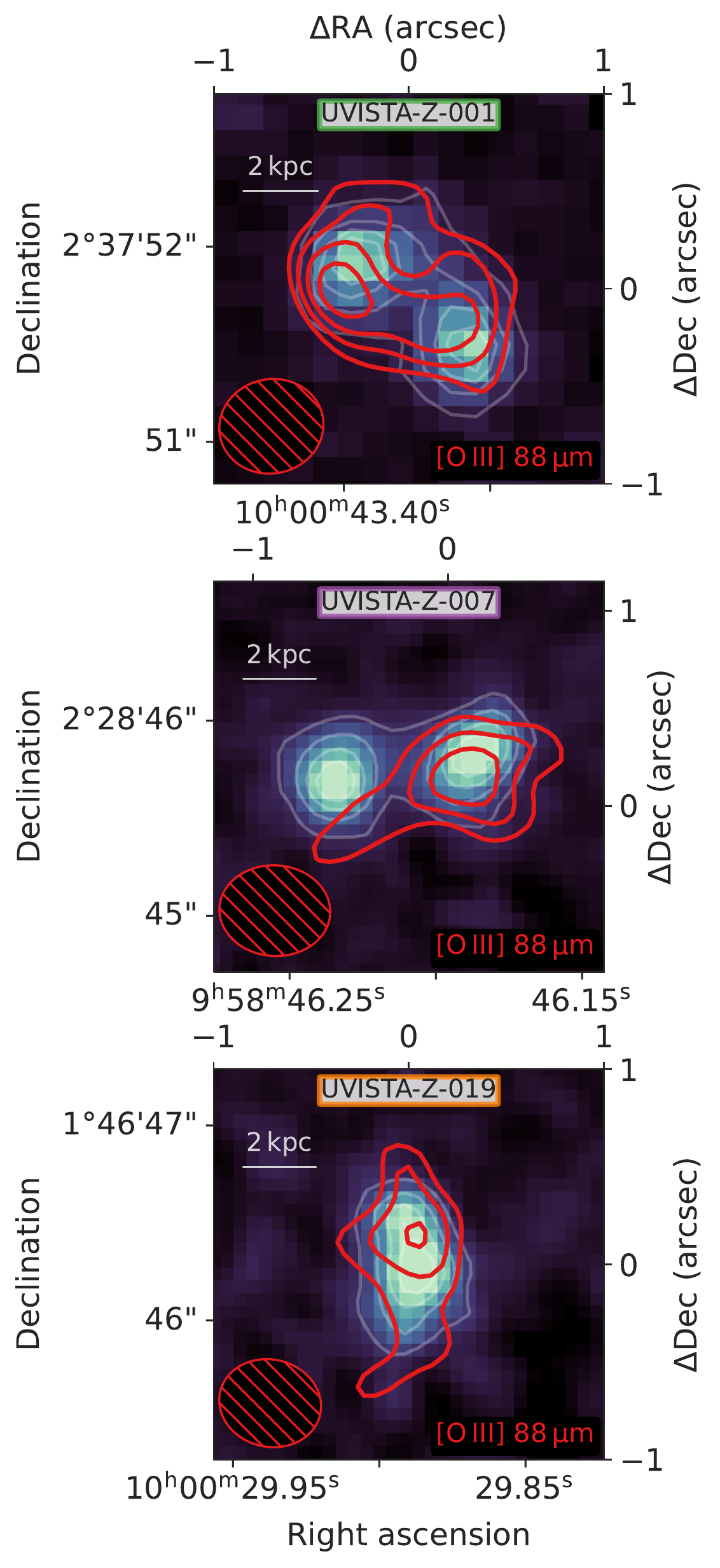}
	\caption[\gls{HST} images of the rest-frame \gls{UV} overlaid by contours of the \OIIILam\ line for UVISTA-Z-001, UVISTA-Z-007, and UVISTA-Z-019.]{\gls{HST} images of the rest-frame \gls{UV} (in the $JH_{140}$ band; \cref{ssec:Observations:HST}) overlaid by contours of the \OIIILam\ line (achieving a $\ssim 0.4 \arcsec$ resolution with natural weighting; see \cref{tab:Observations}) of UVISTA-Z-001 (top panel), UVISTA-Z-007 (middle panel) and UVISTA-Z-019 (bottom panel). Red \OIII\ contours are drawn from $3 \sigma$ and going up in steps of $1 \sigma$. The top left indicates a physical scale of $2 \, \mathrm{kpc}$. Beam sizes for \OIII\ are indicated in the bottom left.
	}
	\label{fig:High-resolution_OIII_maps}
\end{figure}

\begin{figure}
	\centering
	\includegraphics[width=\linewidth]{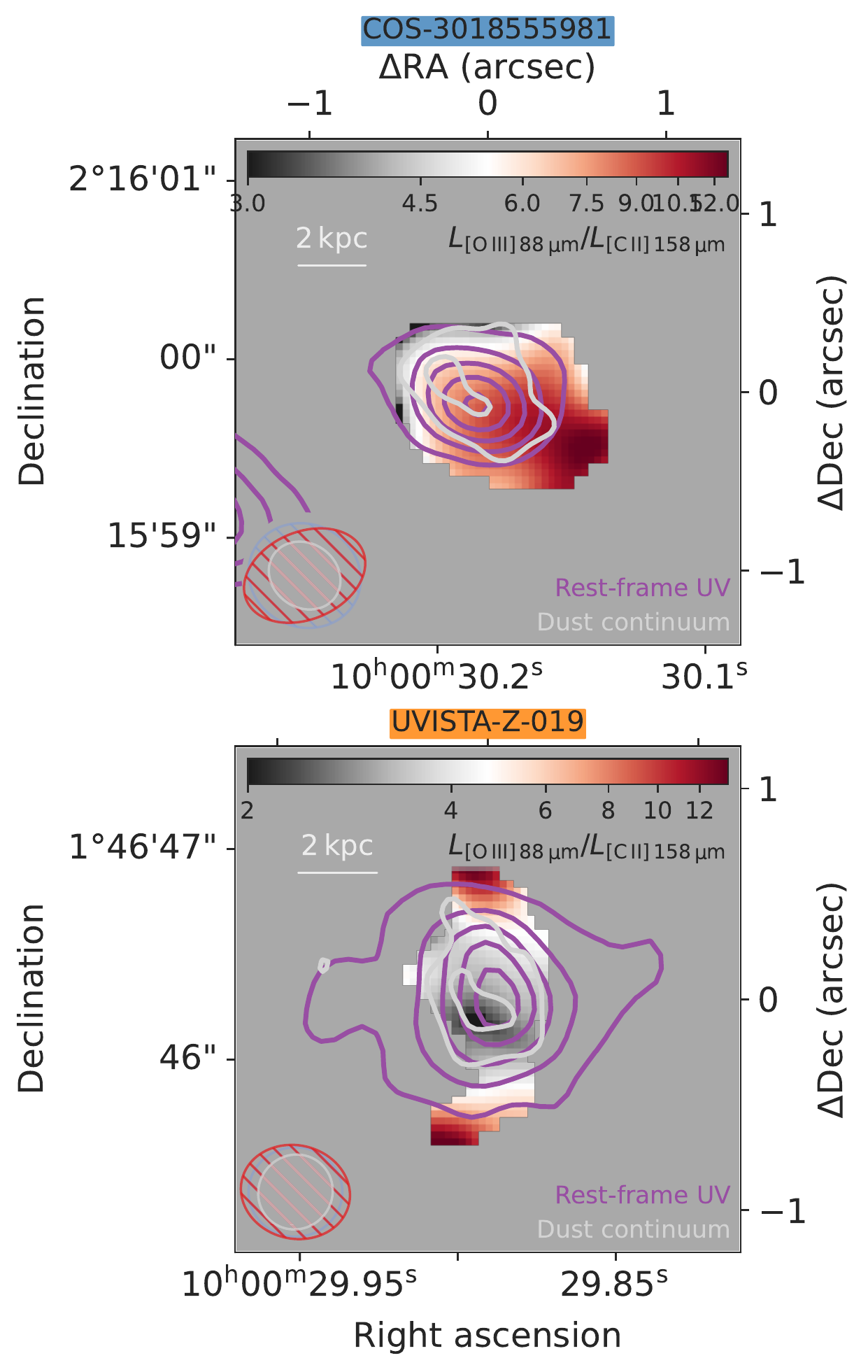}
	\caption[Line ratio map of the \OIIILam\ and \CIILam\ lines for COS-3018555981 and UVISTA-Z-019.]{Line ratio maps of the \OIIILam\ and \CIILam\ lines for COS-3018555981 (top panel) and UVISTA-Z-019 (bottom panel), where imaging parameters have been chosen to match beam sizes (see \cref{tab:Observations}). Light grey contours are drawn for the dust continuum (at $\ssim 160 \, \mathrm{\upmu m}$, with natural weighting scheme; \cref{ssec:Discussion:OIII/CII_spatially_resolved_analysis} for details), starting from $3 \sigma$ and going up in steps of $2 \sigma$. Purple contours are drawn for the \gls{HST} imaging (\cref{ssec:Observations:HST}) of the rest-frame \gls{UV} (convolved to match the dust continuum \gls{PSF}), starting from $5 \sigma$ and going up in steps of $5 \sigma$. Beam sizes are indicated in the bottom left (partially hiding a foreground source in the \gls{HST} imaging of COS-3018555981); the top left indicates a physical scale of $2 \, \mathrm{kpc}$. There is tentative evidence for spatial variation in the \OIII/\CII\ ratio, including a gradient with the lowest \OIII/\CII\ values seeming to align with the location of the dust emission in COS-3018555981.
	}
	\label{fig:OIII/CII_ratio_maps_of_COS-3018555981_and_UVISTA-Z-019}
\end{figure}

\begin{figure}
	\centering
	\includegraphics[width=\linewidth]{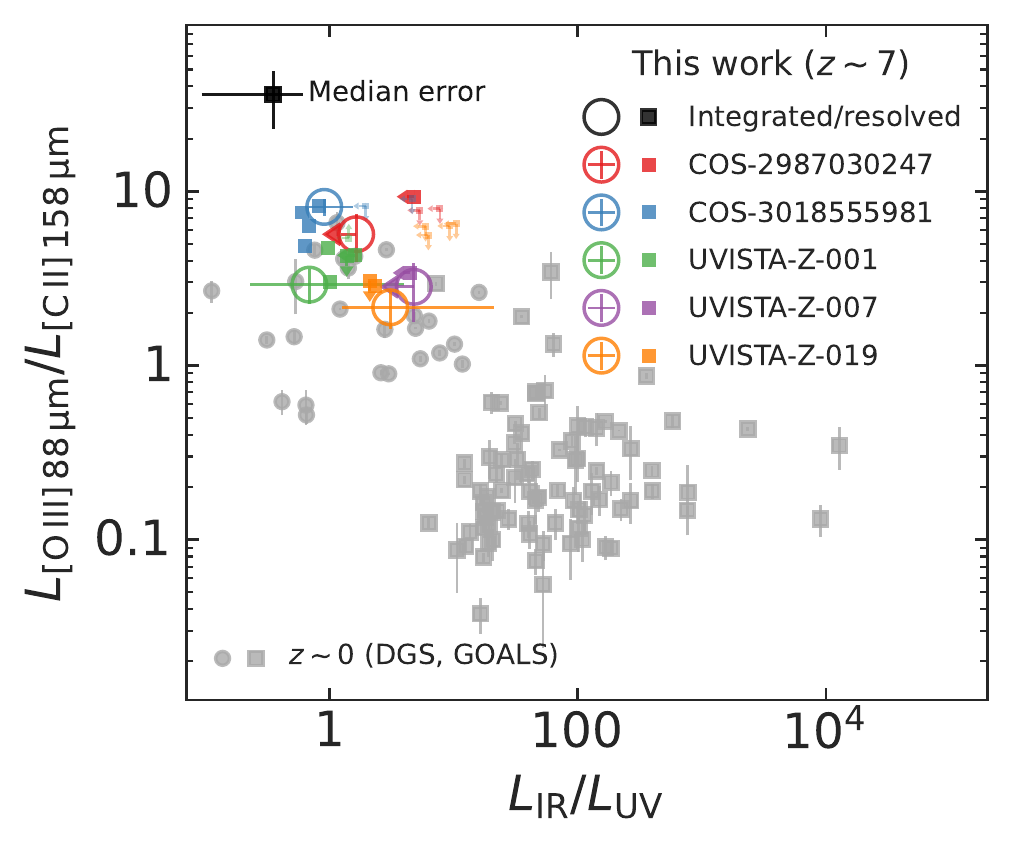}
	\caption[Spatially resolved sequence of \OIIILam\ to \CIILam\ ratio as a function of $L_\text{\gls{IR}}/L_\text{\gls{UV}}$.]{Sequence of the \OIIILam\ to \CIILam\ ratio as a function of $L_\text{\gls{IR}}/L_\text{\gls{UV}}$. Global measurements of the \gls{DGS} and \gls{GOALS} samples are shown, as are the five $z \sim 7$ star-forming galaxies discussed in this work. Furthermore, spatially resolved ratios extracted from individual regions across all five sources (see \cref{ap:Spatially_resolved_analysis} for details) are shown as squares (smaller squares if we can only set $2 \sigma$ limits on both ratios). For clarity, error bars of the spatially resolved measurements are not shown; the median error of these points is indicated by the black square in the top left.
	}
	\label{fig:Spatially_resolved_OIII/CII_vs_L_IR/L_UV_sequence}
\end{figure}

\subsection{Spatially resolved emission line analysis}
\label{ssec:Discussion:OIII/CII_spatially_resolved_analysis}

In the previous sections, we have discussed the interpretation of line strengths and their ratios globally (i.e. integrated across the galaxy). However, the \gls{ISM} is known to be complex and inhomogeneous, especially at high redshift \citep{2021MNRAS.505.5543V, 2022MNRAS.513.5621P}. Here, we will instead briefly discuss detections of the \CIILam\ and \OIIILam\ lines on a spatially resolved scale.

Previous works have already demonstrated significant offsets (as well as different sizes) between the rest-frame \gls{UV} and \CIILam\ emission \citep[e.g.][]{2015MNRAS.452...54M, 2018MNRAS.478.1170C, 2018ApJ...854L...7C}. Here, we show a detailed comparison between the \gls{UV} and \OIIILam\ emission in \cref{fig:High-resolution_OIII_maps} for all three UVISTA sources, where \gls{ALMA} achieves a moderately high spatial resolution (with a beam size of $\ssim 0.4 \arcsec$). Despite good agreement between the compact morphology of both the rest-frame \gls{UV} and the \OIII\ contours, there are signs of minor misalignment: most notably, the observed \OIII\ signal is clearly centred on the western component of UVISTA-Z-007, whereas the eastern component distinctly lacks \OIII\ emission of comparable strength. We quantified this by fitting separate two-dimensional Gaussian components to the \gls{HST} images, resulting in two components with very similar integrated \gls{UV} luminosities of around $7 \cdot 10^{10} \, \mathrm{L_\odot}$ (their ratio being $\ssim 1.1$) at a separation of $\ssim 0.71 \arcsec$ or $3.8 \, \mathrm{kpc}$, whereas the luminosity ratio of the \OIIILam\ line compared to the \gls{UV}, $L_\OIII/L_\text{\gls{UV}}$, differs by almost a factor two (measured as $(1.2 \pm 0.6) \cdot 10^{-3}$ and $(2.2 \pm 0.6) \cdot 10^{-3}$ for the eastern and western components, respectively) in a region encompassing twice the \glspl{FWHM}. As indicated by its kinematics, this is likely an ongoing galaxy merger \citep{2022arXiv220204080S}. A similar double-component \gls{UV} morphology is seen in UVISTA-Z-001, although the two components separated by $\ssim 0.68 \arcsec$ or $3.5 \, \mathrm{kpc}$ display a more equal $L_\OIII/L_\text{\gls{UV}}$ ratio (respectively $(6.3 \pm 0.2) \cdot 10^{-3}$ and $(5.0 \pm 0.2) \cdot 10^{-3}$ for the eastern and western components).

To investigate such apparent peculiarities in the \OIII\ morphology further, we have regridded images of the \OIIILam\ to \CIILam\ lines (integrated over the \gls{FWHM} centred on the line, created from imaging parameters that have been chosen to match beam sizes; see \cref{tab:Observations}) to a common coordinate mesh to create maps of their ratio in regions where both lines are detected with at least $3 \sigma$ so that $\text{\gls{SNR}}_{\OIII/\CII} \gtrsim 2$. On the scales down to which we resolve the lines (around $2$ to $3 \, \mathrm{kpc}$), the line ratio can indeed also vary significantly across the galaxy, as illustrated by the example of COS-3018555981 in the top panel of \cref{fig:OIII/CII_ratio_maps_of_COS-3018555981_and_UVISTA-Z-019}, where the line ratio varies locally from the order of unity up to more than $10$. The ratio map of UVISTA-Z-019, shown in the bottom panel of \cref{fig:OIII/CII_ratio_maps_of_COS-3018555981_and_UVISTA-Z-019}, also reveals indications of a spatially varying ratio (on the order of $2$ to $4$ in the centre). It further shows an enhanced ratio in the northern and southern outskirts; although the \gls{SNR} will be reduced in these regions, this does not appear to be due to noise (demonstrated by spectra extracted from manually placed beams as in \cref{ap:Spatially_resolved_analysis}). It lacks strong evidence for a clear gradient, however, as is the case for the other three sources not shown here (due in part to a lack of resolution and/or overlap in the \OIII\ and \CII\ lines).

The dust continuum and rest-frame \gls{UV} imaging, shown by overlaid contours, are matched in \gls{PSF} by convolving the \gls{HST} imaging with a kernel found by the Richardson-Lucy algorithm. To maximise \gls{SNR}, we show the continuum images created under a natural weighting scheme (see \cref{sssec:Observations:ALMA_data_reduction}), which are therefore at slightly higher resolution than \OIII\ and \CII. Interestingly, the location of the dust continuum seem to mostly coincide with the side of the galaxy which shows a lower \OIII/\CII\ ratio. Given \gls{SNR} of the continuum detection in this case ($\ssim 10 \sigma$; see \cref{fig:Dust_continuum_maps,tab:Continuum_fluxes_and_dust_properties}) and the beam size ($\ssim 0.7 \arcsec$), we estimate the positional accuracy to be $\ssim 0.1 \arcsec$, indicating the observed offset and alignment with the northwestern region is significant.\footnote{For this estimate, we consulted Section 10.5.2 of the \gls{ALMA} Technical Handbook \citep{ALMA_technical_handbook}.} Given the link between a high \OIII/\CII\ ratio on the one hand and a starburst's young age or high ionisation parameter on the other \citep[as shown in \cref{ssec:Discussion:OIII/CII_theoretical_insights}; see also e.g.][]{2021MNRAS.505.5543V}, this suggests the relatively unobscured region in the southeast with high \OIII/\CII\ may be experiencing a more recent and/or intense burst of star formation. In contrast, it appears to be inconsistent with a scenario of UV-bright regions being slightly more evolved sites of star formation, where \glspl{SN} (or other types of feedback) have been able to clear out the dust. A lack of dust implies the gas is likely low in metallicity, potentially linked with an inflow of pristine gas triggering the burst of star formation.

This tentative interplay between the \gls{FIR} line ratio and degree of dust obscuration is further tested in \cref{fig:Spatially_resolved_OIII/CII_vs_L_IR/L_UV_sequence}, where we show measurements of \OIII/\CII\ as a function of the \gls{IRX}, $\text{\gls{IRX}} \equiv L_\text{\gls{IR}}/L_\text{\gls{UV}}$. Global measurements of the \gls{DGS} and \gls{GOALS} samples are shown, as are the five $z \sim 7$ star-forming galaxies discussed in this work. In addition, we placed individual beams across all five sources to measure these quantities on a spatially resolved scale (see \cref{ap:Spatially_resolved_analysis} for details).

We considered measurements only when exceeding the estimated sensitivity by at least $2 \sigma$, setting upper or lower limits ($2 \sigma$) if only one of the lines and/or continua shows a detectable signal. These measurements are clearly pushing the \gls{SNR} of the data since a lot of points are in fact limits, mostly due to a lack of significant dust continuum (see also the large median error bar on non-limits). As may be expected from the data with the highest \gls{SNR}, spatially resolved measurements of COS-3018555981 do yield four significant measurements spanning a \OIII/\CII\ ratio from $5$ to $8$ and \gls{IRX} from $0.5$ to $0.9$ (both with a spread of around $0.2 \, \mathrm{dex}$). Three measurements of UVISTA-Z-001 are similarly spread out ($\ssim 0.2 \, \mathrm{dex}$) at slightly lower \OIII/\CII\ ratio ($3$ to $5$) and higher \gls{IRX} ($1$ to $1.4$). For UVISTA-Z-019, two beams have confidently detected \gls{IRX} values, with one \OIII/\CII\ detection and one upper limit (both approximately $\ssim 2$ and $\ssim 3$, respectively). For several beams placed over COS-2987030247, $\text{\gls{IRX}} \lesssim 5$, while the \OIII/\CII\ ratio in one beam is measured to be $\ssim 9$.

Combined with the data obtained on a global scale, there is a tentative negative correlation where a galaxy (region) with high \gls{IRX} tends to have a lower \OIII/\CII\ ratio. This fits in with the picture sketched above, in which young and/or intense bursts of star formation, accompanied with a large \OIII/\CII\ ratio, occur in unobscured and hence likely metal-poor regions, whereas obscured star formation is linked to more moderate line ratios.

\section{Summary and conclusions}
\label{sec:Summary}

We have presented \gls{ALMA} observations of the \OIIILam\ line in five $z \sim 7$ star-forming (Lyman-break) galaxies that have previously been spectroscopically confirmed by \gls{ALMA} via their \CIILam\ line, each yielding a confident \OIII\ detection ($\text{\gls{SNR}} > 5$). We complement these observations with new \gls{HST} rest-frame \gls{UV} imaging of two of the sources. Furthermore, we have presented a non-detection of \NIILam\ in COS-2987030247 in addition to the corresponding dust continuum measurements around each emission line. We summarise our findings as follows:

\begin{itemize}
    \item For two sources, COS-2987030247 and UVISTA-Z-007, we do not confidently detect dust continuum emission in any ALMA band. For COS-3018555981 and UVISTA-Z-019, however, we have significant detections at $\ssim 160 \, \mathrm{\upmu m}$ yet we do not detect the continuum at $\ssim 90 \, \mathrm{\upmu m}$. In UVISTA-Z-001, we detect compact $\ssim 90 \, \mathrm{\upmu m}$ continuum emission in band 8, in contrast to a more extended dust reservoir observed at $\ssim 160 \, \mathrm{\upmu m}$. The compact component has a seemingly typical dust temperature of $\ssim 60 \, \mathrm{K}$, while the extended component is likely colder. The dust-continuum measurements of COS-3018555981 also favour a low dust temperature coupled with a high dust mass: the dust temperature, nominally $T_\text{dust} = 29_{-5}^{+9} \, \mathrm{K}$ (or $T_\text{dust} < 48 \, \mathrm{K}$ at $95 \%$ confidence) compared to a \gls{CMB} temperature of $21 \, \mathrm{K}$ at $z \sim 7$, may be lower than those of all other \gls{EoR} sources known, while its dust mass implies a high stellar metallicity yield (accompanied by a top-heavy \gls{IMF}) and may point towards the need of other dust production and/or growth mechanisms beyond \glspl{SN}.
    
    \item The non-detection of \NIILam\ in COS-2987030247 allows us to set a lower limit on the \CII/\NII\ ratio ($L_\CII/L_\NII > 4.8$ at $3 \sigma$), which is fully consistent with \CII\ production in a \gls{PDR}-like medium, and renders \HII\ regions the unlikely primary origin of the \CII\ emission (except in a more extreme physical environment with low metallicity, high ionisation parameter, and/or high gas density).
    
    \item We find modest ratios of \OIIILam\ to \CIILam\ (around $2$ to $3$) for the sources with moderate \glspl{EW} of the optical \OIII\ and \Hbeta\ lines (rest-frame $\text{\gls{EW}} \sim 700 \, \Angstrom$) while the ratio is comparatively higher ($6$ to $8$) for the sources with more extreme \glspl{EW} ($\text{\gls{EW}} > \num{1000} \, \Angstrom$), consistent with a positive correlation between \gls{EW} and \OIII/\CII\ seen in local metal-poor dwarf galaxies.
    
    \item Through the photoionisation code \program{Cloudy}, we find that a young stellar population embedded in a nebula of typical density with a high ionisation parameter appears an appropriate model of the physical environment in which the \gls{FIR} emission lines originate. Surprisingly, however, the modelled nebular emission barely reproduces the observed strength of the \OIIILam\ line in sources with high $\text{\gls{EW}} ( \OIII + \Hbeta )$. Moreover, we find the \program{Cloudy} modelling only recovers the observed \OIIILam\ line strength when the $\mathrm{\upalpha/Fe}$ ratio relative to solar is raised by increasing the abundances of $\upalpha$ elements, leading to a relatively cool ionised gas ($T \sim \num{8000} \, \mathrm{K}$), in contrast to direct measurements of potential local analogues (i.e. metal-poor dwarf galaxies). This suggests luminous \glspl{LBG} at $z \sim 7$ might be more chemically enriched than is often assumed, or else, that significant \OIIILam\ emission emerges from a low density, moderately cool medium outside of the modelled \HII\ regions.
    
    \item We find the \OIII/\CII\ ratio shows a tentative anti-correlation with the degree of dust obscuration (measured through the \gls{IRX}) on spatially resolved scales, similar to the trend seen in the local Universe on global scales of metal-poor dwarf galaxies and (U)LIRG starburst galaxies. This suggests a large \OIII/\CII\ ratio accompanies young and/or intense bursts of star formation, which occur in unobscured and hence likely metal-poor regions.
\end{itemize}

\section*{Data availability}

\gls{HST} data underlying this article are available in the MAST archive at \href{https://dx.doi.org/10.17909/6gya-3b10}{10.17909/6gya-3b10} (GO 13793), \href{https://dx.doi.org/10.17909/t9-jhsf-m392}{10.17909/T9-JHSF-M392} (GO 16506) and from \url{https://archive.stsci.edu/prepds/3d-hst/} (the 3D-HST Treasury Program). \textit{Gaia} data may be obtained from \url{https://gea.esac.esa.int/archive/}. The \gls{ALMA} data underlying this article are available in the \gls{ALMA} science archive at \url{https://almascience.eso.org/asax/} under by the following project codes (see also \cref{tab:Observations}):
\begin{itemize}[label={--}]
\item ADS/JAO.ALMA\#2015.1.01111.S
\item ADS/JAO.ALMA\#2018.1.01359.S
\item ADS/JAO.ALMA\#2018.1.00429.S
\item ADS/JAO.ALMA\#2018.1.01551.S
\item ADS/JAO.ALMA\#2017.1.00604.S
\item ADS/JAO.ALMA\#2015.1.00540.S
\item ADS/JAO.ALMA\#2018.1.00085.S
\item ADS/JAO.ALMA\#2018.1.00933.S
\item ADS/JAO.ALMA\#2019.1.01611.S
\item ADS/JAO.ALMA\#2019.1.01524.S
\item ADS/JAO.ALMA\#2018.1.00085.S
\item ADS/JAO.ALMA\#2019.1.01611.S
\end{itemize}

Reduced data underlying this article will be shared on reasonable request to the corresponding author.

\section*{Acknowledgements}

We are grateful to Caitlin Casey and Seiji Fujimoto for insightful discussions on the reduction and interpretation of the \gls{ALMA} measurements. We furthermore thank the anonymous referee for their comments. JW, RS, RM, and GCJ acknowledge support from the ERC Advanced Grant 695671, ``QUENCH'', and the Fondation MERAC. RS acknowledges support from a Netherlands Organisation for Scientific Research (NWO) Rubicon grant, project number 680-50-1518, and an STFC Ernest Rutherford Fellowship (ST/S004831/1). RM also acknowledges funding from a research professorship from the Royal Society. MA acknowledges support from FONDECYT grant 1211951, CONICYT+PCI+INSTITUTO MAX PLANCK DE ASTRONOMIA MPG190030, CONICYT+PCI+REDES 190194 and ANID BASAL project FB210003. JAH gratefully acknowledges support of the VIDI research program with project number 639-042-611, which is (partly) financed by the NWO. This work has extensively used \gls{CASA} \citep{2007ASPC..376..127M}, developed by an international consortium of scientists based at the National Radio Astronomical Observatory (NRAO), the European Southern Observatory (ESO), the National Astronomical Observatory of Japan (NAOJ), the Academia Sinica Institute of Astronomy and Astrophysics (ASIAA), CSIRO Astronomy and Space Science (CSIRO/CASS), and the Netherlands Institute for Radio Astronomy (ASTRON), under the guidance of NRAO. Furthermore, it has used the following packages in \program{python}: the \program{SciPy} library \citep{Jones2001}, its packages \program{NumPy} \citep{2011CSE....13b..22V} and \program{Matplotlib} \citep{Hunter2007}, the \program{Astropy} \citep{2013A&A...558A..33A, 2018AJ....156..123A}, \program{pymultinest} \citep{2009MNRAS.398.1601F, 2014A&A...564A.125B}, and \program{DrizzlePac} packages (\url{https://www.stsci.edu/scientific-community/software/drizzlepac.html}.

This work was based on observations taken by the \glsentryfull{ALMA}. \gls{ALMA} is a partnership of ESO (representing its member states), NSF (USA) and NINS (Japan), together with NRC (Canada), MOST and ASIAA (Taiwan), and KASI (Republic of Korea), in cooperation with the Republic of Chile. The Joint \gls{ALMA} Observatory is operated by ESO, AUI/NRAO and NAOJ.

This work was furthermore partially based on new observations made with the NASA/ESA \glsentryfull{HST}, obtained at the \gls{STScI}, which is operated by the Association of Universities for Research in Astronomy, Inc., under NASA contract NAS 5-26555. These observations are associated with programme \#16506. Additionally, \gls{HST} archival data was obtained from the data archive at the \gls{STScI}. \gls{STScI} is operated by the Association of Universities for Research in Astronomy, Inc. under NASA contract NAS 5-26555.

Finally, this work has made use of data from the European Space Agency (ESA) mission
\textit{Gaia} (\url{https://www.cosmos.esa.int/gaia}), processed by the \textit{Gaia}
Data Processing and Analysis Consortium (DPAC,
\url{https://www.cosmos.esa.int/web/gaia/dpac/consortium}). Funding for the DPAC
has been provided by national institutions, in particular the institutions
participating in the \textit{Gaia} Multilateral Agreement.

\bibliographystyle{mnras}
\bibliography{ALMA_OIII-CII}

\appendix

\section{Dust peak temperature measurements}
\label{ap:Dust peak temperature measurements}

Here, we briefly discuss the significance of the \gls{SED} peak temperature as defined in \cref{eq:T_peak} and various measurements and predictions reported in the literature which are included in \cref{fig:T_peak_evolution}. Importantly, the peak temperature offers a way to compare observations of the dust temperature consistently since this approach avoids degeneracies introduced by the chosen opacity model, a largely unconstrained quantity that is typically assumed to be a fixed value in the greybody spectrum \citep[e.g.][]{2014PhR...541...45C}.

For a perfect blackbody, the intrinsic temperature $T_\text{dust}$ is exactly equal to the peak temperature but notably, for a greybody $T_\text{peak}$ is generally lower \citep{2012MNRAS.425.3094C}. This effect can be understood by considering a simplistic two-component dust model, where the radiation field is driven towards thermal equilibrium through absorption by the colder component in the optically thick regime, resulting in an observed outward spectrum with a peak wavelength shifted to a higher wavelength (i.e. lower $T_\text{peak}$). Vice versa, when fitting a greybody \gls{SED} template, the inferred intrinsic dust temperature will strongly depend on the opacity model \citep[e.g.][]{2020A&A...634L..14C}, while the observed peak temperature should remain the same to best fit the observed data. Indeed, $T_\text{peak}$ derived from our fits is approximately unchanged under the assumption of different opacity models, while the inferred $T_\text{dust}$ can change drastically: the more optically thick the \gls{SED} is (i.e. the higher $\lambda_0$), the higher the resulting intrinsic temperature, $T_\text{dust}$ (see \cref{ssec:Discussion:Dust_SED_fitting_procedure}).

In \cref{fig:T_peak_evolution}, we show the results at lower redshifts ($0 < z < 4$) of \citet{2018A&A...609A..30S}, who fit detailed SED templates built from multiple dust components to stacked spectra. Their reported dust temperatures are thus mass-weighted; however, they find a simple, linear relation where the mass-weighted temperature is roughly $91\%$ of the luminosity-weighted one. This implies temperatures inferred from a greybody, which are necessarily weighted by luminosity, are similar to (although $\ssim 10\%$ higher than) the mass-weighted temperature, effectively setting an upper limit. We also show the (partially extrapolated) linear fit obtained by \citet{2018A&A...609A..30S} and the power-law fit to the peak-temperature evolution of simulated galaxies by \citet{2019MNRAS.489.1397L}.

At intermediate redshifts ($2 < z < 6$), \glspl{SMG} from the \gls{SPT} survey \citep{2020ApJ...902...78R} are shown as squares (all other high-redshift galaxies are circles with errorbars). In addition, results for four star-forming galaxies at $z \sim 6$ with photometric detections in three \gls{ALMA} bands each are included as circles \citep{2020MNRAS.498.4192F}. For five star-forming galaxies at $6 < z < 8$ -- J1211-0118 and J0217-0208 at $z \simeq 6$ \citep{2020ApJ...896...93H}, A1689-zD1 at $z \simeq 7.13$ \citep{2017MNRAS.466..138K, 2020MNRAS.495.1577I, 2021MNRAS.508L..58B}, B14-65666 at $z \simeq 7.15$ \citep{2019PASJ...71...71H, 2021ApJ...923....5S}, and MACS0416-Y1 at $z \simeq 8.31$ \citep{2019ApJ...874...27T, 2020MNRAS.493.4294B} -- we derive dust properties using the same \gls{MERCURIUS} fit described in \cref{ssec:Discussion:Dust_SED_fitting_procedure} for consistency. We allow $\beta_\text{\gls{IR}}$ to vary freely for A1689-zD1, since there are four dust continuum detections; for MACS0416-Y1, we take $\beta_\text{\gls{IR}} = 2$ as this provides a better fit. For A1689-zD1 and B14-65666 we opt for the fiducial self-consistent opacity model (we note this assumption has little impact on the inferred $T_\text{peak}$), while for MACS0416-Y1 we use an optically thin \gls{SED} to obtain a conservative lower limit on the temperature ($95 \%$ confidence). We also adopt an optically thin \gls{SED} for J1211-0118 and J0217-0208 due to the lack of a size measurement.

\begin{figure*}
	\centering
	\includegraphics[width=\linewidth]{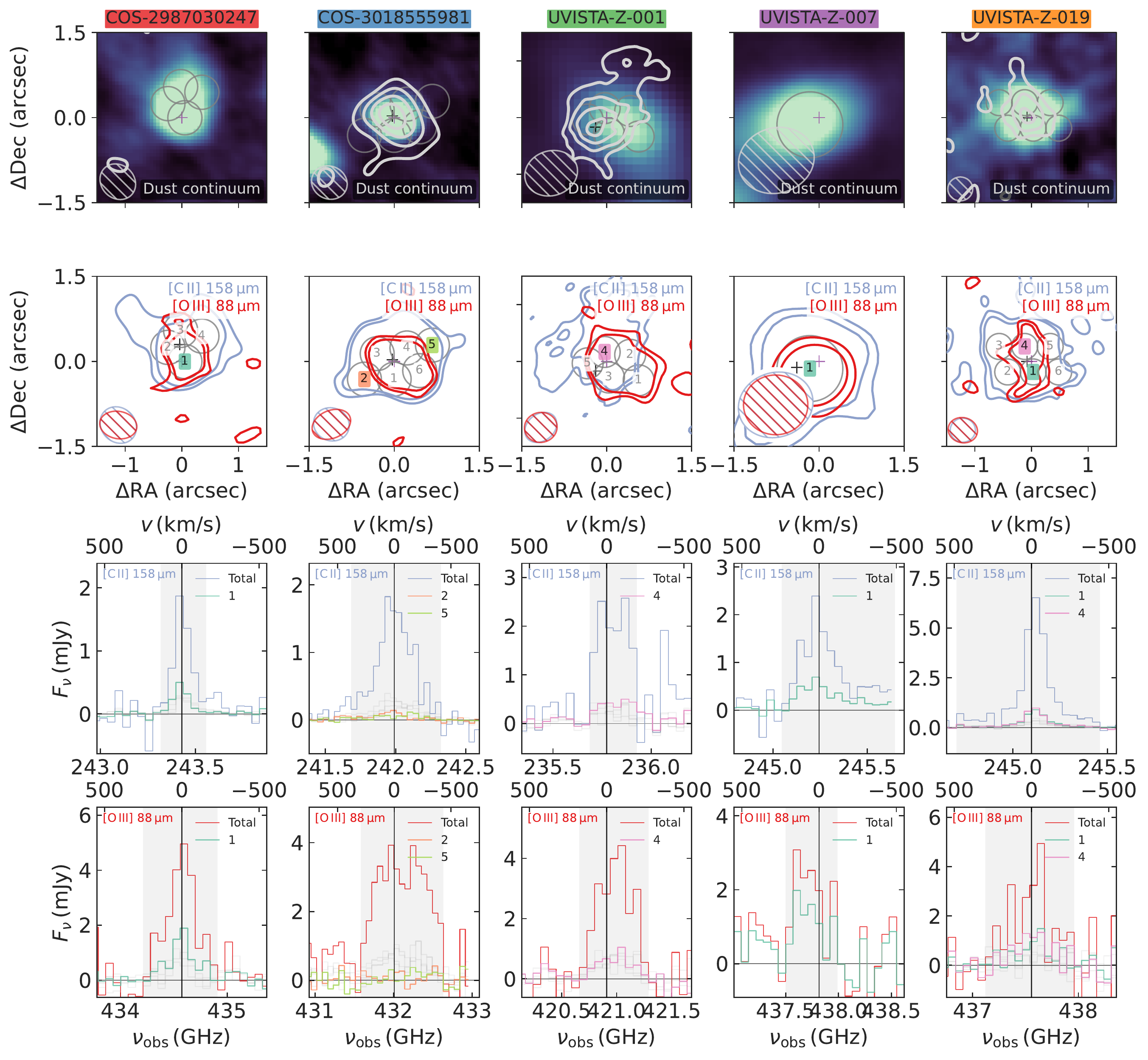}
	\caption[Beam placement for spatially resolved analysis.]{Beam placement for spatially resolved analysis. The top row of images shows the \gls{UV} continuum with contours of the $\ssim 160 \, \mathrm{\upmu m}$ dust continuum (starting from $2 \sigma$ and going up in steps of $1 \sigma$ for COS-2987030247 and UVISTA-Z-019, else $2 \sigma$). The peaks of the \gls{UV} and dust continua (if detected) are indicated with a black and purple cross, respectively. Contours (at $2 \sigma$ and $3 \sigma$) of the \CIILam\ and \OIIILam\ lines are shown in the second row (and again the \gls{UV} and dust peaks). The last two rows contain their spectra, both integrated over the entire $2 \sigma$ region (in the same colour as their contours) as well as for the individual regions. For each source, one or several regions are highlighted in the spectra and correspondingly by their number in the second row of images. The filled-in grey region indicates the spectral channels over which the spectra have been integrated.
	}
	\label{fig:Beam_placement_for_spatially_resolved_analysis}
\end{figure*}

\section{Spatially resolved analysis}
\label{ap:Spatially_resolved_analysis}

In \cref{fig:Beam_placement_for_spatially_resolved_analysis}, we show the placements of individual regions (circles with a radius $1.5$ times that of the mean circularised beam between the \CII\ and \OIII\ observations) that were used in the spatially resolved analysis (\cref{ssec:Discussion:OIII/CII_spatially_resolved_analysis}). The \OIIILam\ and \CIILam\ maps were created from imaging parameters that have been chosen to match beam sizes; see \cref{tab:Observations}; the dust continuum at $\ssim 160 \, \mathrm{\upmu m}$ has the same imaging parameters (and therefore nearly identical beam) as the \CII\ line. \gls{IR} luminosities were calculated similarly as discussed in \cref{sec:Discussion:Dust_properties}. Finally, the \gls{UV} continuum has been convolved with an effective beam found by the Richardson-Lucy algorithm to match the dust continuum \gls{PSF} (see also \cref{ssec:Discussion:OIII/CII_spatially_resolved_analysis}).

For each source, one or several regions are highlighted in the spectra and correspondingly by their number in the second row of images. From the spectra, it for instance becomes clear that, although there still seems to be some residual signal, the \OIII\ flux is weakest in region 2 and 5 of COS-3018555981.

%%%%%%%%%%%%%%%%%%%%%%%%%%%%%%%%%%%%%%%%%%%%%%%%%%

% Don't change these lines
\bsp	% typesetting comment
\label{lastpage}
\end{document}